\documentclass[entropy,preprints,accept,moreauthors,pdftex]{Definitions/mdpi} 
\usepackage{graphicx}
\usepackage{amsmath}
\usepackage{amsfonts}
\usepackage{amssymb}
\usepackage{mathtools}

\firstpage{1} 
\pubvolume{1}
\issuenum{1}
\articlenumber{0}
\pubyear{2021}
\copyrightyear{2020}
\externaleditor{Academic Editor: Firstname Lastname}
\datereceived{} 
\dateaccepted{} 
\datepublished{} 

\Title{Cryptocurrency Market Consolidation in 2020--2021}%MDPI: we changed hyphen to en dash, please confirm.

\TitleCitation{Cryptocurrency Market Consolidation in 2020--2021}

\Author{Jarosław Kwapień $^{1,}$*\orcidC{}, Marcin Wątorek $^{2}$\orcidA{}, Stanisław Drożdż $^{1,2}$\orcidB{}}

\AuthorNames{Marcin Wątorek, Stanisław Drożdż, Jarosław Kwapień}

\AuthorCitation{Kwapień, J.; Wątorek, M.; Drożdż, S.}

\address{%
$^{1}$ \quad Complex Systems Theory Department, Institute of Nuclear Physics, Polish Academy of Sciences, ul.~Radzikowskiego 152, 31-342 Krak\'ow, Poland; Stanislaw.Drozdz@ifj.edu.pl\\
$^{2}$ \quad Faculty of Computer Science and Telecommunications, Cracow University of Technology, ul.~Warszawska 24, 31-155 Krak\'ow, Poland; marcin.watorek@pk.edu.pl}

\corres{Correspondence: jaroslaw.kwapien@ifj.edu.pl}

\abstract{Time series of price returns for 80 of the most liquid cryptocurrencies listed on Binance are investigated for the presence of detrended cross-correlations. A spectral analysis of the detrended correlation matrix and a topological analysis of the minimal spanning trees calculated based on this matrix are applied for different positions of a moving window. The cryptocurrencies become more strongly cross-correlated among themselves than they used to be before. The average cross-correlations increase with time on a specific time scale in a way that resembles the Epps effect amplification when going from past to present. The minimal spanning trees also change their topology and, for the short time scales, they become more centralized with increasing  maximum node degrees, while for the long time scales they become more distributed, but also more correlated at the same time. Apart from the inter-market dependencies, the detrended cross-correlations between the cryptocurrency market and some traditional markets, like the stock markets, commodity markets, and Forex, are also analyzed. The cryptocurrency market shows higher levels of cross-correlations with the other markets during the same turbulent periods, in which it is strongly cross-correlated itself.}

\keyword{financial markets; cryptocurrencies; multiscale analysis; detrended cross-correlations; minimal spanning tree; COVID-19}

\begin{document}

\section{Introduction}

Over the past few years, two processes have had a particularly strong impact on financial markets: the emergence of the cryptocurrency market~\cite{gerlach2018,corbet2019,flori2019,bariviera2021,watorek2021a} and the COVID-19 pandemic~\cite{zhang2020,buszko2021,james2021a,james2021d,jimenez2021,maheu2021,song2022}. Each of these processes alone has already been a topic in numerous pieces of the scientific literature, but they also were studied together~\cite{drozdz2020a,mnif2020,conlon2020,demir2020,kristoufek2020,goodell2021,james2021b,james2021c,watorek2021a,watorek2021b}. Of particular interest in this context is how the ongoing pandemic is changing the cryptocurrency market and how this market position among the other financial and commodity markets undergoes an accelerated evolution. The cryptocurrency market is an interesting object for analysis from the perspective of complex systems, as it is a unique financial market whose establishment and evolution was entirely spontaneous with no intervening government or other regulatory institution. Thus, a process of the market's self-organization can be traced from the very beginning until the present. 

 As the cryptocurrency market properties are constantly evolving and they are still far from being fully identified and understood, there is heavy ongoing related research that points in various directions (see, for example,~\cite{bariviera2021} for comprehensive literature listing and pointing out several significant research voids). On the general level, the cryptocurrency markets are studied at an angle of trading security, the vulnerability to improper trading practices~\cite{gandal2018}, and the formation of demand~\cite{delahorra2019}. On the asset level, the fundamental aspects of the market processes that drive price discovery~\cite{urquhart2018,aalborg2019}, price fluctuations~\cite{drozdz2018,gkillas2018,katsiampa2019}, asset liquidity~\cite{godfrey2017}, and asset–asset correlations~\cite{dyhrberg2016,corbet2018} are studied from the investor's perspective in order to facilitate the optimal portfolio construction both inside the cryptocurrency market and across different markets, including the cryptocurrency one. An associated important direction of research is the possibility of market forecasting, which includes the approach developed in econophysics that is based on a search for evidence of the exogeneous and endogeneous market shocks, speculative bubbles, crashes, and their precursors~\cite{fry2016}. Among the voids, one can count the sparse analyses based on high-frequency data, the exaggerated focus on bitcoin (BTC) alone, and the insufficient attention paid to how different mining protocols can affect the related asset properties and how various legal regulations being (actually or potentially) imposed on the cryptocurrency markets can perturb both the mining and the trade~\cite{bariviera2021}.

From a perspective of their statistical and dynamical properties, the cryptocurrencies neither resemble regular currencies, like the US dollar (USD) or Chinese yuan (CNH), nor commodities, like gold or oil~\cite{baur2018,ferreira2020,manavi2020}. Among the major problems associated with  cryptocurrencies is their significant volatility. In consequence, even the largest and the most capitalized cryptocurrency, BTC, is considered an asset that resides at the interface between a standard financial asset and a speculative one~\cite{kristoufek2015}. Most studies of the cryptocurrency market relations with the traditional markets reported in the literature point  to  relative independence of the cryptocurrencies (see, for example,~\cite{corbet2018,ji2018,drozdz2020a}). However, there were also  some reports concluding that there are temporary or stable cross-correlations or even causality between the major cryptocurrencies and some regular currencies, like TRY~\cite{manavi2020} and some Asian currencies, like BHT, CNH, and TWD~\cite{corelli2018}, as well as between the cryptocurrencies and commodities~\cite{ji2018}.

As a new system, it took several years for the cryptocurrency market to reveal any signatures of maturity, like the market efficiency~\cite{urquhart2016,bariviera2017}. However, already prior to the crash of April 2018, its statistical properties became similar to the properties of the other markets, among which there were the financial stylized facts (the power-law tails of the return distributions, volatility clustering, etc.)~\cite{bariviera2017,katsiampa2019,watorek2021a} and some other complexity traits, like multifractality~\cite{drozdz2018}, and, in some aspects, it started to resemble Forex~\cite{drozdz2018,drozdz2019}. On the other hand, one of the interesting facts about the cryptocurrency market's inner structure is that, unlike other financial markets where, typically, the highly capitalized assets have spillover effects on the less capitalized ones, here the less capitalized assets are able to influence the evolution of the highly capitalized ones. This can lead to more a complex structure than a typical structure of the other markets, where causality is unidirectional~\cite{yi2018,aste2019,ferreira2019,aslanidis2021}.

These and other similarities and differences opened space for a concern, whether bitcoin and other cryptocurrencies may be considered as a safe haven during market turmoils or whether they may be used to hedge against the traditional assets. Although the literature on this issue is growing, the conclusions are mixed: BTC and the other major cryptocurrencies are sometimes indicated as good candidates for a safe haven~\cite{demir2020,conlon2020,corbet2020,goodell2021,mariana2021} but the opposite can also be suggested~\cite{kristoufek2015,shahzad2019,conlon2020,conlon2020b,kristoufek2020,lahmiri2020,grobys2021,jiang2021}, depending on the analyzed data. Sometimes the answer can even be conditional: ``yes'' to a safe haven, ``no'' to a hedge~\cite{wang2019}. An important risk factor of BTC and other cryptocurrencies that acts against their use for hedging is their possible lack of fundamental value~\cite{cheah2015}.

As regards the asset–asset correlations among the cryptocurrencies, it was shown that, besides a trend going towards the stronger market cross-correlations, the cryptocurrencies reveal a cyclic amplification of volatility connectedness during periods of economic instability or external shocks. However, BTC does not play a central role in driving market volatility~\cite{yi2018}. A different study applying different methodologies (principal component analysis, cross-sectional dependence, and vector autoregression framework) confirmed this finding and extended it from volatility to returns~\cite{aslanidis2021}. Another work reported the increased cross-correlations among the cryptocurrencies after the bubble of 2017 as compared to the earlier period by using the detrended fluctuation analysis~\cite{ferreira2019}. The highly capitalized cryptocurrencies show statistically significant time-lagged autocorrelations that may indicate substantial market inefficiency (although not necessarily usable for profit-making)~\cite{ferreira2020}. All these works analyzed very small sets of assets, however, which significantly limited the market insight they were able to provide. A more comprehensive study, which considered over 50 cryptocurrencies, also brought  more diversified results, and identified some assets that were statistically and dynamically different than the others (these were tether, holo, maker, NEM, and nexo)~\cite{james2021c}.

In our former publications we thoroughly analyzed the cryptocurrency market evolution from its early stages of development  to the current, relatively mature phase. In the Ref.~\cite{drozdz2020b} we reported that the cryptocurrency dynamics over the years 2016--2019 displayed signatures of decoupling from dynamics of the regular currencies~\cite{drozdz2020b}. In the Refs.~\cite{drozdz2020a,watorek2021a} we analyzed the cryptocurrency market properties during the pandemic onset (January 2019--October 2020). We showed that before the pandemic, over the years 2018--2019, the evolution of the cryptocurrency market was largely independent from the evolution of the traditional markets. We interpreted this independence as a consequence of a quiet period on the traditional markets and a disparity in the market capitalization: the cryptocurrency market was too small to perturb other markets, while they were too tranquil then to induce any turmoil among the cryptocurrencies. However, in the second half of January 2020, at the moment when the first COVID-19 case was reported in the United States, some cryptocurrencies responded and thus lost their independence. For example, BTC gained positive cross-correlation with JPY, CHF, and gold, which are considered as a financial safe haven, and negative cross-correlation with other major assets, while ETH preserved its independent dynamics longer. Later, during the outburst of the first wave of COVID-19  in April 2020, the cryptocurrencies underwent a crash together with all the major markets, except for a few regular currencies like JPY. This state of cross-market coupling continued in the months that followed, both at the moments of the subsequent pandemic waves and the market rallies. Our analyses ended in the middle of the third pandemic wave before the introduction of anti-COVID vaccines, thus we could not report on how the markets would respond to a decreased pandemic risk. From this angle, our present analysis can be viewed inter alia as a continuation of those previous works based on a new data set.

In the following, we will report on our study of a set of the most liquid cryptocurrencies whose high-frequency price quotes cover the last 21 months. We will apply the generalized detrended cross-correlation analysis~\cite{podobnik2008,zhou2009,zebende2011,kwapien2015} and study the spectral properties of a detrended correlation matrix, as well as the topological properties of its network representation. In the context of the current cryptocurrency research, our main objectives are (1) to look into the most recent data that have not been covered by other works yet, and compare results with the earlier ones, (2) to consider  a  set of assets that is wide as  possible provided the available data quality, and (3) to apply a methodology that is rarely used in this context, that is, the $q$-dependent cross-correlation analysis that is able to filter data according to its magnitude. In Section 2 we will briefly recollect  the related formalism, in Section 3 we present and discuss the main results, and in Section 4 we will present the summary and conclusions. 

\section{Methods}

Data from the cryptocurrency market, which is characterized by volatility that exceeds volatility of the traditional markets, are not well-suited to being studied by means of the standard correlation formalism based on the Pearson correlation~\cite{pearson1895} that requires data stationarity. Thus,  methods based on signal detrending are advised~\cite{peng1994,podobnik2008}.

The $q$-dependent detrended correlation coefficient $\rho_q(s)$ was proposed in~the Ref. \cite{kwapien2015} to quantify the detrended cross-correlations between two, typically non-stationary time series $\{x(i)\}_{i=1,...,T}$ and $\{y(i)\}_{i=1,...,T}$ of length $T$. Let these time series be divided into $M_s$ boxes of length $s$ starting from its opposite ends (thus, there are $2 M_s$ boxes total). In each box, the data points are subject to integration and polynomial trend removal:
\begin{equation}
X_{\nu}(s,i) = \sum_{j=1}^i x(\nu s + j) - P^{(m)}_{X,s,\nu}(i), \qquad
Y_{\nu}(s,i) = \sum_{j=1}^i x(\nu s + j) - P^{(m)}_{Y,s,\nu}(i),
\end{equation}
where the polynomials $P^{(m)}$ of order $m$ are applied. The next step is calculation of the local residual variances and covariance:
\begin{align}
f^2_{\rm XX} (s,\nu) = \sum_{i=1}^s (X_{\nu}(s,i) - \bar{X}_{\nu}(s))^2, \qquad
f^2_{\rm YY} (s,\nu) = \sum_{i=1}^s (Y_{\nu}(s,i) - \bar{Y}_{\nu}(s))^2, \\
f^2_{\rm XY} (s,\nu) = \sum_{i=1}^s (X_{\nu}(s,i) - \bar{X}_{\nu}(s)) (Y_{\nu}(s,i) - \bar{Y}(s)),
\end{align}
where $\bar{X}$ and $\bar{Y}$ denote the local mean of $X$ and $Y$, respectively. These quantities are used to define a family of the fluctuation functions of order $q$:
\begin{align}
\label{eq::fq.zz}
F^{(q)}_{\rm XX} (s) = {1 \over 2 M_s} \sum_{\nu=0}^{2 M_s-1} \left[ f^2_{\rm XX} (s,\nu)\right]^{q/2}, \qquad F^{(q)}_{\rm YY} (s) = {1 \over 2 M_s} \sum_{\nu=0}^{2 M_s-1} \left[ f^2_{\rm YY} (s,\nu)\right]^{q/2}, \\
F^{(q)}_{\rm XY} (s) = {1 \over 2 M_s} \sum_{\nu=0}^{2 M_s-1} \textrm{sign} \left[ f^2_{\rm XY}(s,\nu)\right] |f^2_{\rm XY} (s,\nu)|^{q/2}.
\label{eq::fq.xy}
\end{align}
The sign function in Eq.~(\ref{eq::fq.xy}) preserves the information that is otherwise  lost after taking the modulus of $f^2_{\rm XY}(s,\nu)$, while the modulus itself excludes a possibility of obtaining complex values of the covariance $f^2_{\rm XY}$ raised to a real power $q/2$~\cite{oswiecimka2014,kwapien2015}.
The $q$-dependent detrended correlation coefficient is defined by the following formula:
\begin{equation}
\rho_q^{\rm XY}(s) = {F^{(q)}_{\rm XY}(s) \over \sqrt{F^{(q)}_{\rm XX}(s) F^{(q)}_{\rm YY}(s)}},
\label{eq::rhoq}
\end{equation}
which generalizes for any $q$ the standard ($q=2$) detrended correlation coefficient $\rho_{\rm DCCA}$~\cite{zebende2011}. The parameter $q$ plays the role of a filter weighting the boxes $\nu$ in the sums in Eqs.~(\ref{eq::fq.zz}) and (\ref{eq::fq.xy}) by their variance/covariance magnitudes. For $q>2$, the boxes with large signal fluctuations are given higher weights with respect to the $q=2$ case, while for $q<2$ the boxes with small fluctuations contribute more than for $q=2$. Therefore, by applying $\rho_q$, one can learn which fluctuations are the source of the observed detrended correlation of the time series.

For a set of $N$ parallel time series indexed by $i$, the $q$-dependent correlation coefficient can be calculated for each time series pair $(i,j)$ ($i,j=1,...,N$), and a $q$-dependent detrended correlation matrix ${\bf C}_q(s)$ with the entries $\rho^{(i,j)}_q(s)$ can be created, as well as a $q$-dependent metric distance matrix ${\bf D}_q(s)$ whose entries are
\begin{equation}
d^{(i,j)}_q(s)=\sqrt{2(1-\rho^{(i,j)}_q(s))}.
\label{eq::metric-distance}
\end{equation}
The matrix ${\bf D}_q(s)$ can then be used to create a weighted graph, where nodes labelled by $i=1,...,N$ represent the time series and $N(N-1)/2$ edges connecting the nodes $i,j$ are attributed the weights equal to $d^{(i,j)}_q(s)$. A subset of the complete graph, consisting of all $N$ nodes and only $N-1$ edges that minimize the weight sum, is a $q$-dependent detrended minimum spanning tree ($q$MST)~\cite{kwapien2017}. This tree can be constructed by means of the Prim algorithm, for instance~\cite{prim1957}. However, although the very algorithm is the same, such a tree differs from the standard approach that uses the Pearson correlation coefficient and a corresponding Pearson correlation matrix (see, for example,~\cite{zieba2019,drozdz2020b} for such a standard approach applied to the cryptocurrency market). 

A data set of  interest is the 1 min price quotations of the 80 cryptocurrencies that were among the most actively traded ones on the Binance platform~\cite{binance} over the period from 1 January 2020 to 1 October 2021. The quotes are expressed in USD Tether (USDT) that is a stablecoin linked to the US dollar and its value is \$1.00 by design~\cite{tether}. Each time series of the price quotations is 921,600 points long and covers 640 trading days (the Binance platform is active 24 hours a day and 7 days a week). All the assets used in this study are listed in Appendix (Tab.~\ref{tab::ticker_list}).

\section{Results and Discussion}

The price quotation time series $p_i(t_m)$, where $m=1,...,T$ and $i$ stands for a given cryptocurrency ticker, were first transformed to the time series of logarithmic returns $R_{\rm X}(t_m)=\ln p_i(t_{m+1})-\ln p_i((t_m)$ and then normalized to zero mean and unit variance, which is a standard procedure. Then, for each pair of cryptocurrencies $(i,j)$, the $q$-dependent detrended cross-correlation coefficient $\rho_q^{(i,j)}(s)$ given by Eq.~(\ref{eq::rhoq}) was determined for a number of time scales $s$ from $s=10$ min to $s=360$ min and different values of the filtering parameter $q$. In what follows, we will present results obtained for $q=1$, which corresponds to a situation where the small fluctuation period variances in Eqs.~(\ref{eq::fq.zz}) and (\ref{eq::fq.xy}) are amplified relatively to the large ones, and for $q=4$, which corresponds to the opposite situation. Thus, we can consider the asset cross-correlations for the quiet and turbulent periods in a separate manner.

% Figure 1

\begin{figure}[H]
\centering
\includegraphics[width=0.9\textwidth]{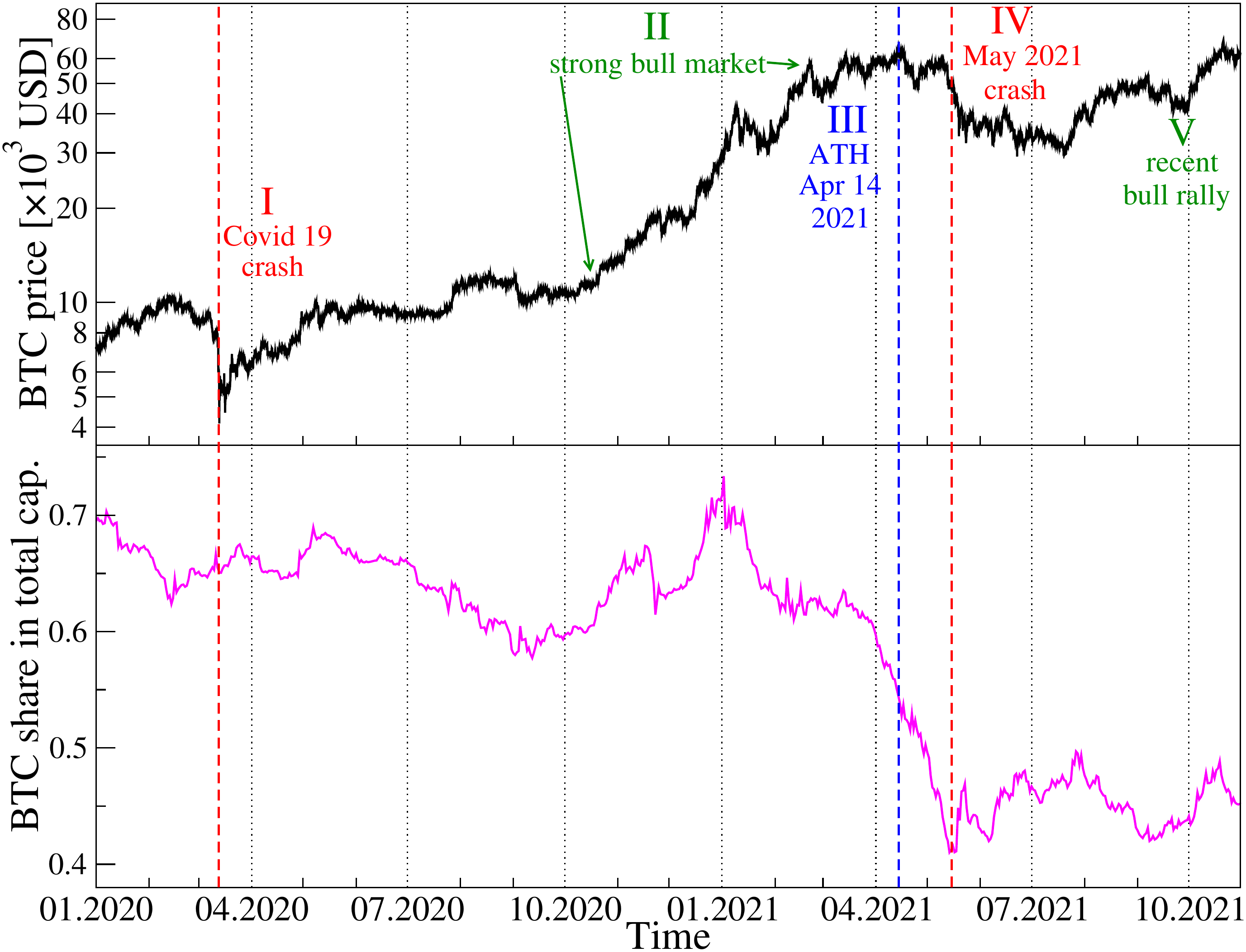}
\caption{Price evolution of bitcoin (BTC) expressed in  US dollars (black) and the BTC share in the total cryptocurrency market capitalization (magenta) over the period from 1 January 2020 to 31 October  2021. Characteristic events are indicated by vertical dashed lines and Roman numerals: COVID-19 crash in March 2020 (event I), strong bull market on cryptocurrency valuation October 2020--April 2021
(event II), all-time high on April 14, 2021  (event III), the May crash on the cryptocurrency market (event IV), and recent rally with new all-time high on 20 October 2021 (event V).}
\label{fig::BTC.evolution}
\end{figure}

Before we start a presentation of our results, in Fig.~\ref{fig::BTC.evolution} we show the historical data of the BTC price in USD in the years 2020--2021 together with the BTC share in the total cryptocurrency market capitalization over the same period. Among the most characteristic events for BTC was the crash on Mar 13, 2020 related to the COVID-19 pandemic onset in the United States, when BTC surged below 4107 USD, a long rally that started in October 2020 and ended on Apr 14, 2021 with then the all-time-high equal to 64,830 USD, a subsequent drop-down phase that ended on Jul 20, 2021 at 29,324 USD, and the next all-time-high on 20 October 2021 equal to 66,961 USD. As the BTC has been priced higher and higher, its share in the total market capitalization drops down steadily from about 70\% in January 2020 to below 45\% in October 2021, which seems to be inevitable if the number of the actively traded cryptocurrencies grows quickly.

\begin{figure}[H]
\widefigure
\includegraphics[width=0.8\textwidth]{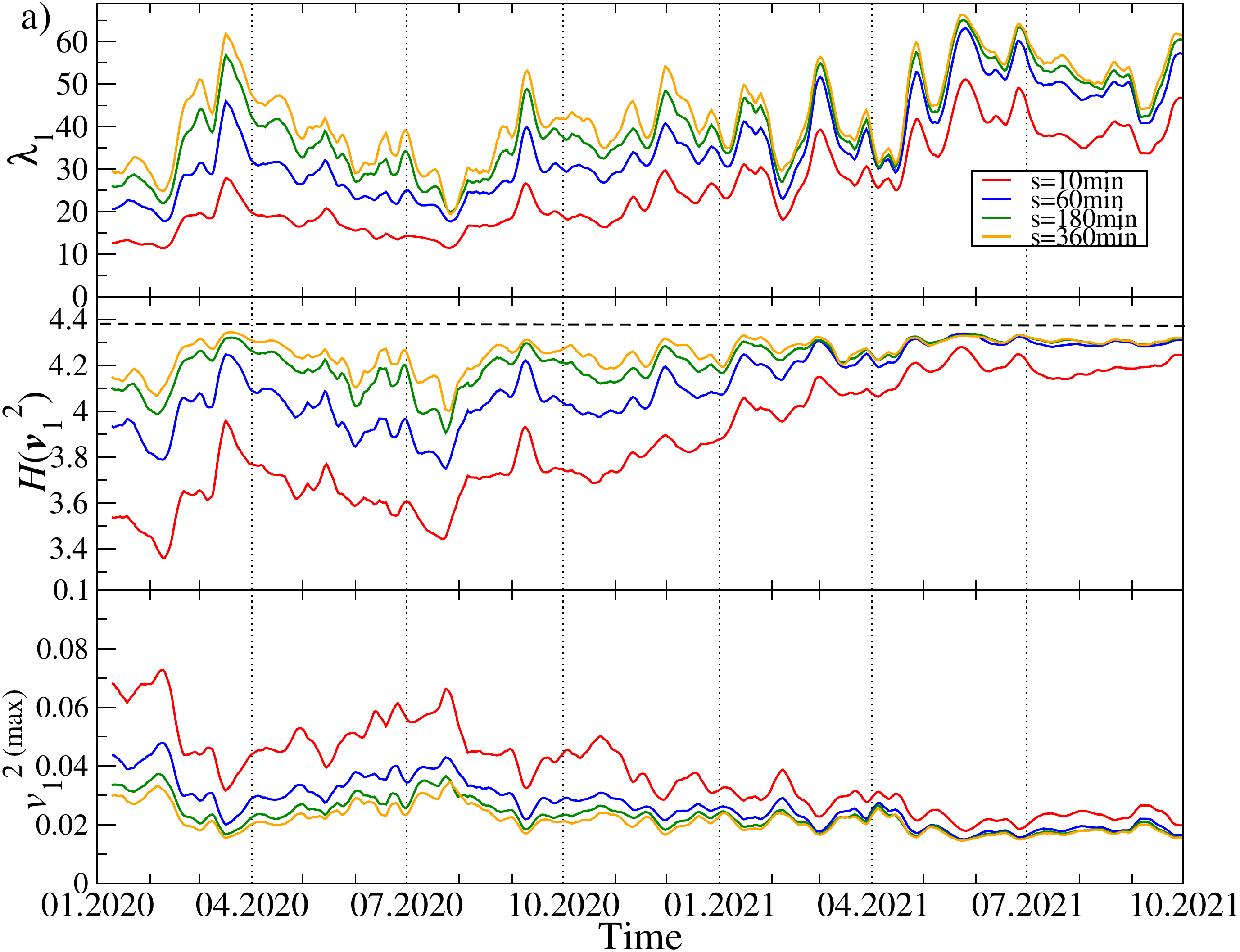}

\vspace{0.2cm}
\includegraphics[width=0.8\textwidth]{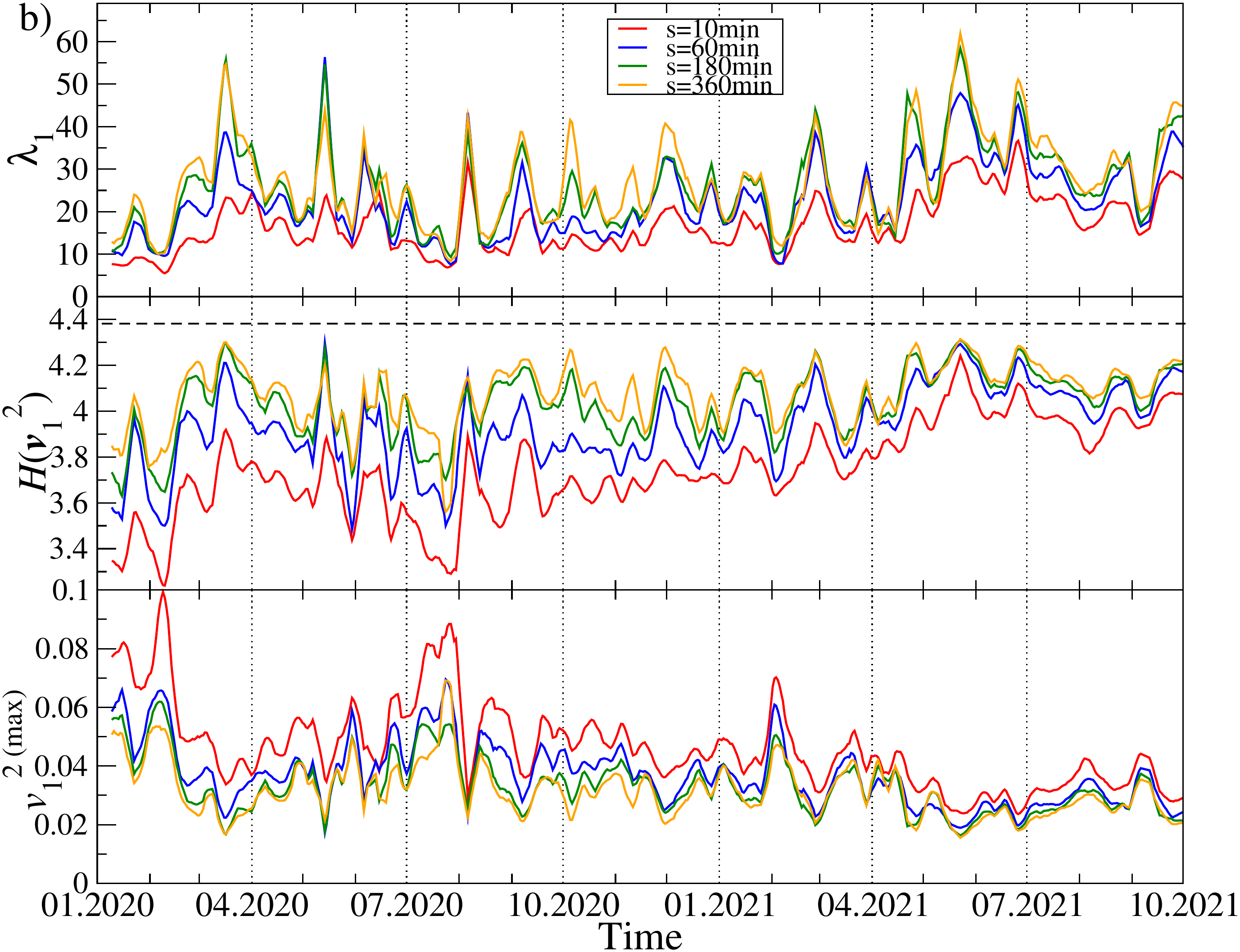}
\caption{Time evolution of the selected spectral characteristics of the $q$-dependent detrended correlation matrix ${\bf C}_q(s)$ for $q=1$ (a) and $q=4$ (b). A moving window of a length of 7 days shifted by 1 day was applied for sample values of the scale: $s=10$ min (red), $s=60$ min (blue), $s=180$ min (green), and $s=360$ min (orange). The largest eigenvalue $\lambda_1$ (top panels in (a) and (b)), the Shannon entropy $H({\bf v}_1)$ of the squared eigenvector components $v_1(j)$ with $j=1,...,N$ (middle panels), and the squared maximum component of the eigenvector ${\bf v}_1$ associated with $\lambda_1$ (bottom panels) are shown. The cryptocurrency prices are expressed in USDT.}
\label{fig::eigenspectrum.lambda1.complete}
\end{figure}

\begin{figure}[H]
\widefigure
\includegraphics[width=0.8\textwidth]{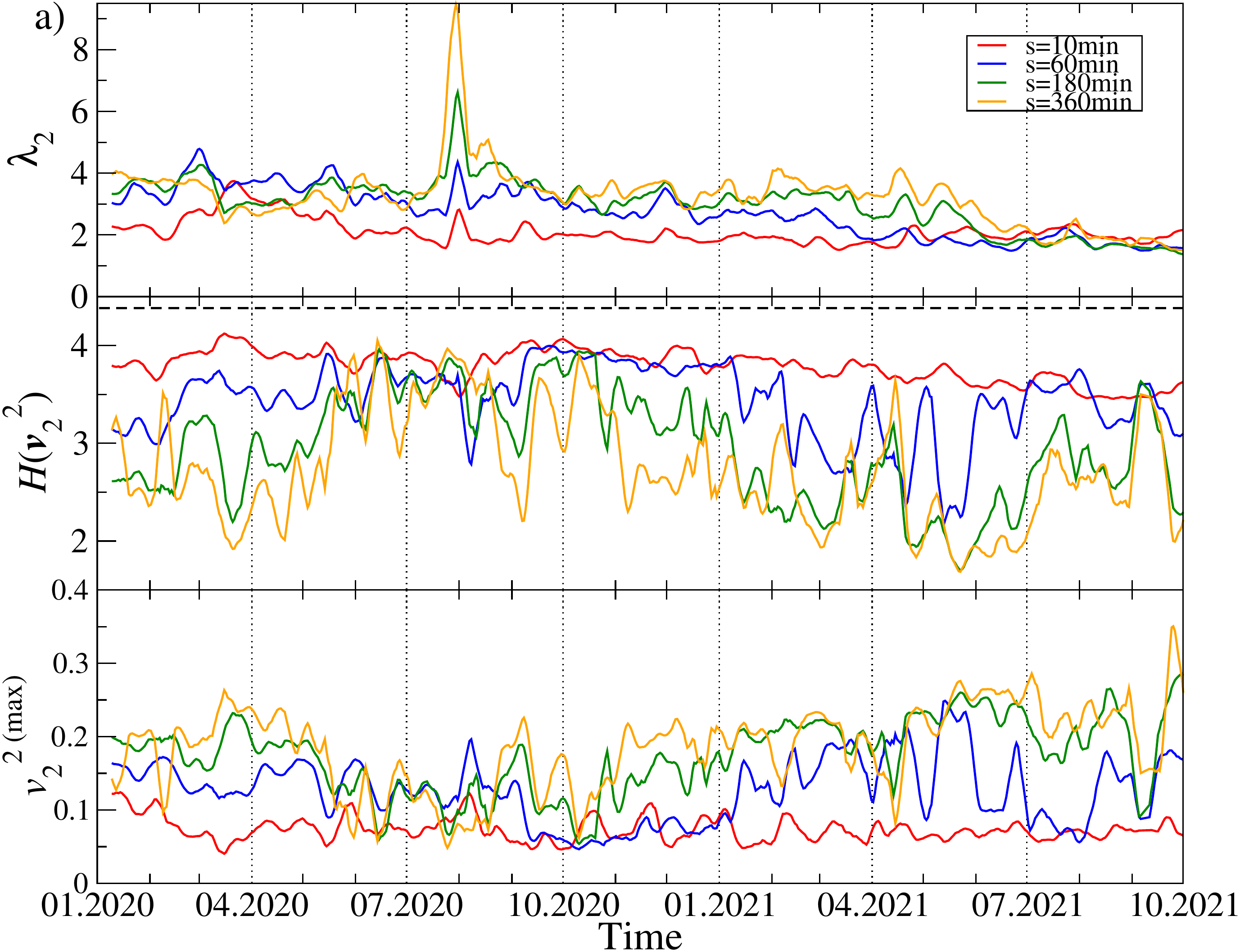}

\vspace{0.2cm}
\includegraphics[width=0.8\textwidth]{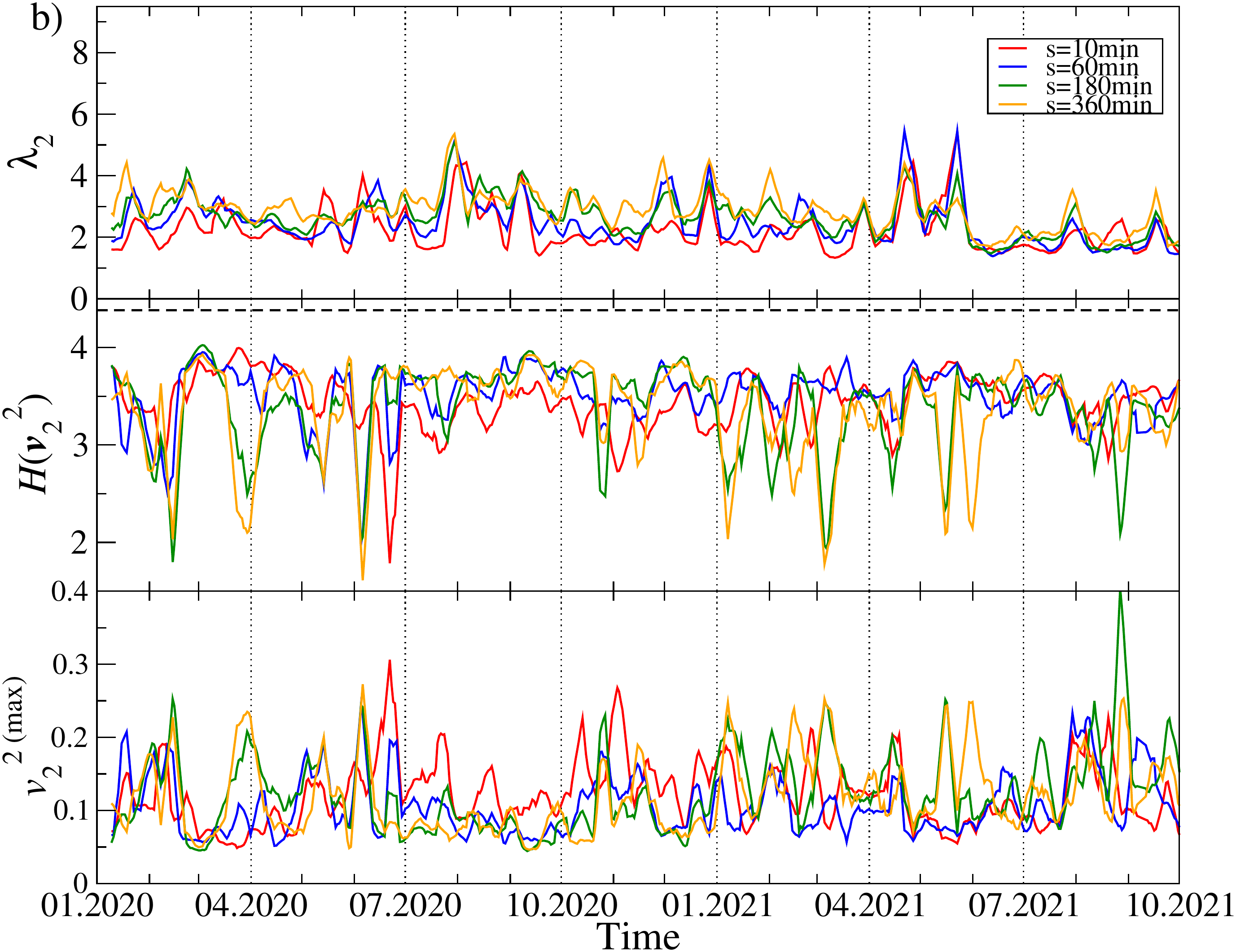}
\caption{Time evolution of the selected spectral characteristics of ${\bf C}_q(s)$ (continuing). As in Fig.~\ref{fig::eigenspectrum.lambda1.complete}, two cases are shown: $q=1$ (a) and $q=4$ (b). A moving window of length 7 days shifted by 1 day was applied for sample values of the scale: $s=10$ min (red), $s=60$ min (blue), $s=180$ min (green), and $s=360$ min (orange). The second largest eigenvalue $\lambda_2$ (top panels), the Shannon entropy $H({\bf v}_2)$ of the squared eigenvector components $v_2(j)$ with $j=1,...,N$ (middle panels), and the squared maximum component of the eigenvector ${\bf v}_2$ associated with $\lambda_2$ (bottom panels) are shown.}
\label{fig::eigenspectrum.lambda2.complete}
\end{figure}

% start a new page without indent 4.6cm
\clearpage
\end{paracol}
\nointerlineskip
\begin{paracol}{2}
%\linenumbers
\switchcolumn

Since for $N=80$ cryptocurrencies there are $\mathcal{N}=N(N-1)/2=3160$ cryptocurrency pairs that have to be considered, it is convenient to analyze the whole set collectively by means of the spectral analysis of the $N \times N$ $q$-dependent detrended correlation matrix ${\bf C}_q(s)$, whose entries are the coefficients $\rho_q^{(i,j)}(s)$. We can diagonalize it and calculate its eigenvalues $\lambda_i$ and eigenvectors ${\bf v}_i$ (with $i=1,...,N$):
\begin{equation}
{\bf C}_q(s) {\bf v}_i^{(q)}(s) = \lambda_i^{(q)}(s) {\bf v}_i^{(q)}(s).
\label{eq::eigenspectrum}
\end{equation}
The eigenvalues are ordered typically from the largest one ($i=1$) to the smallest one ($i=N$).
(For simplicity, from now on we will omit the parameters $q$ and $s$ when dealing with the eigenvalues and eigenvectors of ${\bf C}_q(s)$. Their value will be known from the context.)

For the financial markets, a typical eigenvalue spectrum of the Pearson-coefficient-based correlation matrix consists of a large $\lambda_1$ that is separated from the remaining eigenvalues by a considerable gap and corresponds to the average behaviour of the considered assets (the so-called market factor), a few elevated non-random eigenvalues that correspond to subsets of related assets (e.g., representing companies from the same industry or currencies from the same geographical region), and a bulk of mean eigenvalues that correspond to random fluctuations and, essentially, carry no genuine information. Here we use the detrended correlation coefficient $\rho_q$ instead of the Pearson coefficient~\cite{pearson1895}, but our experience shows that the corresponding matrix ${\bf C}_q$ reveals similar spectral properties~\cite{kwapien2017}. The largest eigenvalue $\lambda_1$ is associated with a maximally delocalized eigenvector ${\bf v}_1$ with many significant components, while the eigenvectors representing smaller eigenvalues are more localized, that is,  few components are significant. The eigenvector structure is usually expressed by the inverse participation ratio or the localization length~\cite{kwapien2012}, but here we apply the Shannon entropy defined by
\begin{equation}
H({\bf v}_i) = - \sum_{j=1}^N p_i(j) \ln p_i(j),
\end{equation}
with $p_i(j)=v_i^2(j)$ (the eigenvectors are normalized to unit length, so that $\sum_{j=1}^N v_i^2(j)=1$). If the 
eigenvector is maximally delocalized and all its components are equal to each other, the Shannon entropy assumes its maximum value: $H({\bf v}_i)=\ln N$, while if there is only a single non-zero component, the entropy vanishes: $H({\bf v}_i)=0$. Entropy can thus serve as a measure of vector localization.

% Figure 2

In order to track the evolution of the asset–asset detrended cross-correlations, we apply a moving window of size of 7 days (10,080 data points), which was shifted by a daily step (1440 data points) along the time series. For each window position $t$, based on the 80 time series of price returns, we create a detrended correlation matrix ${\bf C}_q(s,t)$ for a few selected values of $q$ ($q=1$ and $q=4$) and $s$ ($s=10$ min, $s=60$ min, $s=180$ min, and $s=360$ min). Next we diagonalize $C_q(s,t)$ and derive a complete set of the eigenvalues $\lambda_i(t)$ and eigenvectors ${\bf v}_i(t)$. Fig.~\ref{fig::eigenspectrum.lambda1.complete} exhibits $\lambda_1(t)$, $H({\bf v}_1(t))$, and the largest squared component $v_1^{({\rm max})}(t)$ of the eigenvector ${\bf v}_1(t)$ for different time scales $s$ and different values of the filtering parameter $q$. By increasing $s$, we also obtain a systematically increasing $\lambda_1(t)$, which reflects the increasing strength of the mean asset–asset detrended cross-correlations for the longer time scales $s$. This is a well-known property of the financial and commodity markets and it is called the Epps effect~\cite{epps1979,kwapien2005,drozdz2010,drozdz2019}. This effect has already been observed on the cryptocurrency market and reported, for example, in the Ref.~\cite{watorek2021a}. It is a consequence of the fact that what dominates the price evolution on short time scales is noise: it takes time to spread a piece of information among the assets, especially if the asset liquidity is small like in the case of the cryptocurrencies. Therefore, only on the sufficiently long time scales, the cross-correlations are able to be built up to a full extent. 

Another observation is that the difference in correlation strength between $s=10$ min and $s=360$ min is much stronger for $q=1$ than for $q=4$; the correlation strength for large scales is also significant then. The behavior of $\lambda_1$ is also different: in the case of $q=1$, periods with a large value of $\lambda_1$ are accompanied by periods of moderate value, but there are also few periods with relatively small values of the largest eigenvalue. In turn, for $q=4$ the $\lambda_1(t)$ evolution consists of large, but short "bursts" separated by small background values. In the latter case, $\lambda_1$ is more sensitive to changes.
Looking at the $\lambda_1(t)$ chart for $q=1$ and the shorter $s$ time scales, two characteristic epochs can be distinguished: (1) more or less until October 2020, we observe a horizontal trend, where the average value of $\lambda_1$ does not change much, and (2) from October 2020 to mid-2021, a strong upward trend is noticeable. This is confirmed by looking at the Shannon entropy panel, where the behavior of this quantity is very similar. This means that from the fall of 2020 to mid-2021, there was a gradual increase in the strength of the market correlation and more cryptocurrencies began to behave in a similar way. It can be said that the market has consolidated. In the third quarter of 2021, this trend was halted, $\lambda_1$ began to decrease slightly, and $H({\bf v}_1)$ was saturated close to its maximum allowed value of approximately 4.38. Understandably, as the delocalization of the vector ${\bf v}_1$ increases, the value of its largest component $v_1^{({\rm max})}$ decreases (see Fig.~\ref{fig::eigenspectrum.lambda1.complete}). 

% start a new page without indent 4.6cm
\clearpage
\end{paracol}
\nointerlineskip
\begin{paracol}{2}
%\linenumbers
\switchcolumn

% Figure 3

Fig.~\ref{fig::eigenspectrum.lambda2.complete} shows the changes over time of the second largest eigenvalue $\lambda_2$, the entropy of the components of the corresponding eigenvector ${\bf v}_2$ and the changes in the value of the largest component $v_2^{({\rm max})}$ of this vector. For both $q=1$ and $q=4$, the value of $\lambda_2$ is much lower than the value of $\lambda_1$, which results from a much smaller number of significant eigenvector components: entropy is lower than 4, and for $q=1$, in the vast majority of windows, its value decreases as $s$ increases, which is the opposite of the $\lambda_1$ case. For $q=4$, we do not observe such an effect. With $q=1$, the global maximum of $\lambda_2$ falls in July 2020, when its value more than doubled if compared to other time intervals. Simultaneously, $\lambda_1$ reached one of its lowest values, as did $H({\bf v}_1)$. At the same time, the entropy for ${\bf v}_2$ did not change much from its typical value, but then and in the preceding period $H({\bf v}_2$ was similar for different time scales. For large fluctuations ($q = 4$), the maximum $\lambda_2$ also occurred in the same period, but was not as unique as for the smaller fluctuations ($q=1$), because $\lambda_2$ reached equally high magnitude in April and May 2021. However,  $\lambda_1$ for $q=4$ also had its maxima at the same moments. This means that briefly in July 2020, there was  a strong correlation of a small group of cryptocurrencies, and this mainly concerned small and medium fluctuations in their price, while the market as a whole was in a decoupling stage. In turn, in April and May 2021 there was a stronger than usual correlation of the entire market, with large fluctuations being particularly strongly correlated. As for the largest component of the vector ${\bf v}_2$ and $q=1$, we do not observe systematic changes in its value for the short time scales, while for the long ones, starting from autumn 2020, there is a growing trend that ends in mid-2021. This increase in $v_2^{({\rm max})}$ suggests that one of the cryptocurrencies increased its dominance over other cryptocurrencies at that time. This behavior differs from the behavior of the analogous measure described above in the case of the vector ${\bf v}_1$, where there was a clear decrease.

% start a new page without indent 4.6cm
\clearpage
\end{paracol}
\nointerlineskip
\begin{paracol}{2}
%\linenumbers
\switchcolumn

Since a sum of all the eigenvalues must equal trace of ${\bf C}_q$ with ${\rm Tr}{\bf C}_q=N$, the high values of $\lambda_1$ take a significant part of each time series variance. This can suppress all the other eigenvalues with $\lambda_2$ in particular and can also have a strong impact on the eigenvector ${\bf v}_2$. We thus prefer to look at these quantities once more after removing the variance contribution of $\lambda_1$ from the original time series of returns. In order to accomplish this, we created an eigensignal representing $\lambda_1$ as a sum of the original time series weighted by the corresponding eigenvector components $z_1(t_m)=\sum_{j=1}^N v_1(j) r_j(t_m)$, where $r_j(t_m)$ are the normalized returns of a cryptocurrency $j$ at time $t_m$, $m=1,...,T$. We then least-square fit the eigensignal $\{z_1(t_m)\}$ to each original time series $\{r_j(t_m)\}$ and subtract the fitted component from $\{r_j(t_m)\}$. What remains then is a residual signal $\{r_j^{({\rm res})}\}$, which does not comprise any contribution from $\{z_1(t_m)\}$ and, thus, also from $\lambda_1$:
\begin{equation}
r_i^{({\rm res})}(t_m) = r_i(t_m) - \alpha_i z_1(t_m) - \beta_i,
\end{equation}
where $\alpha_i,\beta_i$ are the parameters of a linear fit. Finally, we calculate the coefficients $\rho_q^{(i,j)}(s)$ for all the cryptocurrency pairs $(i,j)$ and form a residual $q$-dependent detrended correlation matrix ${\bf C}_q^{({\rm res})}(s)$. After diagonalising it, we obtain its eigenvalues $\lambda_i^{({\rm res})}$ and eigenvectors ${\bf v}_i^{({\rm res})}$. We repeat this procedure a few times for different scales $s$ and filtering parameters $q$. Fig.~\ref{fig::eigenspectrum.lambda1.nomax} collects the results.

% Figure 4

Now the largest eigenvalue $\lambda_1^{({\rm res})}$, which inherits some information stored previously in $\lambda_2$ but without the former clear impact of $\lambda_1$, is not suppressed any more and, for $q=1$, it shows richer behaviour with more fluctuations and more pronounced maxima (see Fig.~\ref{fig::eigenspectrum.lambda1.nomax}(a)). Interestingly, the large maximum of $\lambda_2$ observed in Fig.~\ref{fig::eigenspectrum.lambda2.complete}(a) in July 2020 disappeared almost completely here and was replaced by a series of pronounced maxima in February, March, September, and December 2020, and a smaller one in May 2021. They are the more visible the longer time scale is considered. From a present perspective, the unique maximum of $\lambda_2$ in July 2020 might solely be a product of a relatively small value of $\lambda_1$ in that moment, which was unable to suppress $\lambda_2$ to its overall level of 4. 

As regards the Shannon entropy, three phases can be distinguished: (1) from January to May 2020, (2) from May 2020 to April 2021, and (3) from May to October 2021. In the first and third phases there is no difference in $H({\bf v}_1^{({\rm res})})$ if we consider different scales $s$, while during the second phase, which largely overlapped with the bull market, the entropy fluctuates in time and increases with increasing $s$. However, its saturation level for $s=360$ min in this phase is comparable with the analogous level in the other phases -- this is because $H({\bf v}_1^{({\rm res})})$ for $s=10$ min can be much smaller in phase (2) than in phases (1) and (3). Dissimilarity between the phases is observed also for $v_1^{({\rm res})(max)}$: in phase (2) its value is substantially elevated as compared with the phases (1) and (2). These outcomes suggest that the eigenvector ${\bf v}_1^{({\rm res})}$ became delocalised and some cryptocurrency used to contribute more to this eigenvector during phase (2) than the other cryptocurrencies did.

For $q=4$ (Fig.~\ref{fig::eigenspectrum.lambda1.nomax}(b)), both $H({\bf v}_1^{({\rm res})})$ and $v_1^{({\rm res})(\rm max)}$ fluctuate over the whole analysed period more than it is observed for $q=1$. The largest residual eigenvalue for $q=4$ displays local maxima in the same moments as for $q=1$, but their height varies. Apart from the maxima, typical fluctuations of $\lambda_1^{({\rm res})}$ are smaller in 2021 than they used to be in 2020.

% start a new page without indent 4.6cm
\clearpage
\end{paracol}
\nointerlineskip
\begin{figure}[H]
\widefigure
\includegraphics[width=0.75\textwidth]{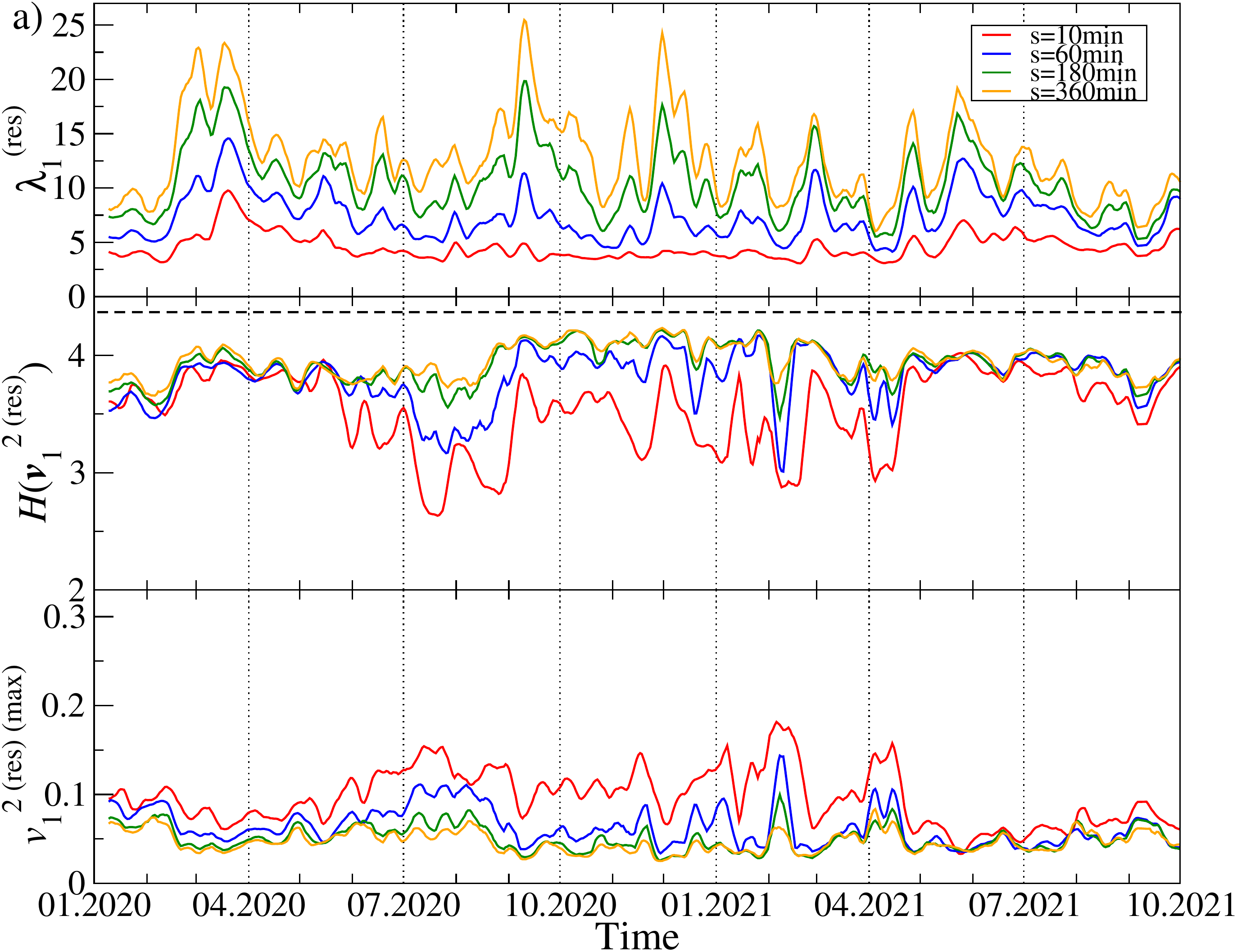}

\vspace{0.2cm}
\includegraphics[width=0.75\textwidth]{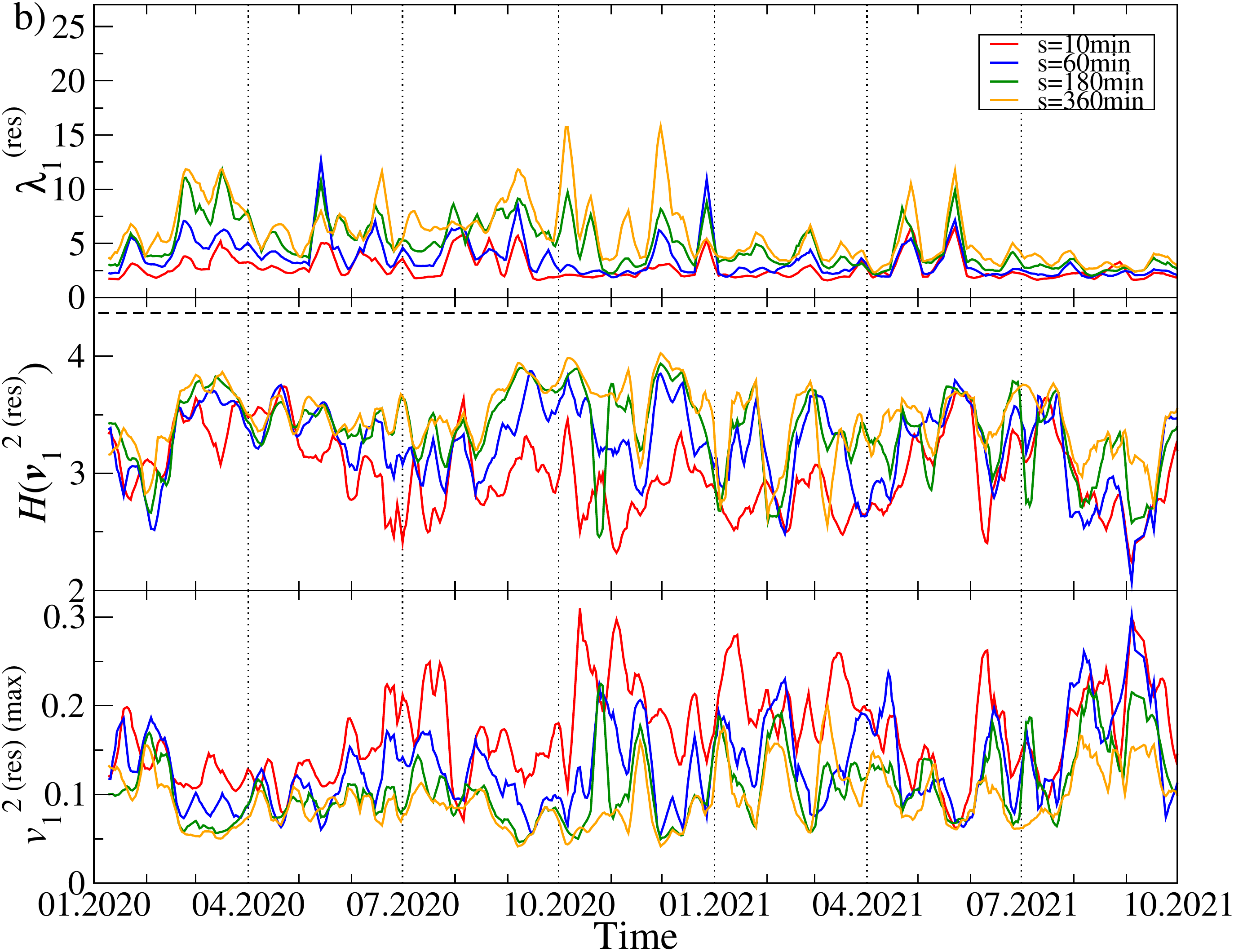}
\caption{Time evolution of the selected spectral characteristics of the residual $q$-dependent detrended correlation matrix ${\bf C}_q^{({\rm res})}(s)$ after filtering out the component corresponding to $\lambda_1$. As in Fig.~\ref{fig::eigenspectrum.lambda1.complete}, two cases are shown: $q=1$ (a) and $q=4$ (b). A moving window of length 7 days shifted by 1 day was applied for sample values of the scale: $s=10$ min (red), $s=60$ min (blue), $s=180$ min (green), and $s=360$ min (orange). The largest residual eigenvalue $\lambda_1^{({\rm res})}$ (top panels), the Shannon entropy $H({\bf v}_1^{({\rm res})})$ of the squared eigenvector components $v_1^{({\rm res})}(j)$ with $j=1,...,N$ (middle panels), and the squared maximum component of the eigenvector ${\bf v}_1^{({\rm res})}$ associated with $\lambda_1^{({\rm res})}$ (bottom panels) are shown.}
\label{fig::eigenspectrum.lambda1.nomax}
\end{figure}
\begin{paracol}{2}
%\linenumbers
\switchcolumn

Some deeper insight into the cross-correlation structure of the cryptocurrency market can be gained by transforming the $q$-dependent detrened correlation matrix ${\bf C}_q(s)$ into a related distance matrix ${\bf D}_q(s)$, whose elements are defined by Eq.~(\ref{eq::metric-distance}). The latter is used as a basis for creating a minimum spanning tree, in which each node represents a particular cryptocurrency and each weighted edge represent the metric distance between a pair of assets or, equivalently, the detrended cross-correlation coefficient. To facilitate comprehension of the MST pictures, the edge weights between the nodes ($i,j$) are proportional to the coefficients $\rho_q^{(i,j)}(s)$ even though the metric distances $d_q^{(i,j)}(s)$ were used to determine the MST edges in this work.

% Figure 5

We created an MST for each moving window position and for the same values of $s$ and $q$ as before. Owing to this, we are able to observe the evolution of the MST topology along the considered time span. The first topological characteristics we discuss here is the probability that a given node has a degree $k$. Its cumulative distributions $P(X\ge k)$ for a few sample window positions are shown in Fig.~\ref{fig::node_degree_distr} for $q=1$ (top) and $q=4$ (bottom) and for $s=10$ min (red line) and $s=360$ min (blue line). The MST topology expressed by these characteristics varies between different time intervals from a centralized graph with a single dominant node playing the role of a hub, that is,  when there is a significant gap between the largest degree $k_{\rm max}$ and the second largest degree, to a distributed graph with a small $k_{\rm max}$ and a small difference in the degrees of the most connected nodes. The former situation is more typical for the short time scales ($s=10$ min) and the periods with small return fluctuations ($q=1$), while the latter situation occurs frequently for the long time scales ($s=360$ min) and both the small and large fluctuation periods ($q=1$ and $q=4$); see Fig.~\ref{fig::node_degree_distr}.

While increasing the scale from $s=10$ min to $s=360$ min, for $q=1$ we observe a systematic change of the MST topology from centralized towards more distributed. For $q=4$ there is no such a change and the topology is largely preserved. From the network perspective, this means that the detrended cross-correlations during the strong volatility periods are already well-developed at the 10-min time scale and, possibly, one has to consider even shorter scales to detect any topological transition (this would require a higher frequency of the price quotations than 1 min considered here, however). It is also worth noting that the cumulative probability distributions in some windows show a scale-free decay with $k$ (the almost-straight lines in double logarithmic plots). This conclusion supports the results reported earlier for the data covering the years 2016--2019~\cite{watorek2021a} and 2017--2018~\cite{polovnikov2020}.

% start a new page without indent 4.6cm
\clearpage
\end{paracol}
\nointerlineskip
\begin{figure}[H]
\centering
\includegraphics[width=0.8\textwidth]{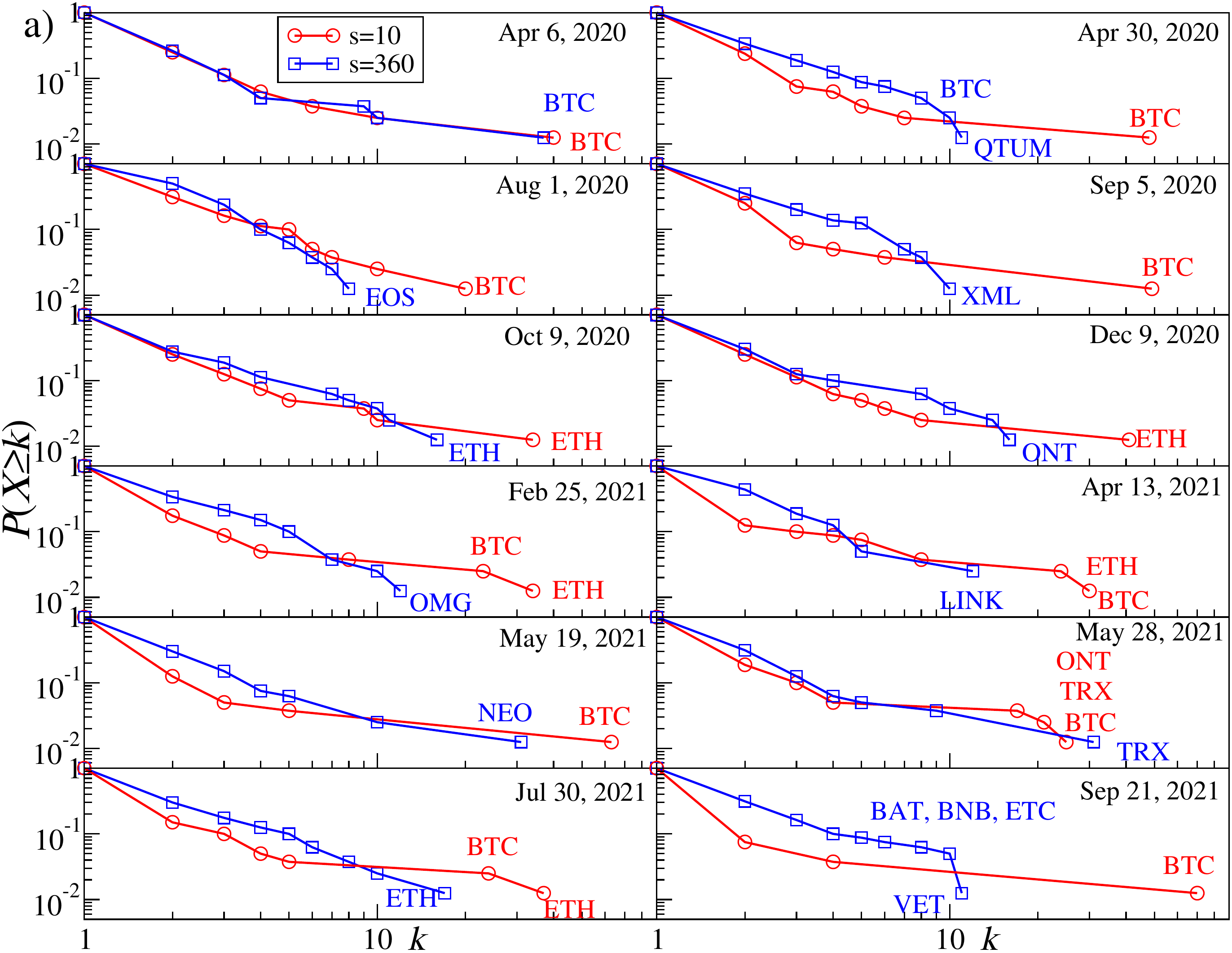}

\includegraphics[width=0.8\textwidth]{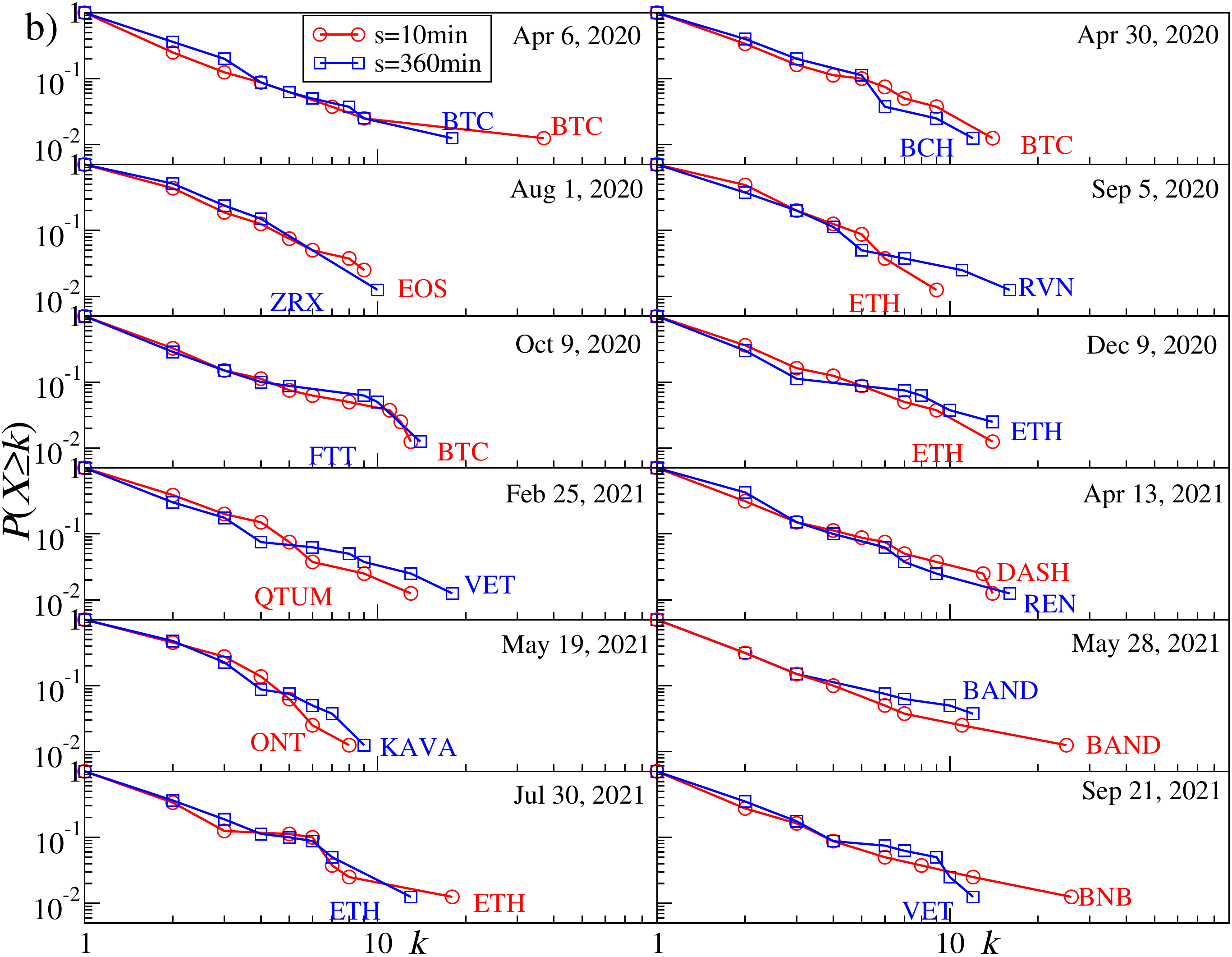}
\caption{Node degree cumulative distribution $P(X\ge k)$ of the MSTs created for the cryptocurrency prices expressed in USDT. Results for sample moving windows are shown for $q=1$ (a) and $q=4$ (b). In each panel the distributions for two temporal scales are displayed: $s=10$ min (red) and $s=360$ min (blue). The nodes with the highest degree $k$ are labelled by the corresponding cryptocurrency ticker.}
\label{fig::node_degree_distr}
\end{figure}
\begin{paracol}{2}
%\linenumbers
\switchcolumn

% Figure 6

Topological changes of the MSTs while going from past to present can be expressed by the time evolution of the node degree $k_i(t)$ for the most connected nodes representing the cryptocurrencies $i$. The results for the MSTs created based on three distinct data sets are presented in Fig.~\ref{fig::kmax_evol}: (1) the original time series of the price returns, (2) the residual time series obtained after filtering out the contribution of $\lambda_1$ from the original data (both are based on the quotes given in USDT), and (3) the time series of the price returns based on the quotes given in BTC. The latter case allows us for effective filtering out the impact of BTC on the other assets' detrended cross-correlations.

There are the following observations:

(1) Exactly as expected from the above discussion related to Fig.~\ref{fig::node_degree_distr}, for each data type, the degree of the most connected nodes tends to decrease with increasing $s$ and the degree gap between $k_{\rm max}$ and the smaller values of $k_i$ decreases as well. For longer time scales, the topology becomes less centralized and more ``democratic'' with a few hubs of a comparable connectivity.

(2) As the most capitalized cryptocurrency, BTC remains the most connected node over the longest time for $s=10$ min and, to a lesser extent, for $s=60$ min. However, for $s=360$ min, it ceases to play such a role in August 2020, when the MST becomes decentralized permanently and the most connected node can be a cryptocurrency of moderate capitalization (see, for example,~\cite{papadimitriou2020} for a similar observation).

(3) It happened for $s=10$ min that the periods when ETH was the most connected node as frequently as BTC prevailed between September 2020 and February 2021. For $s=60$ min also some other assets like ONT and TRX are represented by the most connected nodes from time to time, but it happens more because of a temporarily diminished degree of BTC and ETH than because of their own importance.

(4) In the residual data, BTC does not play so substantial role as in the original data, because its dominating role was largely wiped out by filtering out the $\lambda_1$ contribution. It remains, however, a hub with the second largest connectivity throughout the whole analyzed interval for $s=10$ min. If $s$ is increased to 60 min, BTC is degraded further on to be among a few secondary hubs with a few connections only. For both the scales, the most connected node is FTT, but its distinguished structural position vanishes almost completely after April 2021. For $s=360$ min the MSTs always show a decentralized topology.

(5) If the prices are expressed in BTC, $k_{\rm max}(t)$ is typically smaller ($k_{\rm max} < 30$ out of 68) than when they are expressed in USDT ($k_{\rm max} < 70$ out of 80). This is the expected property as BTC is the most connected hub in the case of the prices given in a stable coin. For any scale, a typical situation in this case is that there is frequent alternation of the most connected nodes: ETH, BNB, LINK, ONT, LTC, XRP, DASH, and so forth are among the assets that have the largest degree in certain time intervals, but none of them is able to substantially centralize the network. For long time scales, it even occurs that the largest degree nodes are switched almost random.

\begin{figure}[H]
\centering
\includegraphics[width=0.5\textwidth]{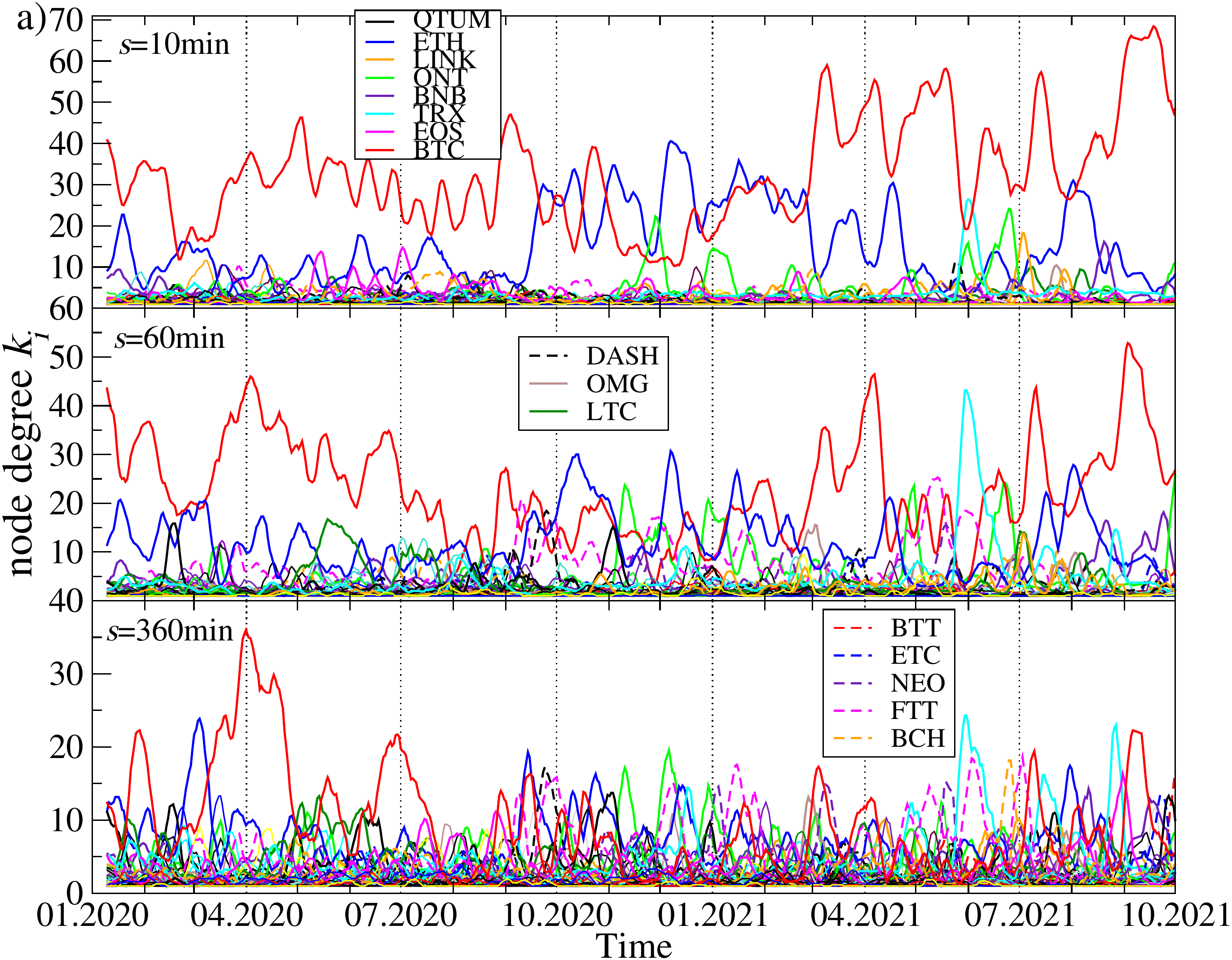}
\includegraphics[width=0.5\textwidth]{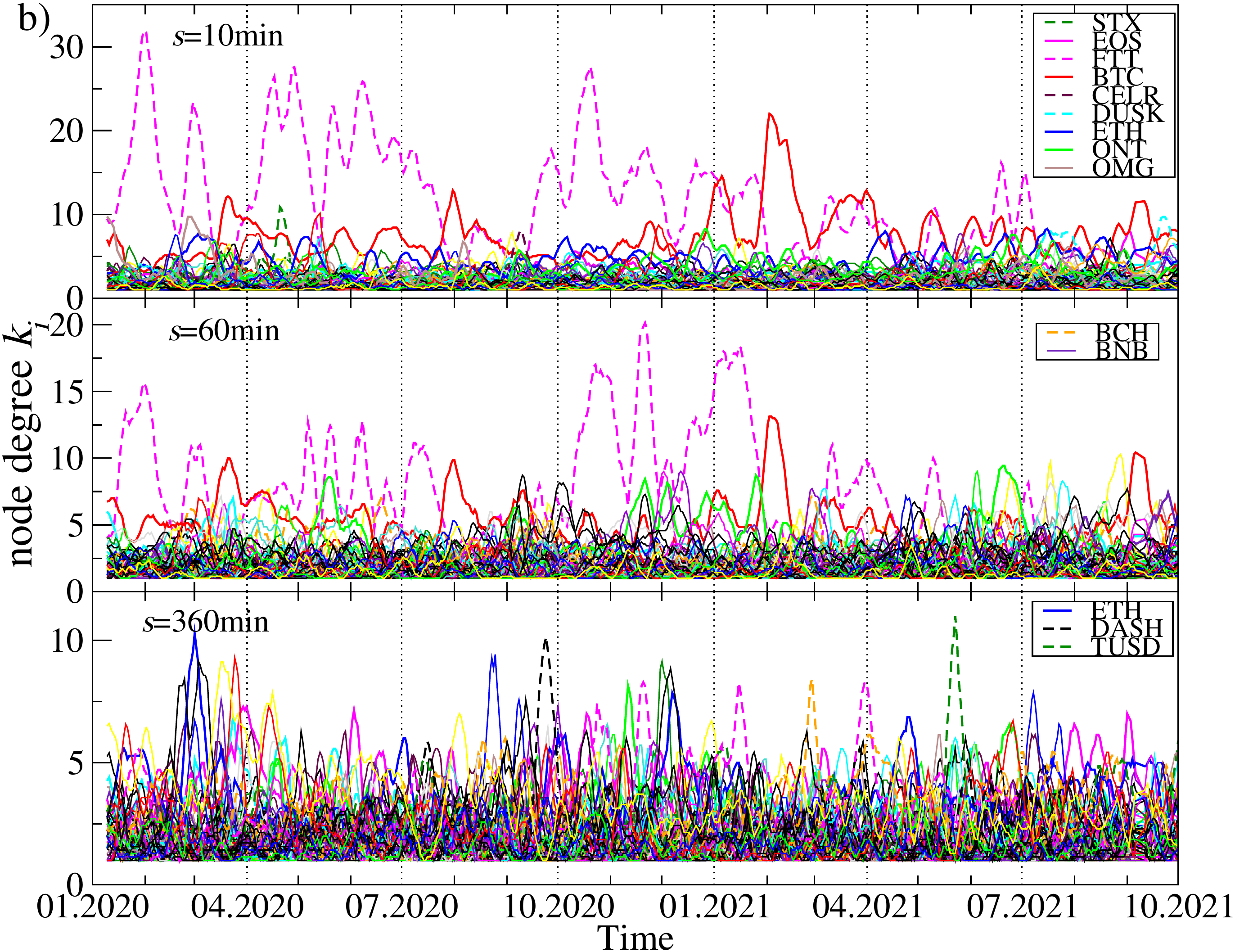}

\includegraphics[width=0.5\textwidth]{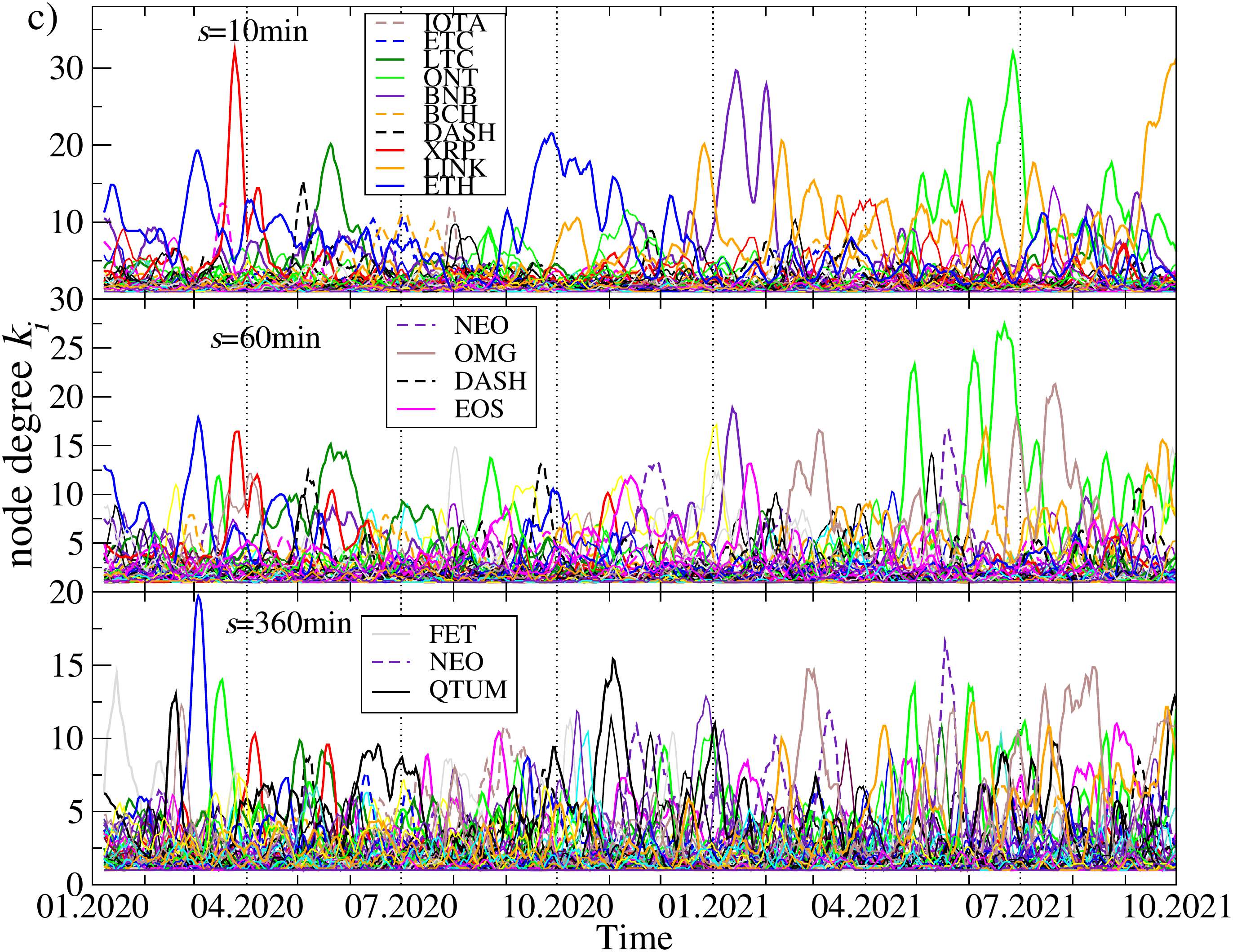}
\caption{Evolution of the node degree $k_i$ for the most connected nodes of the MST calculated in the seven-day-long moving window with a step of 1 day. For the prices expressed in USDT, two cases are shown: (a) the results for the complete data set without any filtering and (b) the results for the residual signals after filtering out a contribution from the component represented by the largest eigenvalue $\lambda_1$. The results for (c) - the prices expressed in BTC , which corresponds to filtering out any BTC-related contribution to other assets' evolution, are also shown. In each case, three exemplary scales are shown: $s=10$ min (top graph in each panel), $s=60$ min (middle graph), and $s=360$ min (bottom graph). Different colors and line styles denote the node degree for different cryptocurrencies.}
\label{fig::kmax_evol}
\end{figure}

% Figure 7

A variety of the MST topologies that can be observed in the cryptocurrency market in different periods is illustrated in Figs.~\ref{fig::MST.complete.q1.s10} and~\ref{fig::MST.complete.q1.s360}. The top left MST in Fig.~\ref{fig::MST.complete.q1.s10} has largely a star-like structure with BTC being its central node and ETH being a secondary hub. All other nodes are peripheral in respect to these two. The bottom left tree is also significantly centralized but now the most connected node is ETH, while BTC, FTT, and BAT are secondary hubs. A mixed type of topology is shown in the bottom right MST, where there are two primary hubs that are almost equivalent topologically (BTC and ETH) and a single secondary hub (BCH). However, despite this interesting dual centrality, the network has a part that is rather distributed. A largely distributed structure can be seen in the top right MST, in which only BTC possesses a significant number of the satellite nodes, while the overall network structure is distributed and almost random.

% start a new page without indent 4.6cm
%\clearpage
\end{paracol}
\nointerlineskip
\begin{figure}[H]
\widefigure
\includegraphics[width=0.45\textwidth]{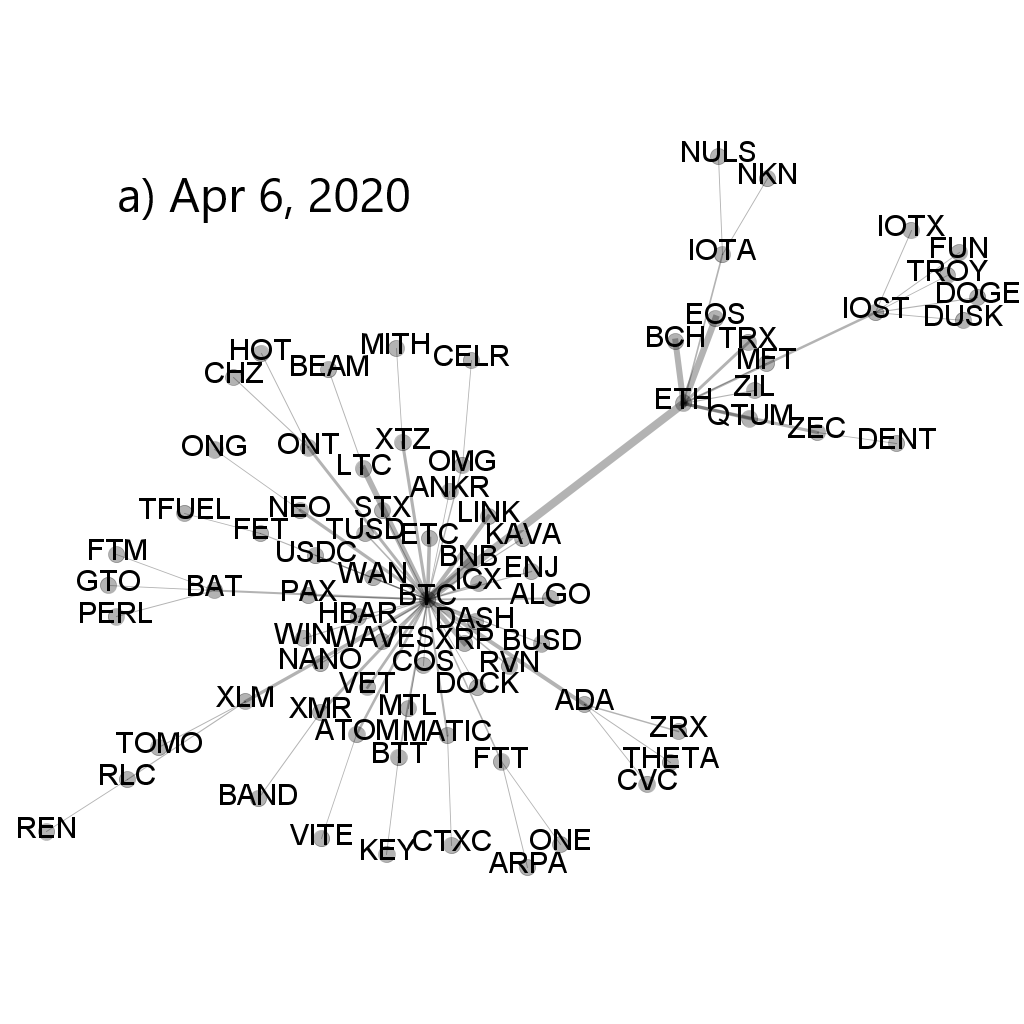}
\includegraphics[width=0.45\textwidth]{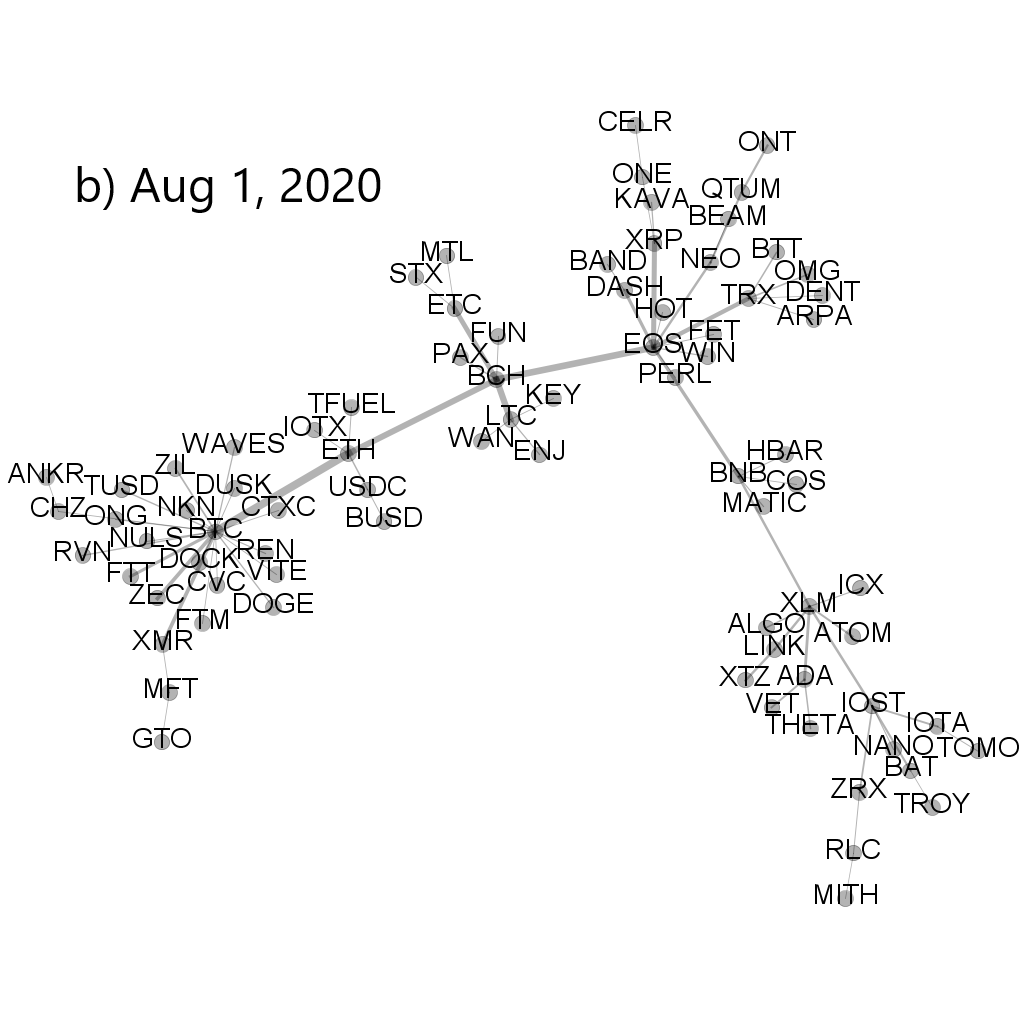}

\includegraphics[width=0.45\textwidth]{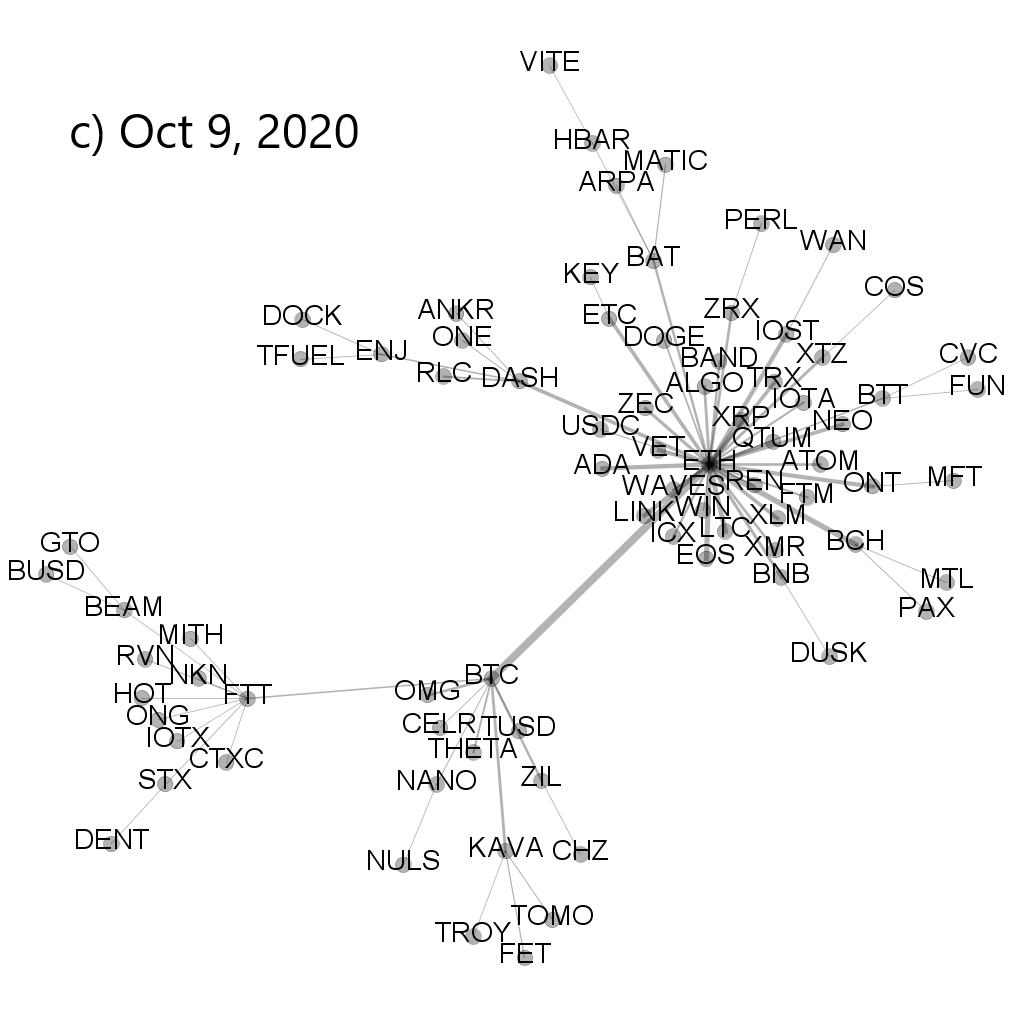}
\includegraphics[width=0.45\textwidth]{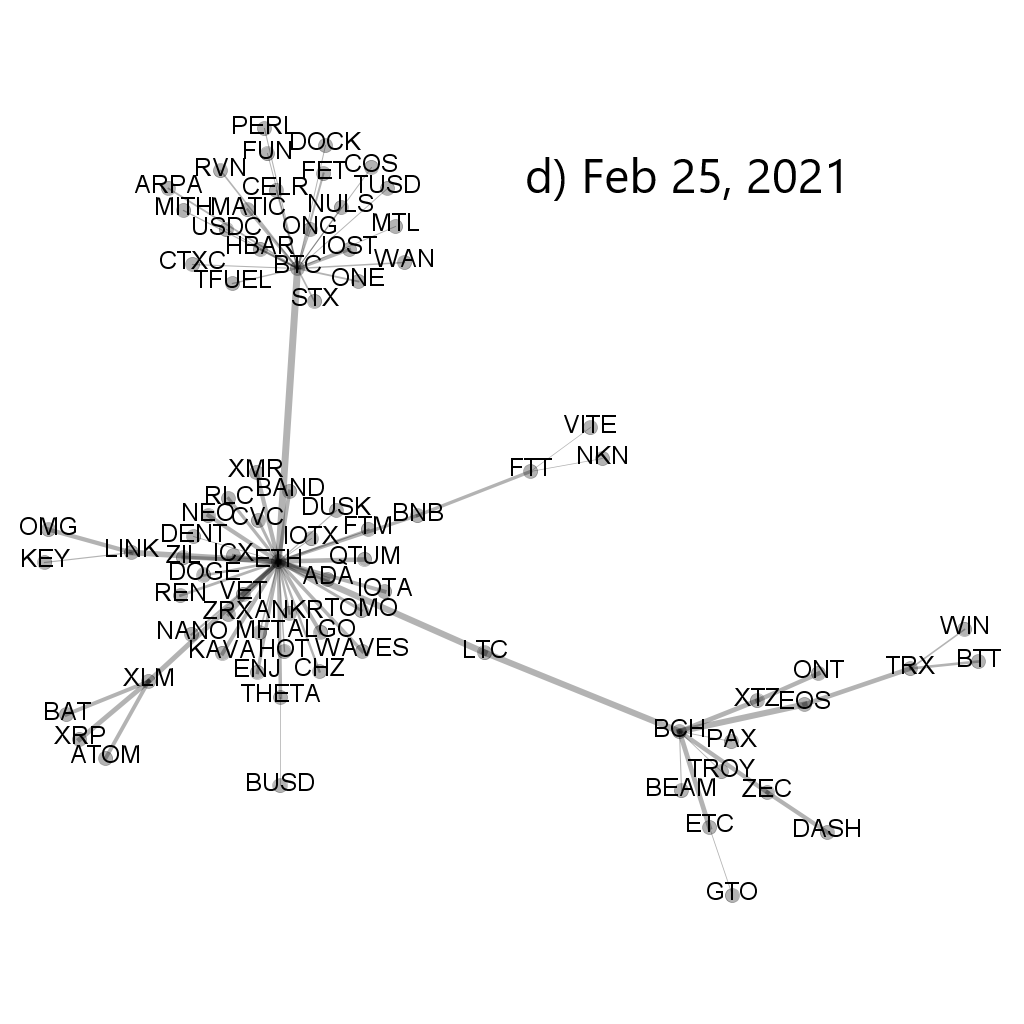}
\caption{Minimal spanning trees calculated from a distance matrix ${\bf D}_q(s)$ based on $\rho_q(s)$ for $q=1$ and $s=10$ min. Each node represents a cryptocurrency and the edge widths are proportional to value of the corresponding coefficient $\rho_q(s)$. Each MST was created for moving window of length 7 days ended at specific dates: (a) 6 April 2020, (b) 1 August 2020, (c) 9 October 2020, and (d) 25 February 2021.}
\label{fig::MST.complete.q1.s10}
\end{figure}
\begin{paracol}{2}
%\linenumbers
\switchcolumn

% Figure 8

While the asset–asset correlation strength can be amplified by increasing scale $s$, Fig.~\ref{fig::MST.complete.q1.s360} shows that this operation weakens at the same time the centralized topology of the associated MST, which can show the signatures of a decentralized network. This can be seen by comparing the trees corresponding to the same windows in Figs.~\ref{fig::MST.complete.q1.s10} and~\ref{fig::MST.complete.q1.s360}.

% start a new page without indent 4.6cm
%\clearpage
\end{paracol}
\nointerlineskip
\begin{figure}[H]
\widefigure
\includegraphics[width=0.45\textwidth]{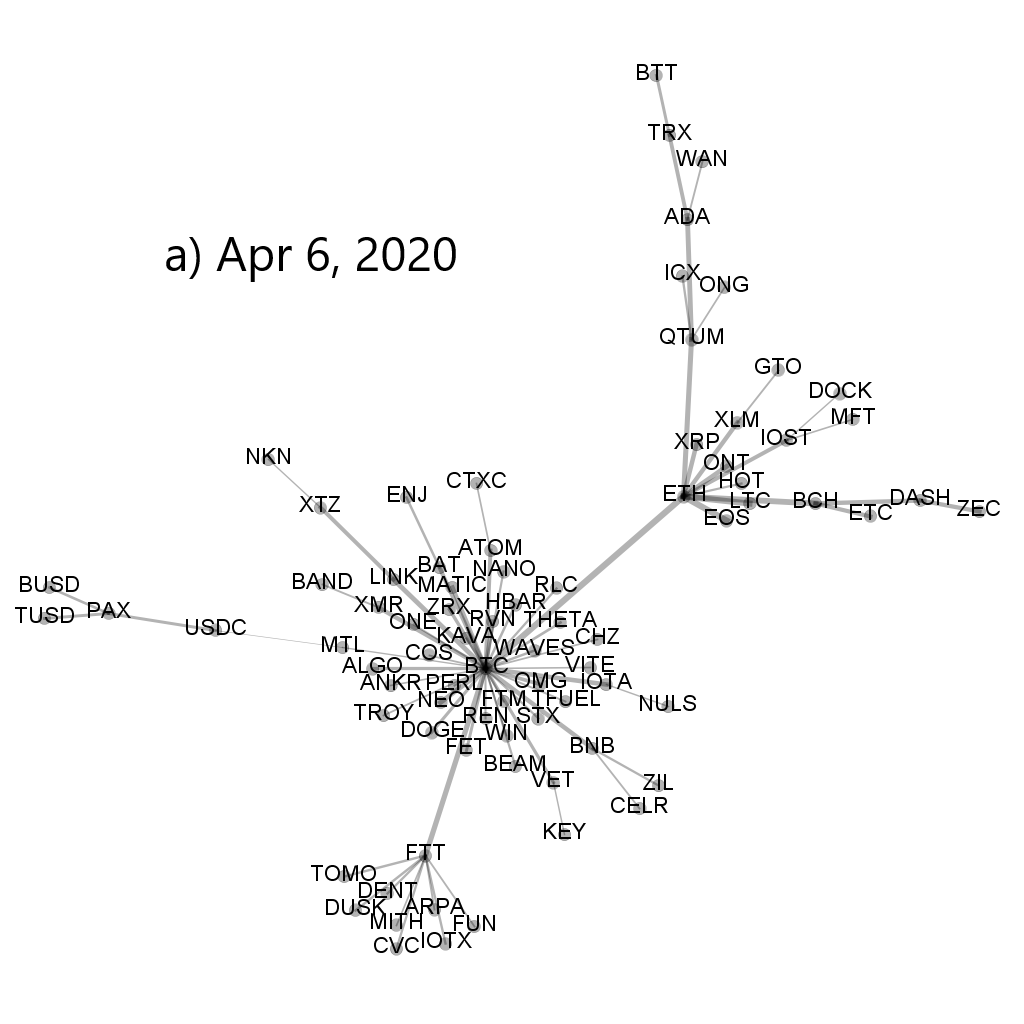}
\includegraphics[width=0.45\textwidth]{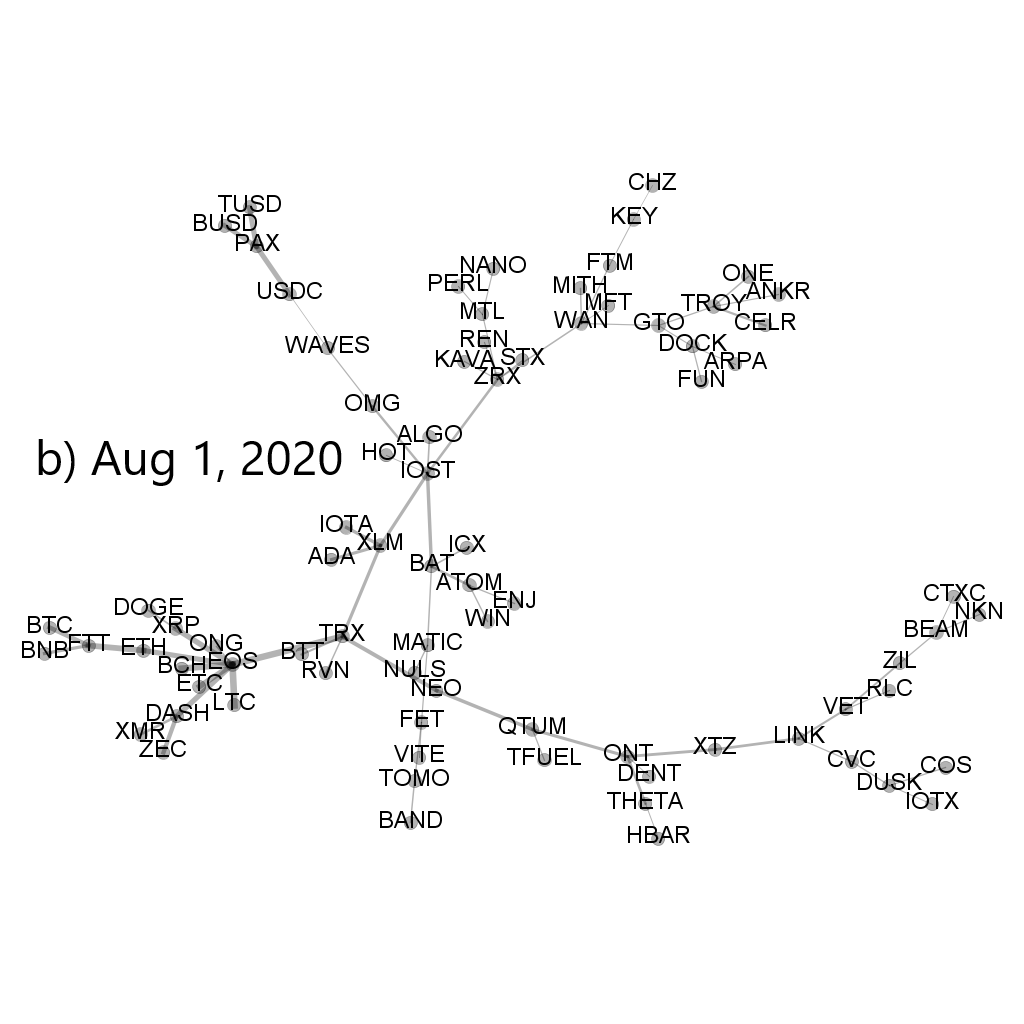}

\includegraphics[width=0.45\textwidth]{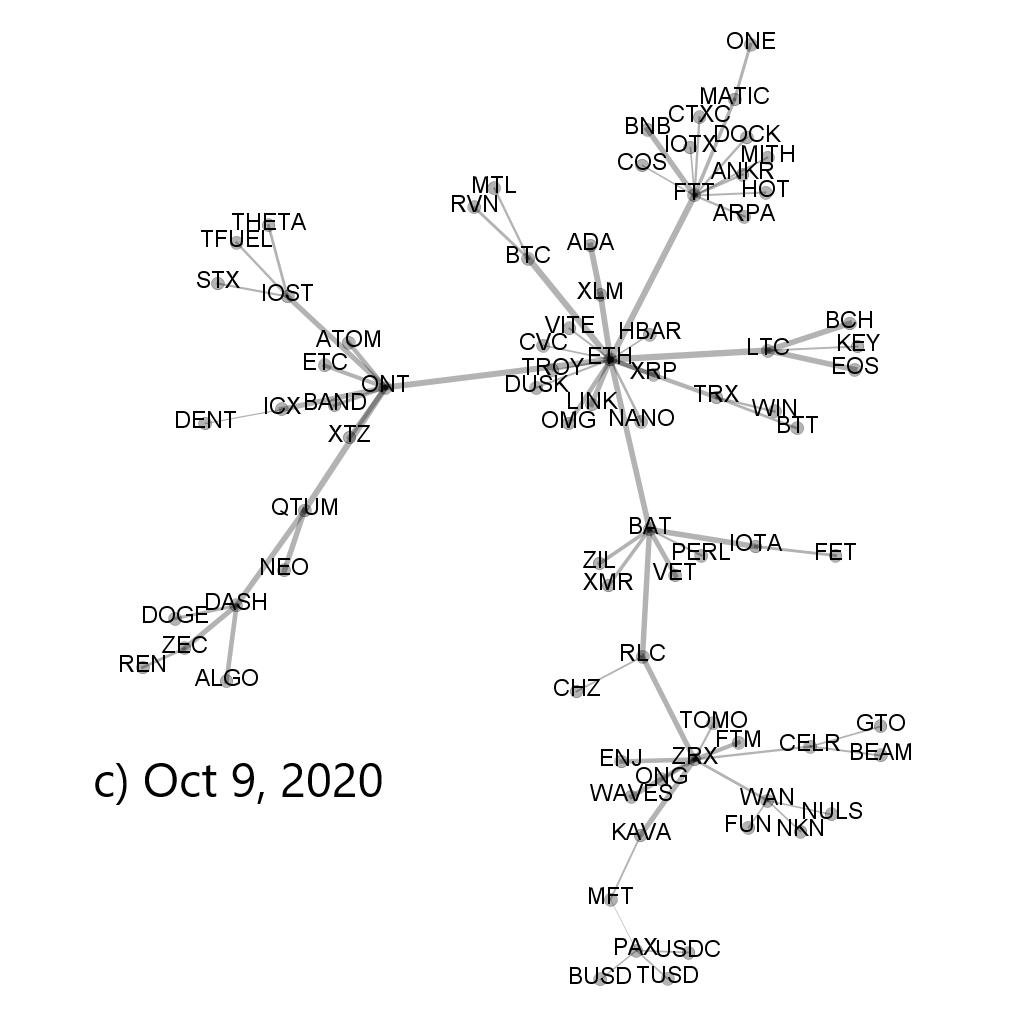}
\includegraphics[width=0.45\textwidth]{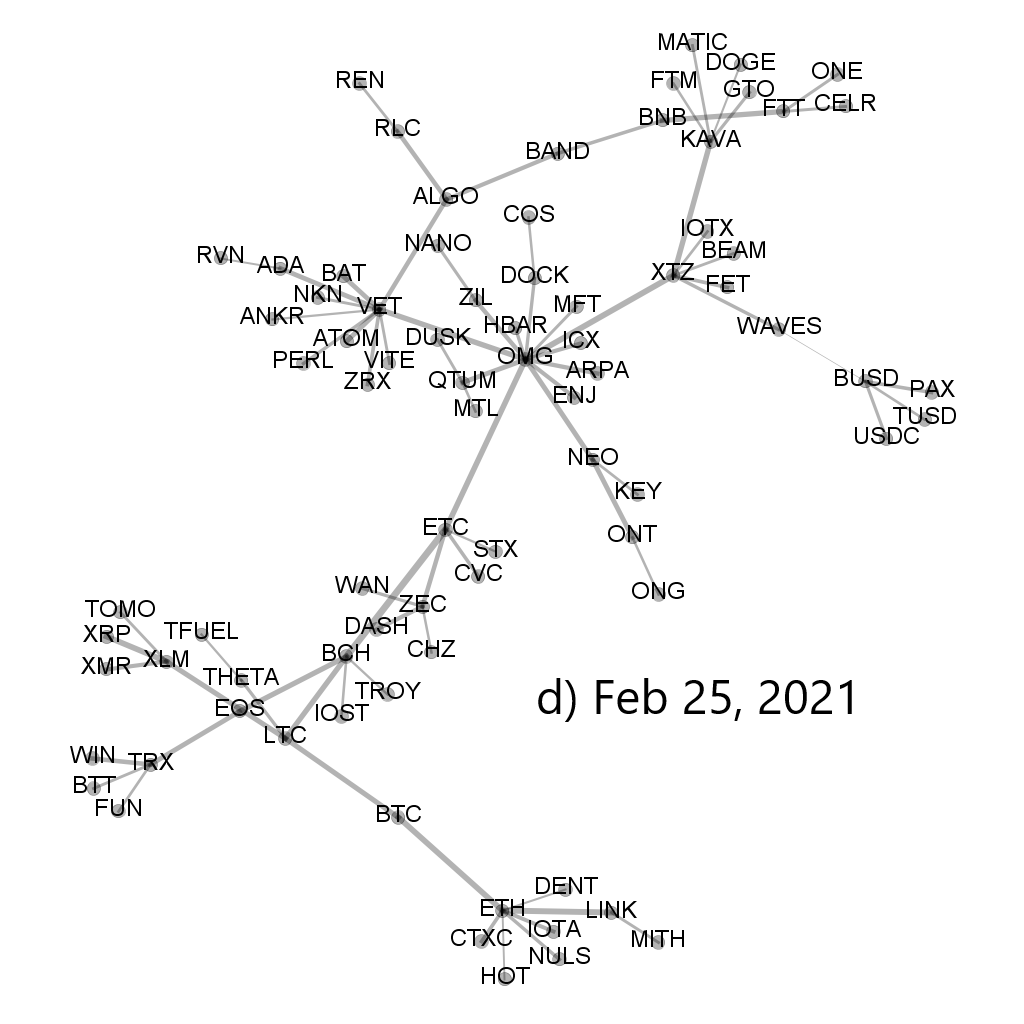}
\caption{Minimal spanning trees calculated from a distance matrix ${\bf D}_q(s)$ based on $\rho_q(s)$ for $q=1$ and $s=360$ min. Each node represents a cryptocurrency and the edge widths are proportional to value of the corresponding coefficient $\rho_q(s)$. Each MST was created for moving window of a length of 7 days ended at specific dates:  (a) 6 April 2020, (b) 1 August 2020, (c) 9 October 2020, and (d) 25 February 2021.}
\label{fig::MST.complete.q1.s360}
\end{figure}
\begin{paracol}{2}
%\linenumbers
\switchcolumn

% Figure 9

This conclusion receives additional support from the top panels of Fig.~\ref{fig::network-topology.complete.USDT}(a)(b) presenting the mean path length as a function of time. It is defined by the following formula:
\begin{equation}
\langle L(q,s,t) \rangle = {1 \over N(N-1)} \sum_{i=1}^N \sum_{j=i+1}^N L_{ij}(q,s,t),
\end{equation}
where $L_{ij}$ is the length of the path connecting nodes $i$ and $j$. The larger $\langle L(q,s,t) \rangle$ is, the more distributed is the corresponding MST. Indeed, by considering a given window, this quantity systematically increases with increasing $s$. The smallest values of the mean path length ($2 < \langle L_{ij}(q,s,t) \rangle < 3$) can be seen in April-May 2020 (see also~\cite{garcia-medina2020}), in August--September 2020, between March and May 2021, in May 2021, and in September--October 2021 for $s=10$ min. These are the periods of the most centralised market, where a vast majority of the nodes is connected to a central hub. In each of these periods, the maximum node degree $k_{\rm max}$ assumes high values as well (see Fig.~\ref{fig::kmax_evol}(a)). In contrast, the elevated values of $\langle L_{ij}(q,s,t) \rangle$ ($L_{ij}(q,s,t) > 5$) are observed in February 2020, July 2020, and between February and May 2021.

The power-law exponent $\gamma(q,s,t)$ describing slope of the cumulative probability distribution of the node degree is shown in the middle panel of Fig.~\ref{fig::network-topology.complete.USDT}(a) for $q=1$ and it is accompanied by the standard error of its least-square fit (the lower panel). It is an unstable quantity that fluctuates between 0.5 and 2 (see also Fig.~\ref{fig::node_degree_distr}) for the results in sample windows. The smaller $\gamma(q,s,t)$ is, the more distant $k_{\rm max}$ can be from the smaller values of $k_i$, but this relation does not always hold. The same quantities are shown in Fig.~\ref{fig::network-topology.complete.USDT} (b) for the case of $q=4$. Now we see smaller differences between the network characteristics for different time scales. This is the same rule as the one observed in Fig.~\ref{fig::node_degree_distr} for $q=4$.

% start a new page without indent 4.6cm
\clearpage
\end{paracol}
\nointerlineskip
\begin{figure}[H]
\widefigure
\includegraphics[width=0.8\textwidth]{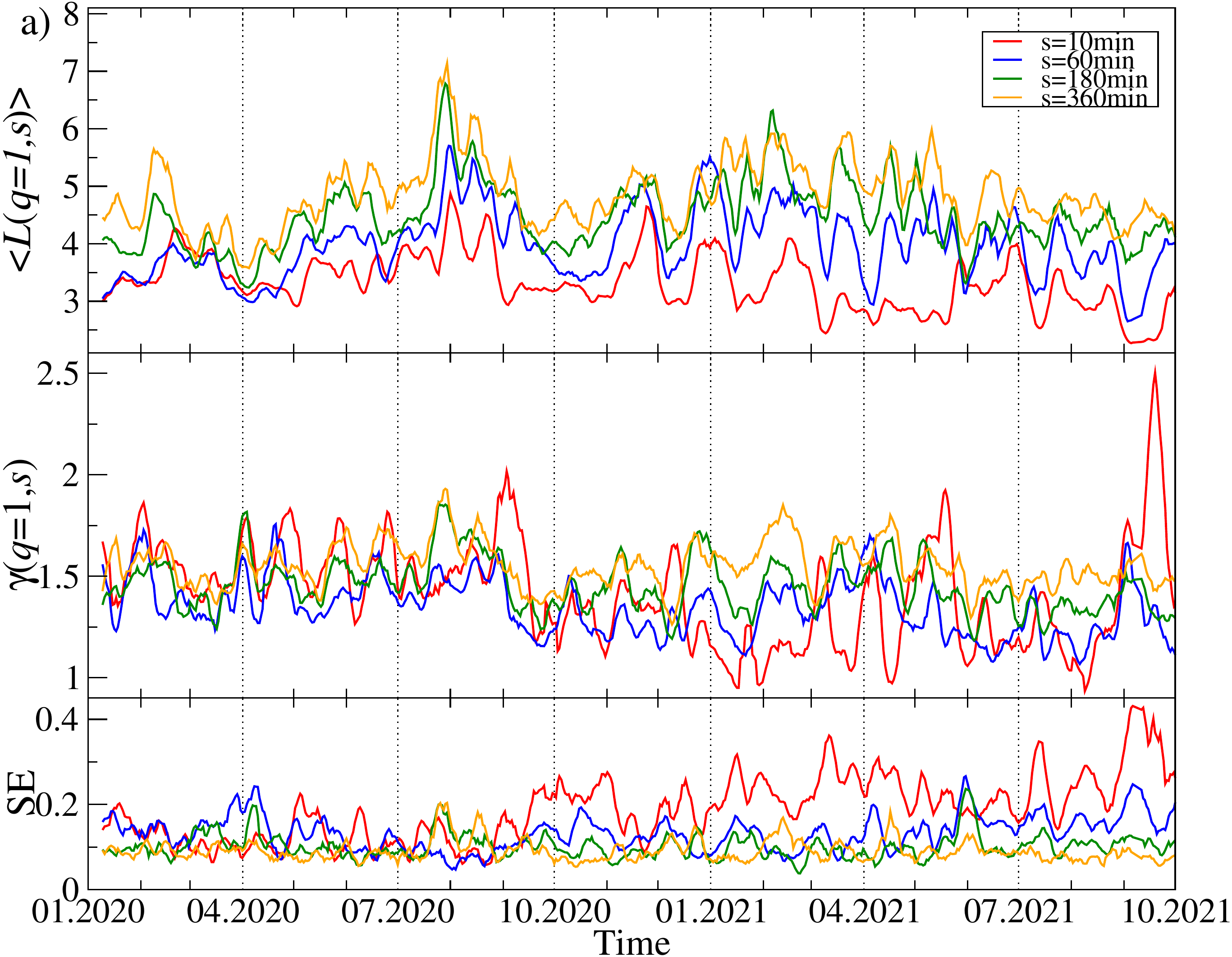}

\vspace{0.2cm}
\includegraphics[width=0.8\textwidth]{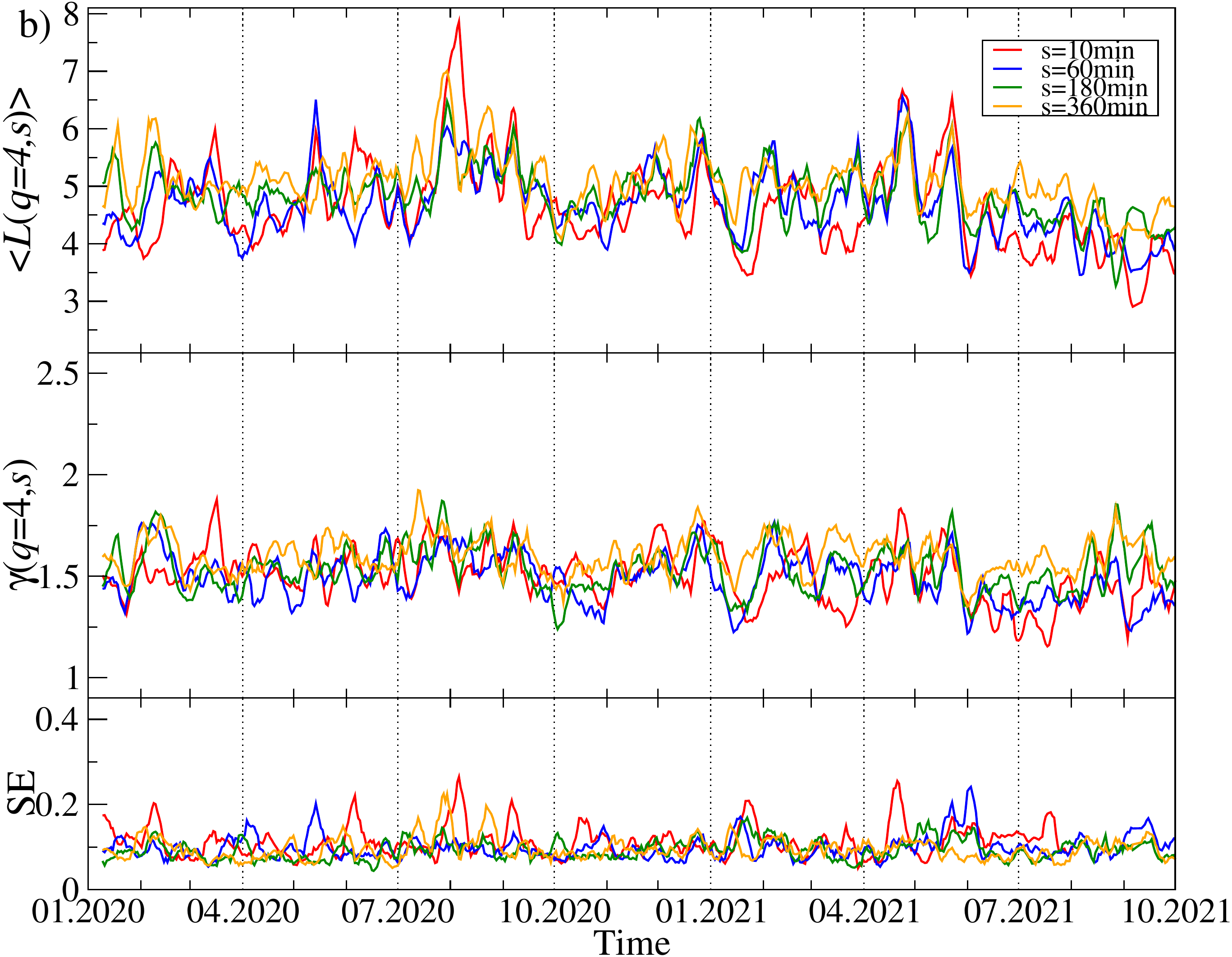}
\caption{Time evolution of the selected network characteristics of the MST created from a distance matrix ${\bf D}_q(s)$. Two cases are shown: $q=1$ (a) and $q=4$ (b). In each case, a moving window of length 7 days shifted by 1 day was applied for the scales: $s=10$ min (red), $s=60$ min (blue), $s=180$ min (green), and $s=360$ min (orange). The mean path length $\langle L(q,s,t) \rangle$ (top panels), the node degree cumulative probability distribution $P(X\ge k)$ power-law slope exponent $\gamma(q,s,t)$ (middle panels) together with its standard error (SE, bottom panels). The cryptocurrency prices are expressed in USDT.}
\label{fig::network-topology.complete.USDT}
\end{figure}
\begin{paracol}{2}
%\linenumbers
\switchcolumn

% Figure 10

Topology of the MSTs representing the residual time series $\{r_i^{({\rm res})}(t_m)\}$ differs from the original time series $\{r_i(t_m)\}$ significantly. Because the removed component representing $\lambda_1$ is connected with the strength of the average detrended cross-correlation coefficient $\langle \rho_q(s) \rangle$, a lack of this component weakens the detrended cross-correlations and can thus destroy the star-like structures within the MST. This must obviously lengthen many inter-node paths and increase $\langle L_{ij}(q,s,t) \rangle$. In fact, Fig.~\ref{fig::network-topology.nomax.USDT}(a,b) shows that $\langle L_{ij}(q,s,t) \rangle > 5$ over almost the whole analyzed period for both $q=1$ and $q=4$. It happens sometimes that its value reaches 10, which indicates a distributed network topology. The slope exponent $\gamma(q,s,t)$ behaves even more erratically than for the original, complete data in Fig.~\ref{fig::network-topology.complete.USDT}, and the standard error of the fitted values is much larger.

% start a new page without indent 4.6cm
\clearpage
\end{paracol}
\nointerlineskip
\begin{figure}[H]
\widefigure
\includegraphics[width=0.8\textwidth]{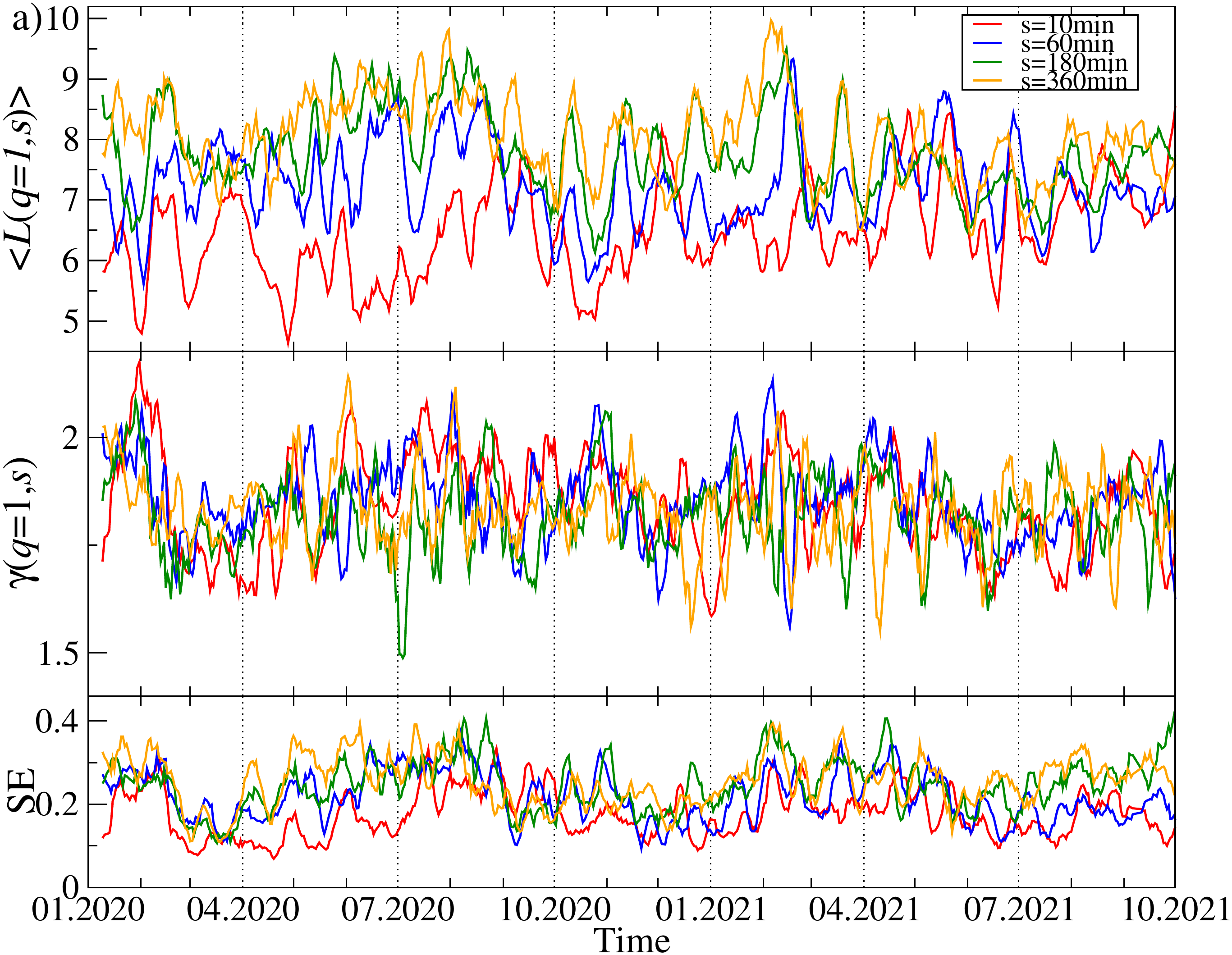}

\vspace{0.2cm}
\includegraphics[width=0.8\textwidth]{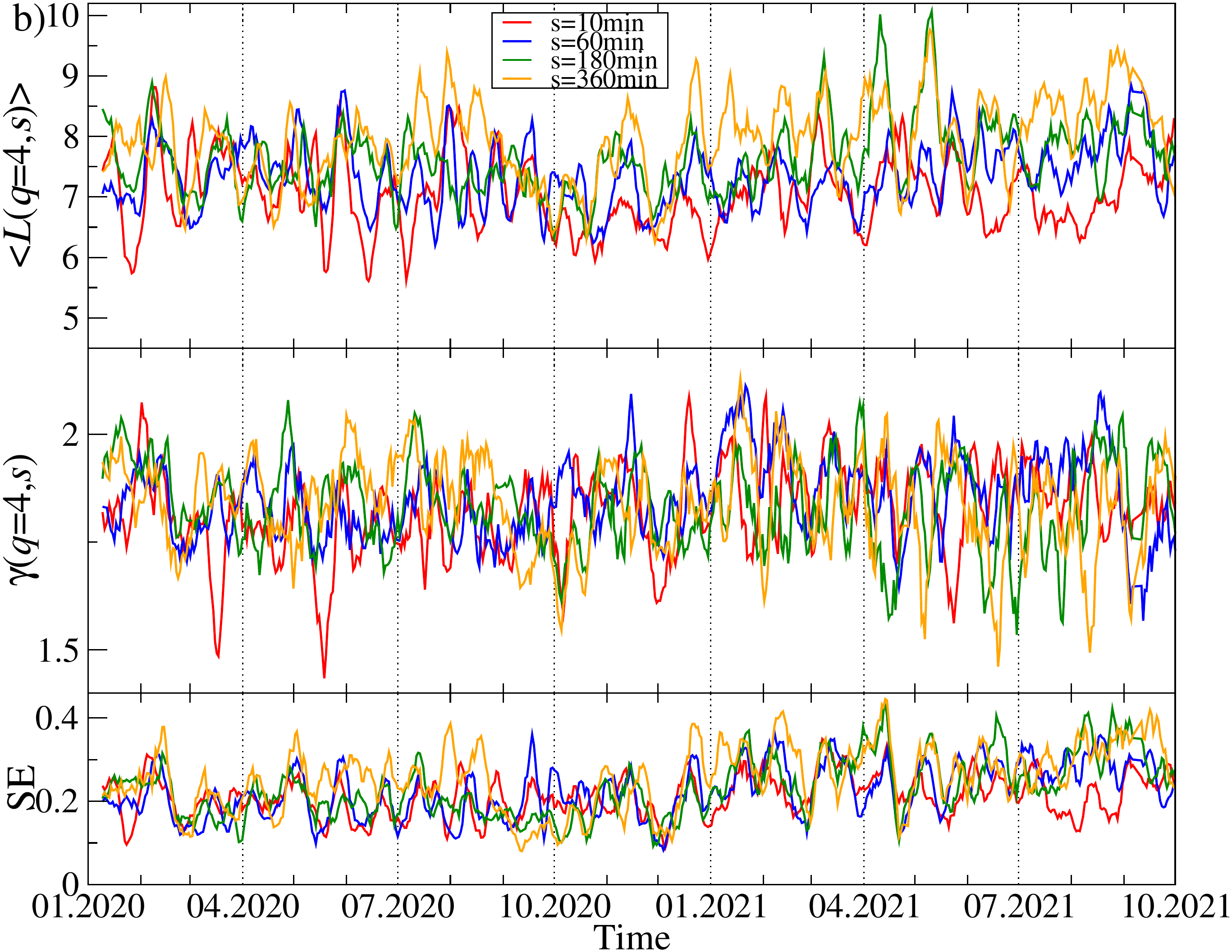}
\caption{The same quantities as in Fig.~\ref{fig::network-topology.complete.USDT} but here obtained from the residual MSTs calculated for ${\bf D}_q^{({\rm res})}(s)$ after filtering out the component corresponding to $\lambda_1$. The cryptocurrency prices are expressed in USDT.}%MDPI: please add the explanation for sub-figure (a) and (b) in the caption.
\label{fig::network-topology.nomax.USDT}
\end{figure}
\begin{paracol}{2}
%\linenumbers
\switchcolumn

% Figure 11

The same topological characteristics for the MSTs created from the time series of price quotations expressed in BTC are presented in Fig.~\ref{fig::network-topology.complete.BTC}(a,b). Their temporal evolution seems to be less random than in Fig.~\ref{fig::network-topology.nomax.USDT} and resembles the picture for the USDT-based data shown in Fig.~\ref{fig::network-topology.complete.USDT}. For $q=1$, the mean path length fluctuates along a horizontal line at $\langle L_{ij} \rangle \approx 5$ until April 2021. Then the trend line starts to decrease towards a level of 4 or even below this value. This suggests that the MST topology has gradually become more centralized in the recent months. Such an effect is hardly visible for $q=4$. A rather high values of $\gamma(q,s,t)$ above 1.5 for $q=4$ confirm a more compact topology of the corresponding MSTs than in the case of the prices expressed in USDT.

% start a new page without indent 4.6cm
\clearpage
\end{paracol}
\nointerlineskip
\begin{figure}[H]
\widefigure
\includegraphics[width=0.8\textwidth]{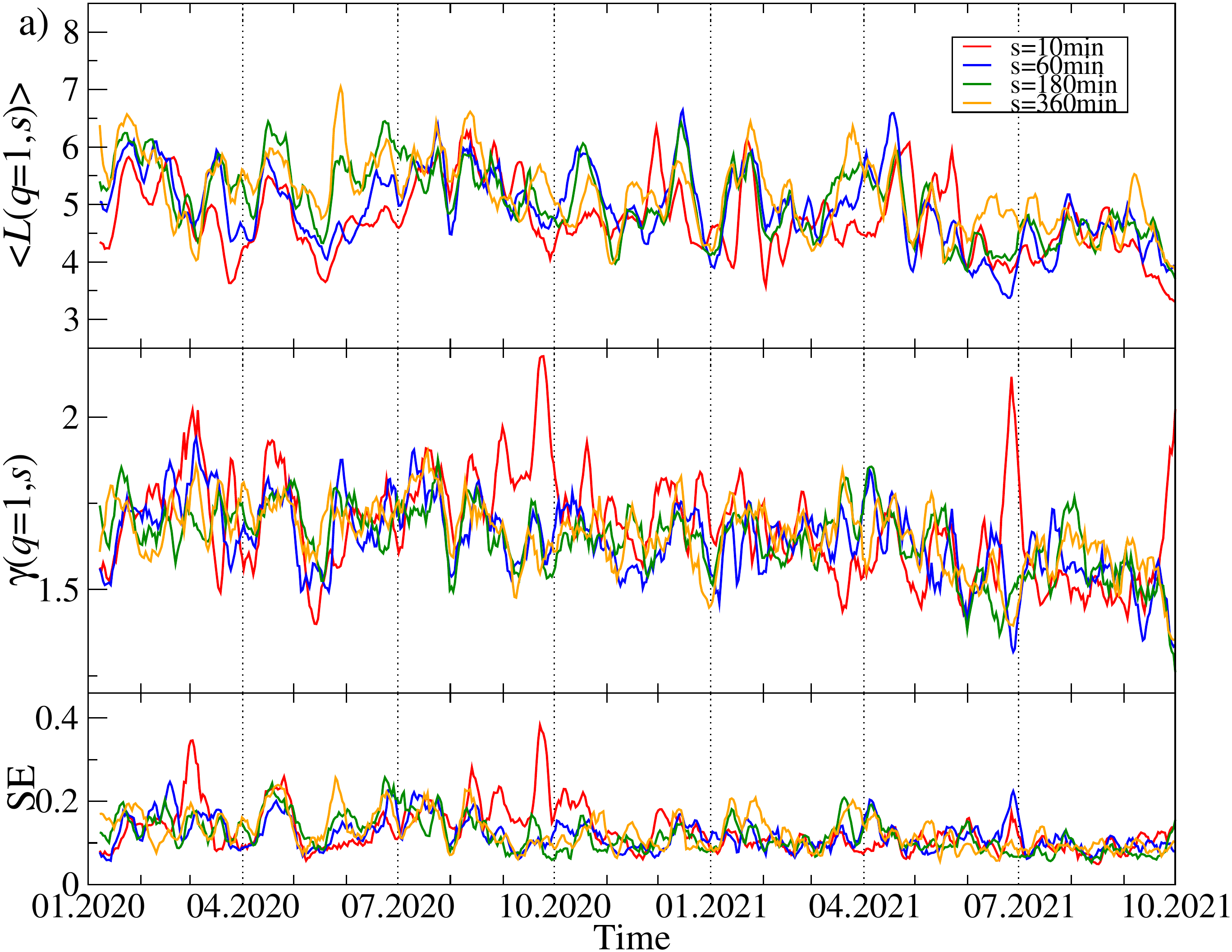}

\vspace{0.2cm}
\includegraphics[width=0.8\textwidth]{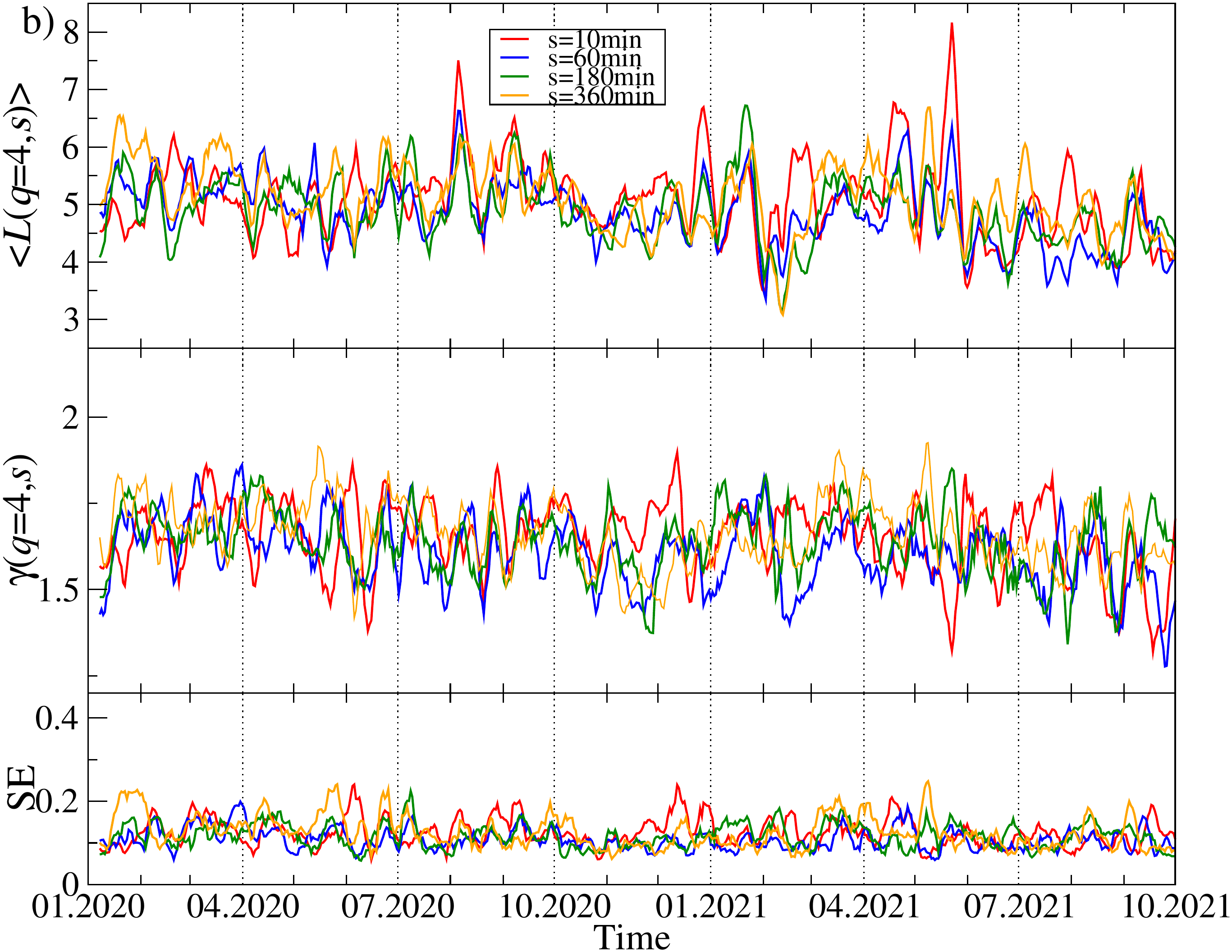}
\caption{The same quantities as in Fig.~\ref{fig::network-topology.complete.USDT} and~\ref{fig::network-topology.nomax.USDT} but obtained from the cryptocurrency prices expressed in BTC.}
\label{fig::network-topology.complete.BTC}%MDPI: please add the explanation for sub-figure (a) and (b) in the caption.
\end{figure}
\begin{paracol}{2}
%\linenumbers
\switchcolumn

Our study of the cryptocurrency network topology can be completed with an analysis of the network cluster structure. Obviously, in this case we have to consider the complete weighted networks defined by the matrix ${\bf C}_q(s)$ instead of the MSTs. In order to identify node clusters, we exploit the Louvain algorithm of community detection, whose performance is counted among the best methods~\cite{blondel2008}.

% Figure 12

For the most moving window positions, the algorithm detects a few cryptocurrency clusters, but their composition fluctuates among the windows. To show how the clusters vary in time, we select a few significant nodes and associate them with a set of nodes they share a given cluster with. Among the distinguished nodes that frequently play a role of the MST cluster centers are BTC, ETH, LINK, TRX, ONT, BNB, and others. In the case of BTC, we consider a network of all 80 cryptocurrencies expressed in USDT, while for the other nodes, we consider a limited set of 68 cryptocurrencies expressed in BTC and that are not pegged to US dollar. Some of the related clusters consist of a few nodes only throughout the whole period under study, but there are also clusters consisting of a variable number of nodes. Here we show the examples of the latter group of clusters: the clusters to which BTC, ETH, BNB, or ONT belong. It should be noted, however, that (1) a node representing a given cluster might not necessarily be its center in the MST representation, (2) some clusters are merged in some windows, while they remain separate in the other windows, and (3) the nodes can jump between clusters.

In Figs.~\ref{fig::clusters.BTC}--\ref{fig::clusters.ONT} we present the time evolution of the cluster composition for different time scales: $s=10$ min, $s=60$ min, and $s=360$ min. and for $q=1$. For example, a full point in the plot depicting the BTC cluster indicates that a respective cryptocurrency shares a cluster with BTC in a particular time window. The more dense points are seen along a horizontal line representing that cryptocurrency, the more stable is the coexistence of these two cryptocurrencies within the same cluster. On the other hand, the more numerous are the points along a vertical line, the larger is the cluster at that particular moment.

A cluster, to which BTC belongs, is typically the largest cluster in the network. By looking at Fig.~\ref{fig::clusters.BTC}, we see that, on the shortest time scale of 10 min, the BTC cluster's size increases substantially in March 2021 and remains such till the end of the analyzed time interval. This is in agreement with the increase of the Shannon entropy $H({\bf v}_1)$ observed in Fig.~\ref{fig::eigenspectrum.lambda1.complete} and it indicates that the market network has become more compact recently. A situation looks different for $s=60$ min, because apart from the BTC cluster growth observed in the $s=10$ min case, only a slightly smaller cluster structure was seen before mid-2020. Thus, for $s=60$ min the BTC cluster shrunk considerably over the period from July 2020 to February 2021 and it was larger outside that period. There are nodes that accompany BTC regularly, like STX, RVN, NANO, and BEAM, and there are nodes that fall into the BTC cluster only few times, like TRX and ETH, or even never do this, like ETH. For $s=360$ min we do not detect any comparably large cluster and the BTC cluster is much smaller. It also tends to shrink even more after mid-2020.

% start a new page without indent 4.6cm
\clearpage
\end{paracol}
\nointerlineskip
\begin{figure}[H]
\widefigure
\includegraphics[width=0.8\textwidth]{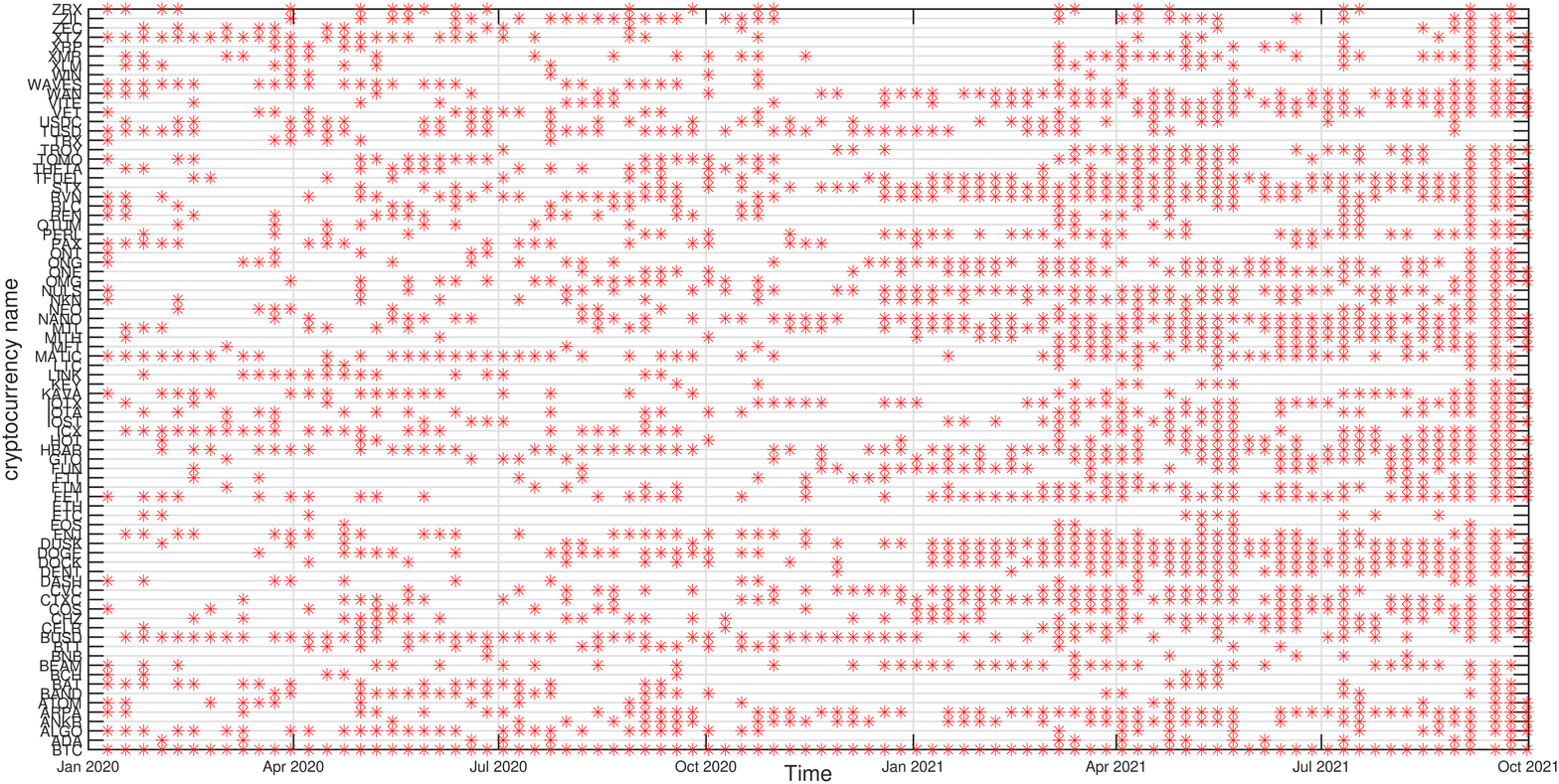}

\includegraphics[width=0.8\textwidth]{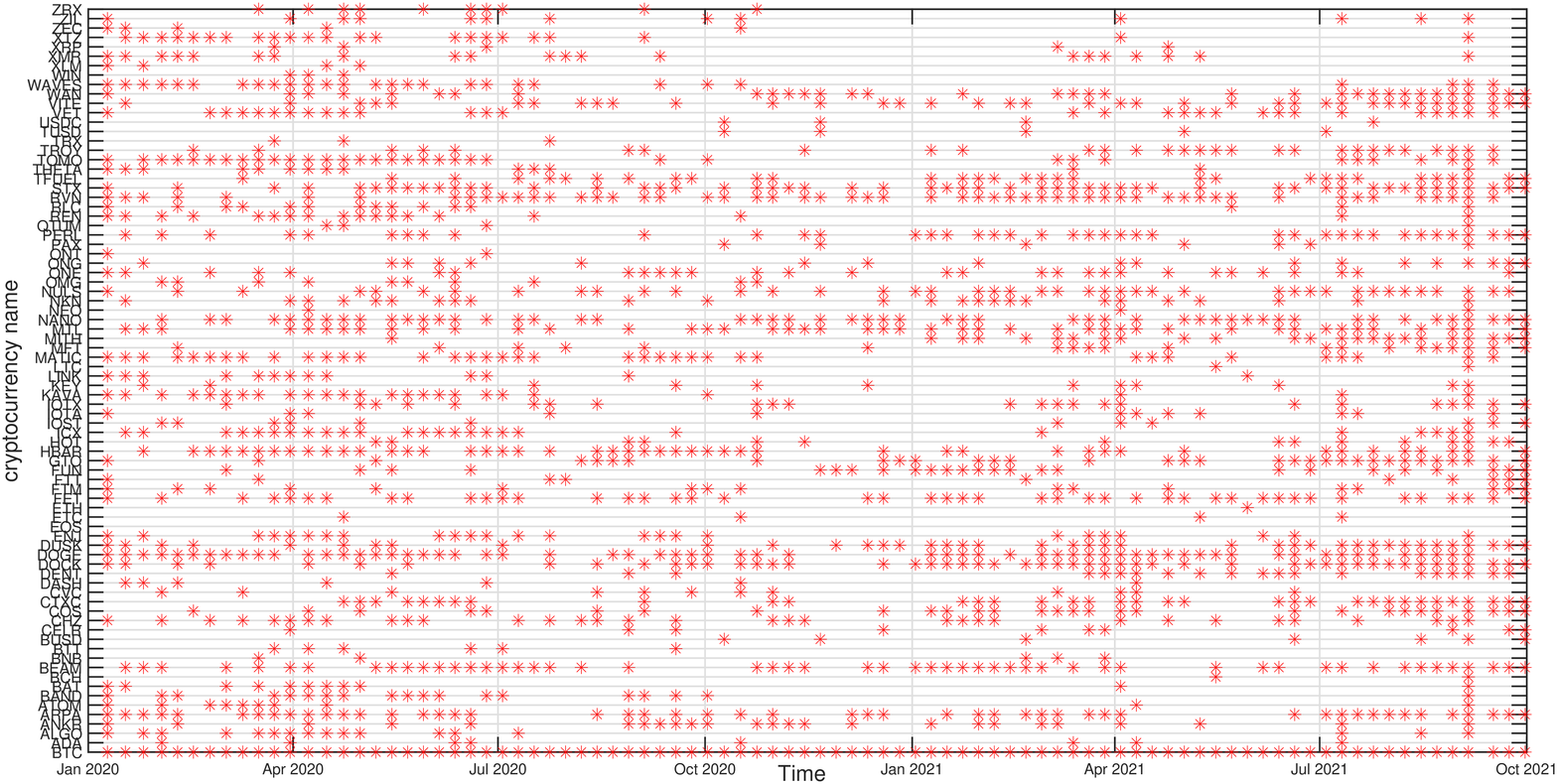}

\includegraphics[width=0.8\textwidth]{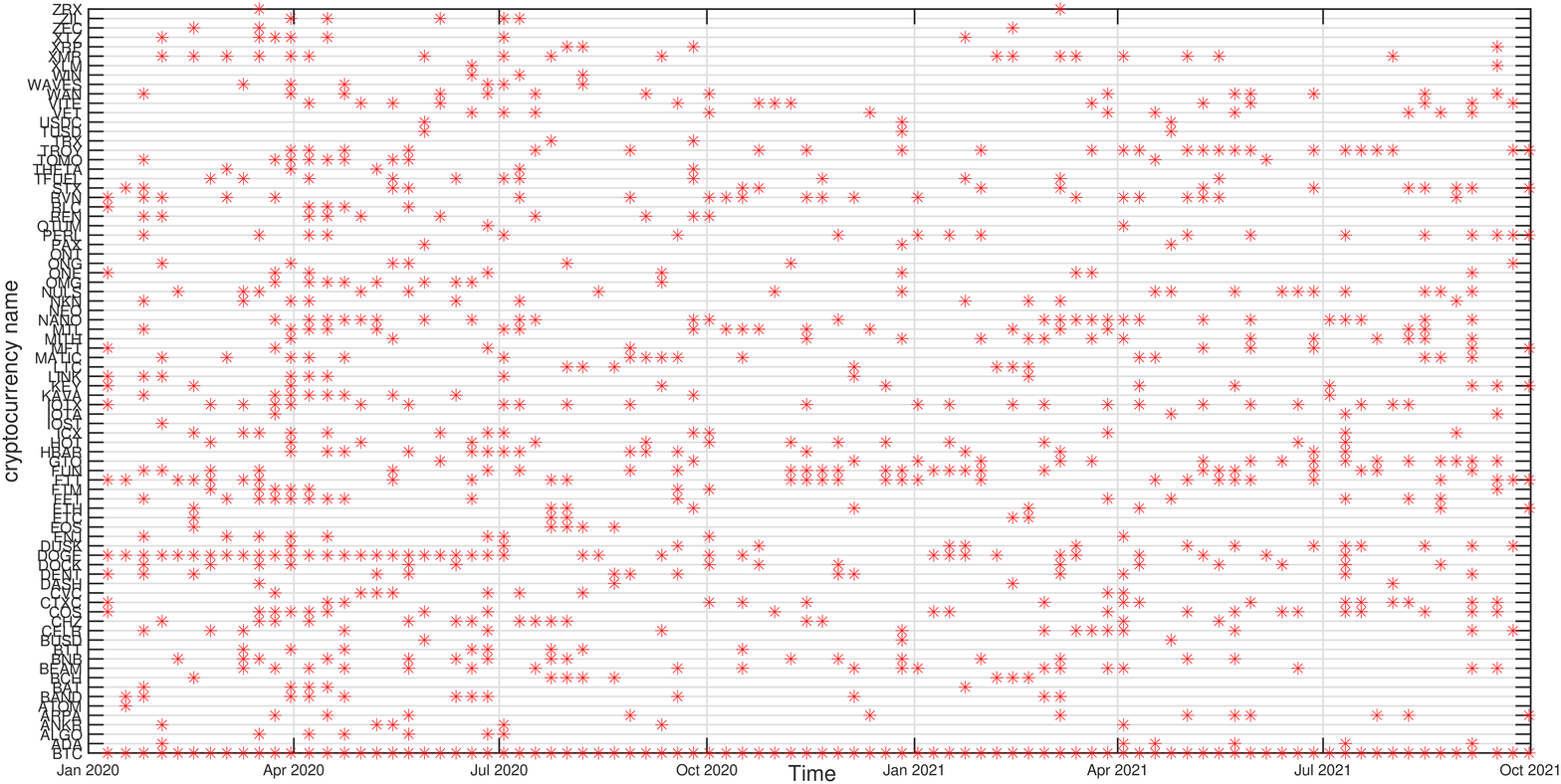}
\caption{Composition of the BTC-related cryptocurrency cluster as a function of time for sample temporal scales: $s=10$ min (top), $s=60$ min (middle), and $s=360$ min (bottom). Each point on the horizontal axis represents a non-overlapping seven-day-long moving window. Asset prices have been expressed in USDT.}
\label{fig::clusters.BTC}
\end{figure}
\begin{paracol}{2}
%\linenumbers
\switchcolumn

% Figure 13

The ETH cluster is much less numerous that the BTC one, which is partially due to a smaller number of the analyzed assets, but also to the properties of this cluster. Despite this, however, some long-term trend can be seen for $s=10$ min that resulted in the temporary cluster growth in the latter half of 2020 followed by its shrinking that lasts till the end of the analyzed period. Such an effect cannot be noticed for the longer scales, where the density of points remains at the same level throughout the years 2020-2021. Among the nodes that frequently accompany ETH are BNB, LTC, BCH, and LINK.

% start a new page without indent 4.6cm
\clearpage
\end{paracol}
\nointerlineskip
\begin{figure}[H]
\widefigure
\includegraphics[width=0.8\textwidth]{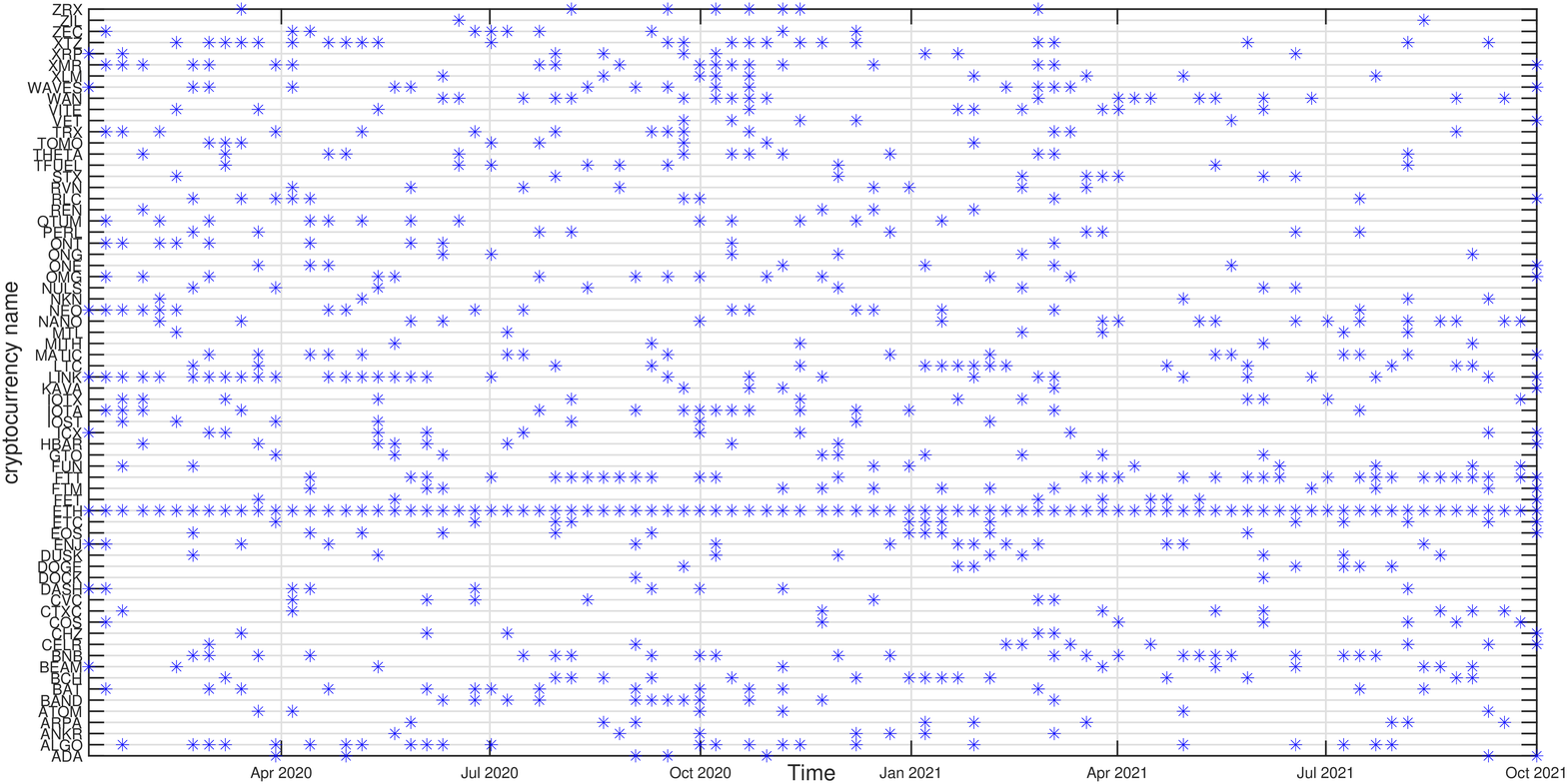}

\includegraphics[width=0.8\textwidth]{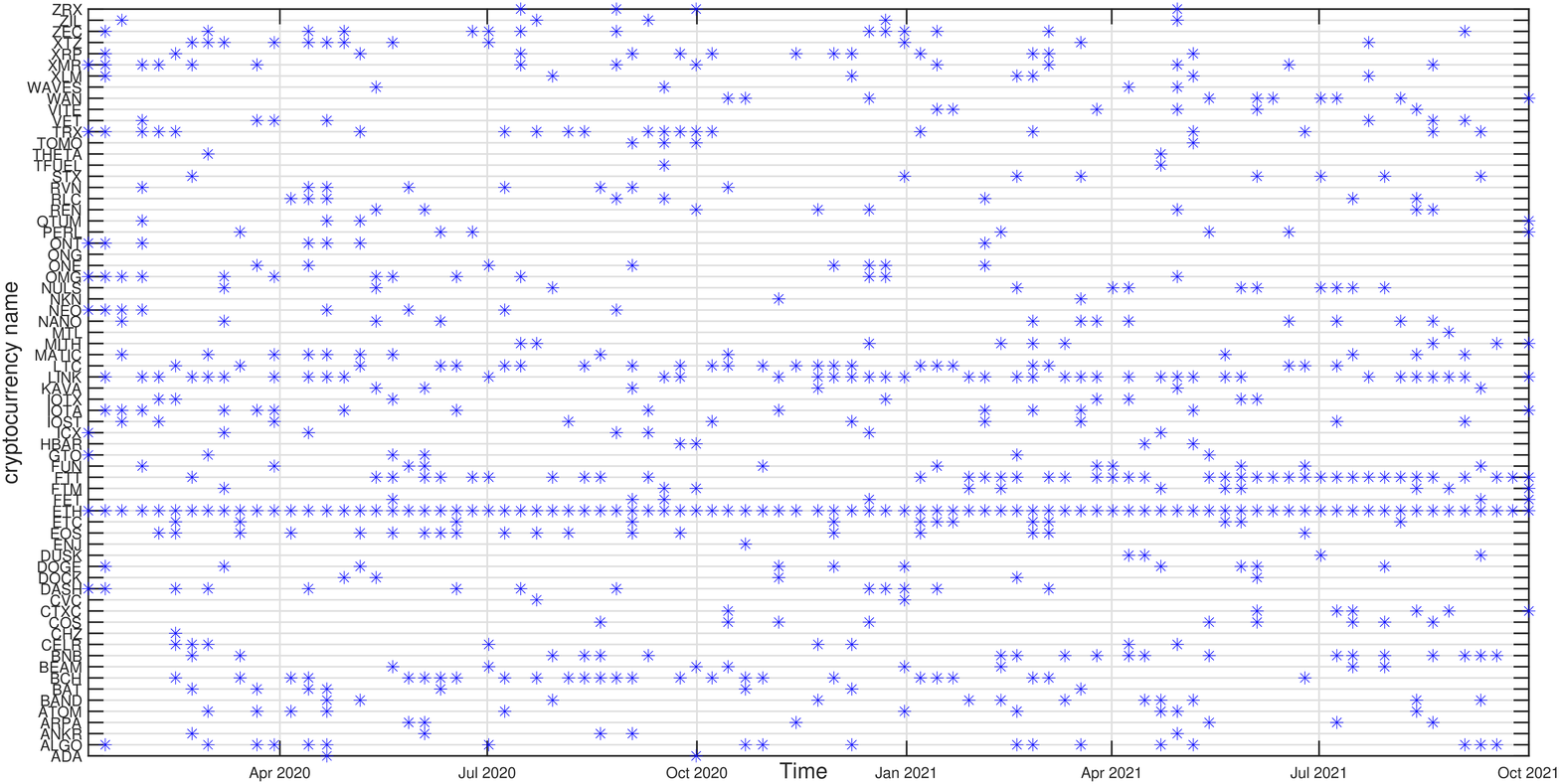}

\includegraphics[width=0.8\textwidth]{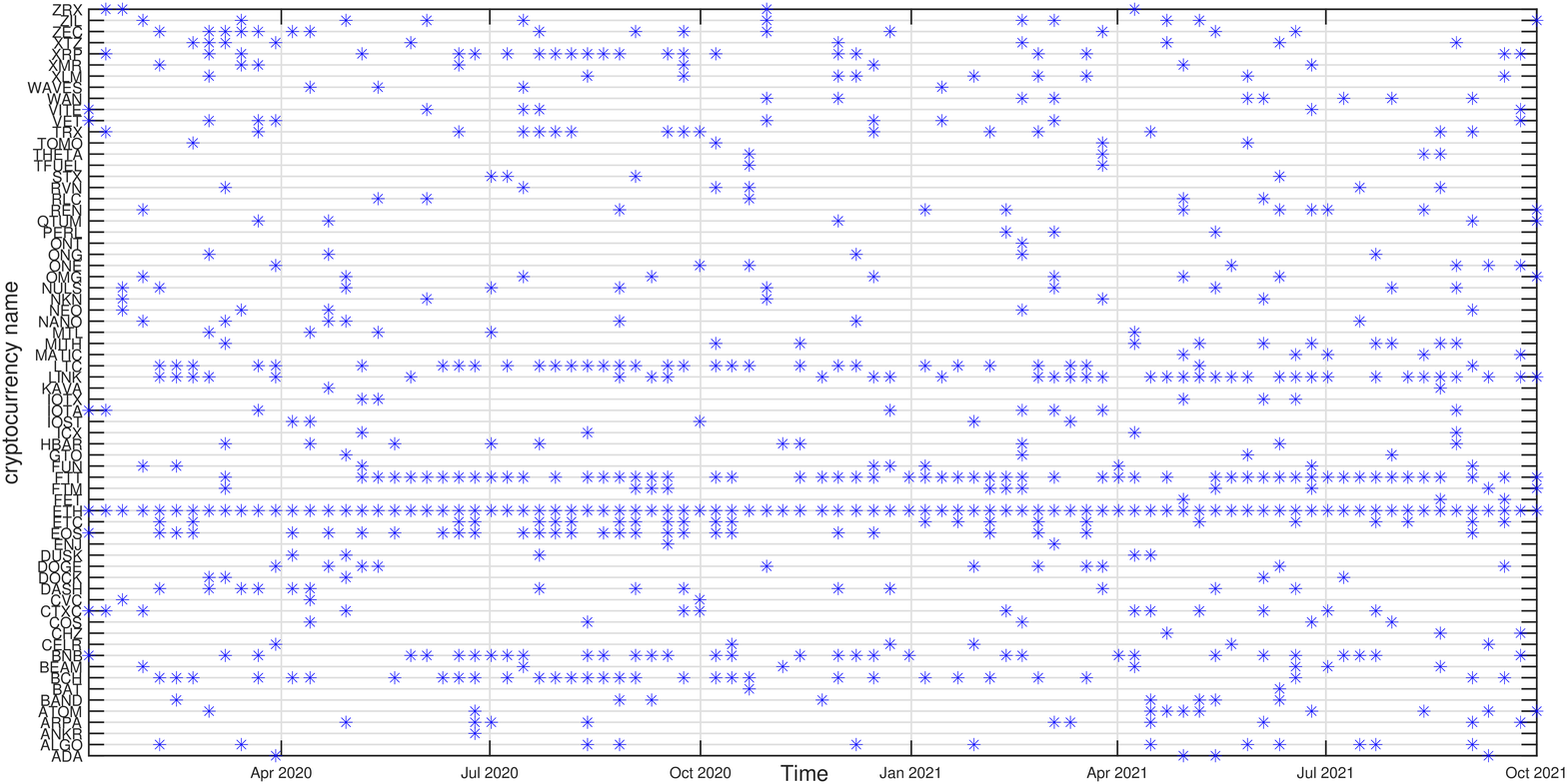}
\caption{Composition of the ETH-related cryptocurrency cluster as a function of time for sample temporal scales: $s=10$ min (top), $s=60$ min (middle), and $s=360$ min (bottom). Each point on the horizontal axis represents a non-overlapping seven-day-long moving window. Asset prices have been expressed in BTC, therefore any BTC-related contribution has been filtered out.}
\label{fig::clusters.ETH}
\end{figure}
\begin{paracol}{2}
%\linenumbers
\switchcolumn

% Figure 14

The BNB cluster can be counted among the most numerous clusters on a par with the ETH cluster. For $s=10$ min we also observe its interim growth between September 2020 and January 2021, which overlaps with the ETH growth phase. It also overlaps with the BTC cluster shrinking phase, which suggests that these events can be related with each other. No significant trends can be seen for $s=60$ min and $s=360$ min. The nodes that share the cluster with BNB most frequently are FTT and ETH.

% start a new page without indent 4.6cm
\clearpage
\end{paracol}
\nointerlineskip
\begin{figure}[H]
\widefigure
\includegraphics[width=0.8\textwidth]{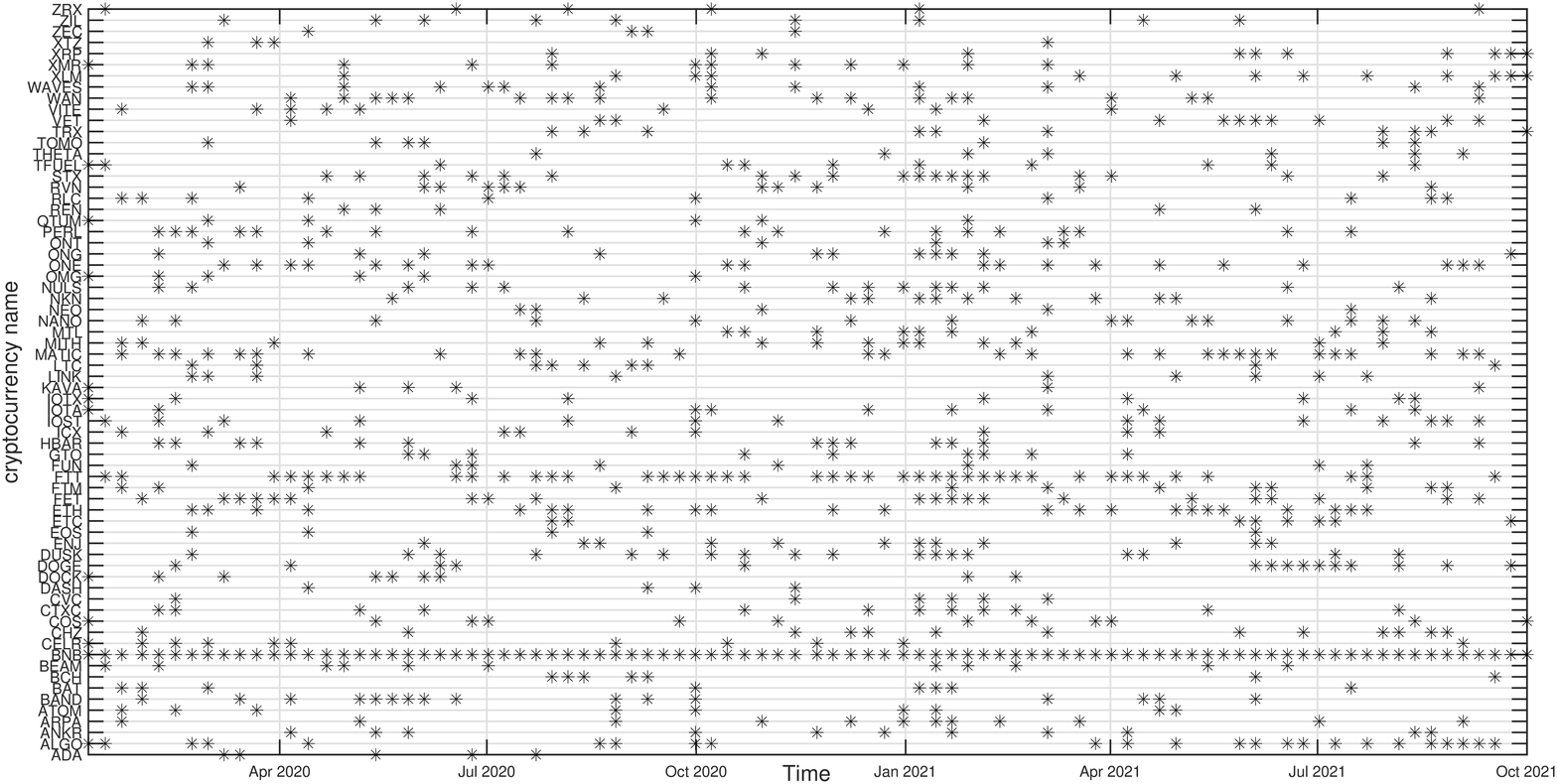}

\includegraphics[width=0.8\textwidth]{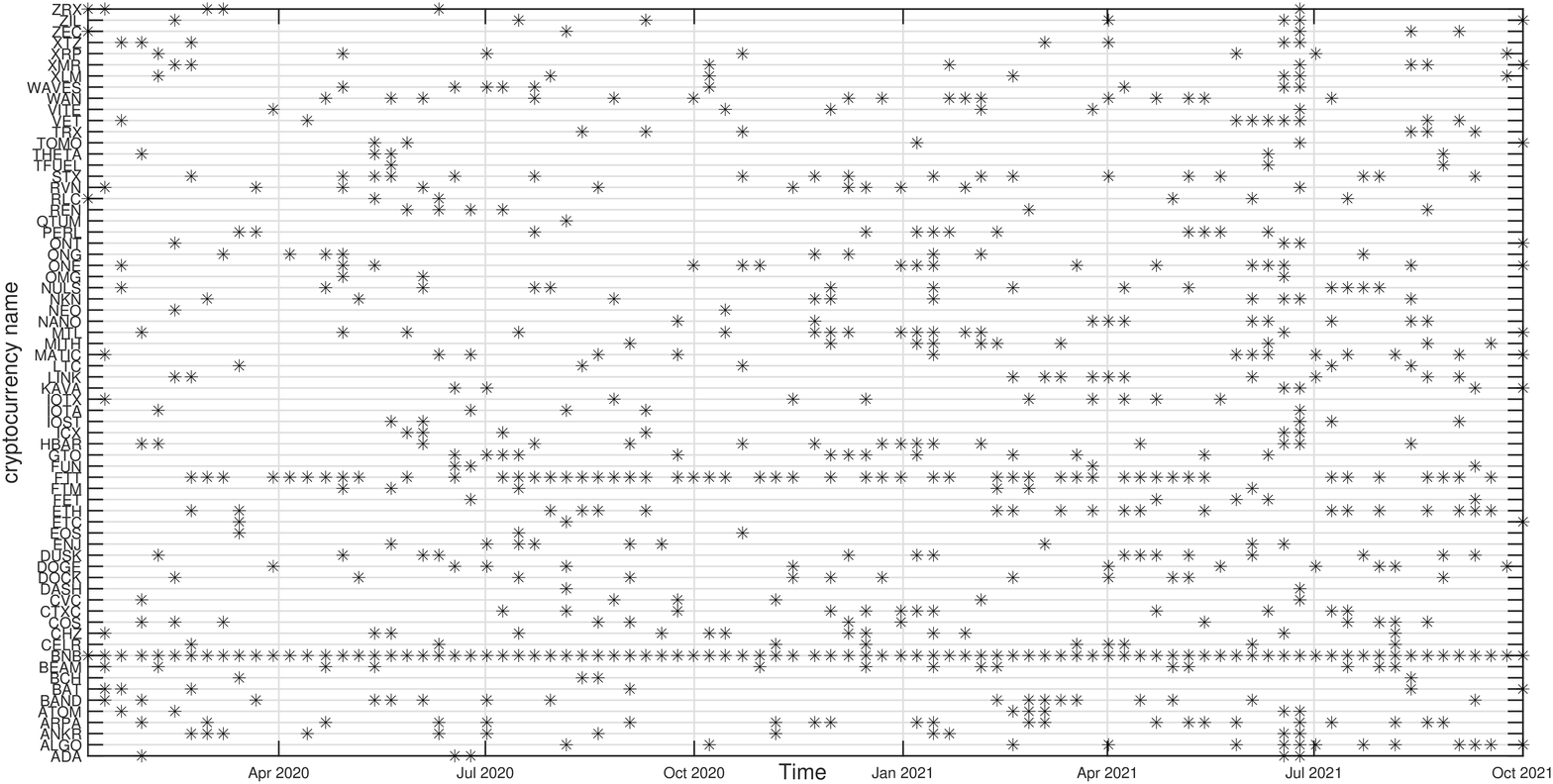}

\includegraphics[width=0.8\textwidth]{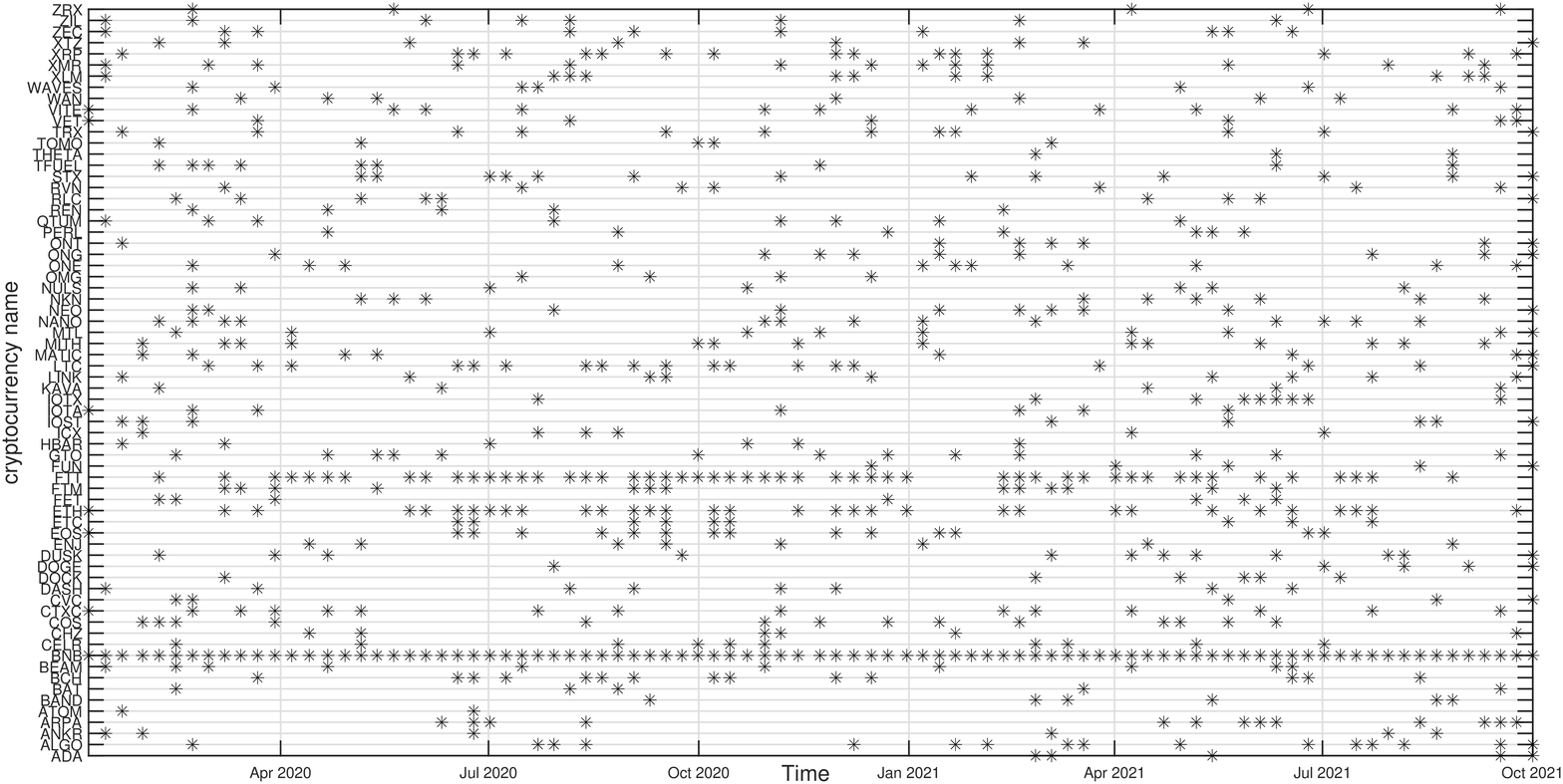}
\caption{Composition of the BNB-related cryptocurrency cluster as a function of time for sample temporal scales: $s=10$ min (top), $s=60$ min (middle), and $s=360$ min (bottom). Each point on the horizontal axis represents a non-overlapping seven-day-long moving window. Asset prices have been expressed in BTC, therefore any BTC-related contribution has been filtered out.}
\label{fig::clusters.BNB}
\end{figure}
\begin{paracol}{2}
%\linenumbers
\switchcolumn

% Figure 15

Finally, the ONT cluster also shows its specific growth phase between May and July 2021 ($s=10$ min and $s=60$ min), outside of which no significant trend can be seen. NEO and IOTA are the nodes that appear the most frequently in the same cluster with ONT. In general, Figs.~\ref{fig::clusters.BTC}--\ref{fig::clusters.ONT} show highly unstable composition of the analyzed clusters. This outcome differs from the results of some earlier studies based on data from more a distant past that reported stability of the cryptocurrency clusters (e.g.,~\cite{stosic2018}). Additionally, the identified community structure of the market differs from the result of another study, where a core-periphery structure was identified instead~\cite{polovnikov2020}.

% start a new page without indent 4.6cm
\clearpage
\end{paracol}
\nointerlineskip
\begin{figure}[H]
\widefigure
\includegraphics[width=0.8\textwidth]{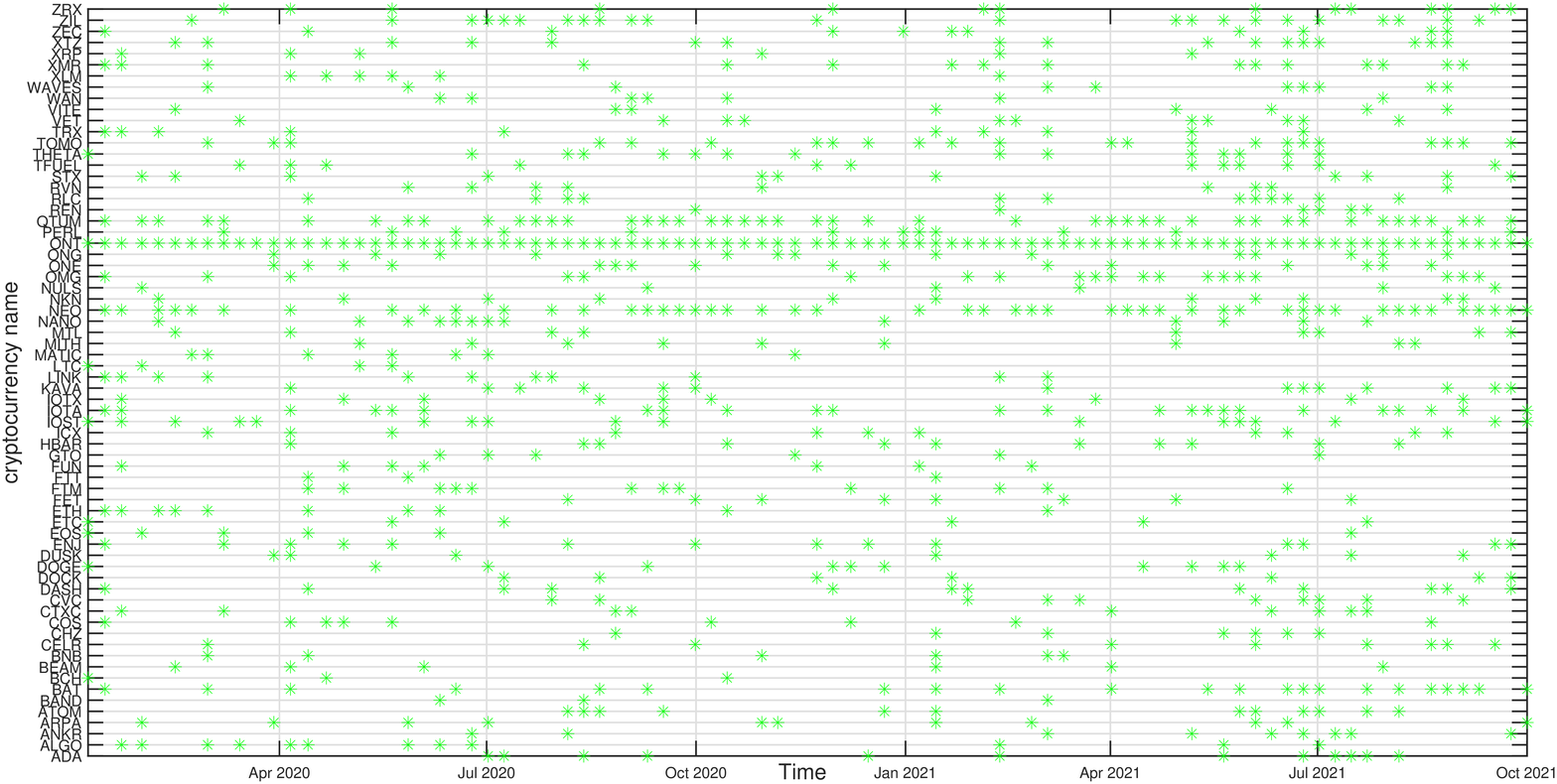}

\includegraphics[width=0.8\textwidth]{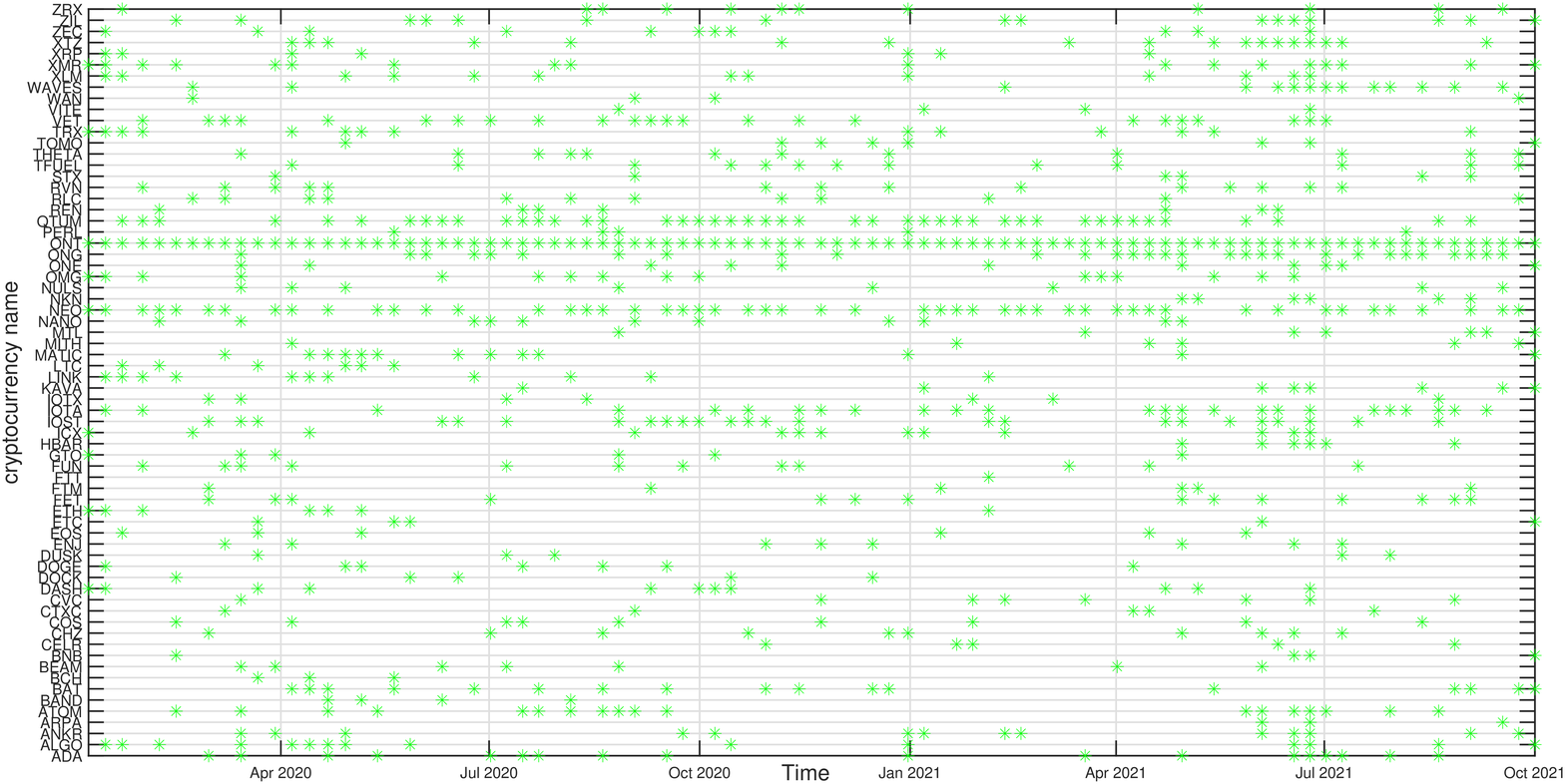}

\includegraphics[width=0.8\textwidth]{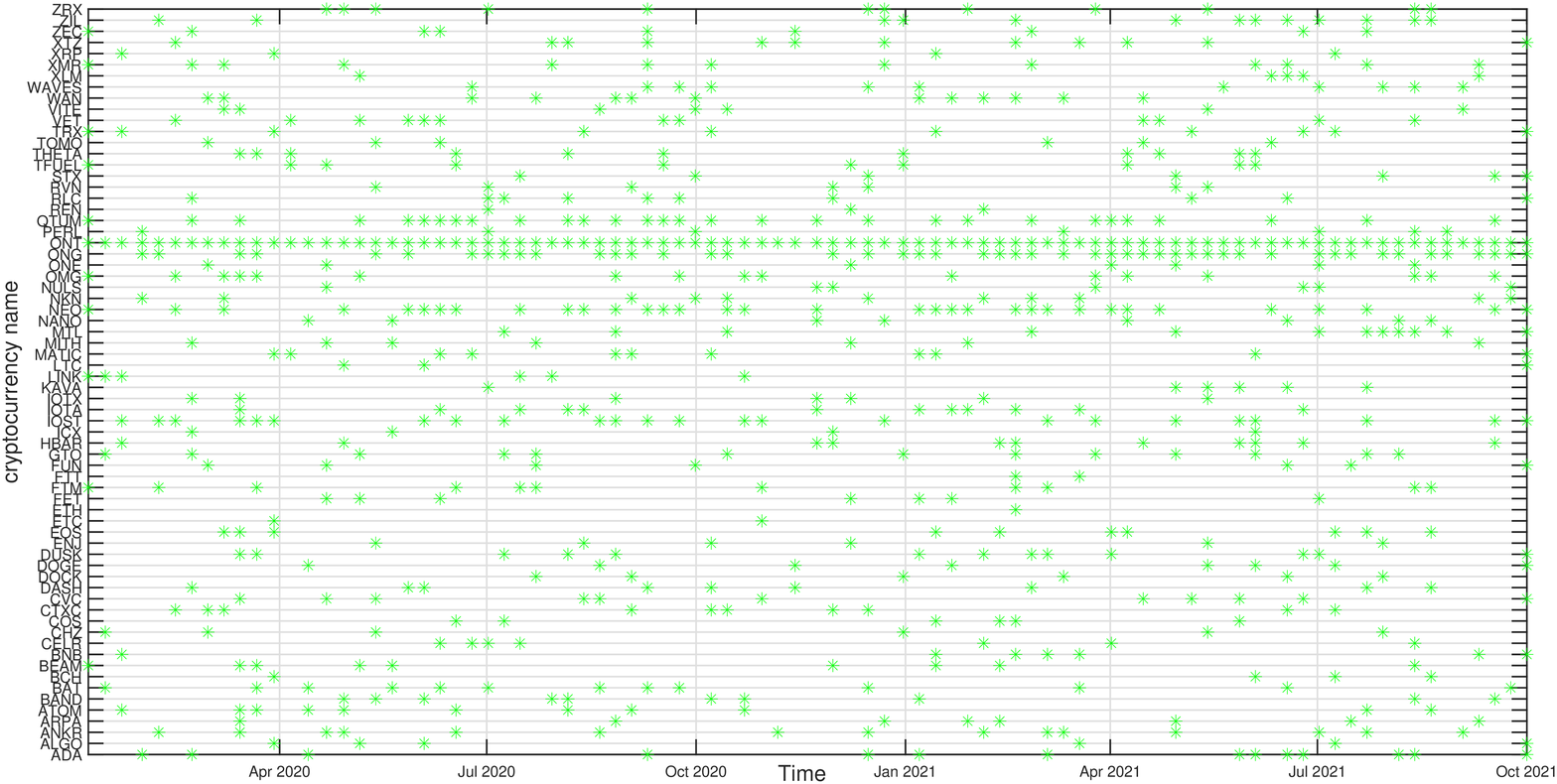}
\caption{Composition of the ONT-related cryptocurrency cluster as a function of time for sample temporal scales: $s=10$ min (top), $s=60$ min (middle), and $s=360$ min (bottom). Each point on the horizontal axis represents a non-overlapping seven-day-long moving window. Asset prices have been expressed in BTC, therefore any BTC-related contribution has been filtered out.}
\label{fig::clusters.ONT}
\end{figure}
\begin{paracol}{2}
%\linenumbers
\switchcolumn

Our discussion hitherto is focused on the simultaneous time series without delays between them. However, there is an interesting question whether the most capitalized and liquid cryptocurrencies like BTC and ETH drive the remaining ones, which can generate the delayed cross-correlations that can be observable. In order to address this question, we calculated the coefficients $\rho_q^{({\rm BTC,X})}(s,\tau)$ for all the cryptocurrency pairs (BTC,X) and (ETH,X), where X stands for any cryptocurrency other than BTC and ETH. A time lag $\tau$ that can assume two values: $\tau=-1$ min and $\tau=1$ min, defines whether the BTC (ETH) time series is advanced or lagged relatively to the second time series. For these two cases, we calculate the average coefficients $\langle \rho_q(s,\tau) \rangle$ for BTC and ETH (the averaging is carried out over all other cryptocurrencies X).

% Figure 16

Fig.~\ref{fig::BTC.ETH.lagged} shows the results for $q=1$ and $q=4$ and for the shortest scale $s=10$ min (a potential effect of 1 minute delay can be too weak to be detectable on longer scales). If the time series of the BTC returns is considered, $\langle \rho_q(s,\tau) \rangle$ is significantly larger for $\tau=0$ than for $\tau=\pm 1$. For $q=1$ the advanced BTC time series produces larger $\langle \rho_q(s,\tau) \rangle$ than the lagged one. This difference is statistically significant. For $q=4$ both shifted time series produce $\langle \rho_q(s,\tau) \rangle$ with comparable magnitude for a vast majority of windows with a few exceptions, where the advanced BTC time series produces slightly stronger cross-correlations than the lagged one does. The qualitatively similar results are obtained for the advanced and lagged ETH time series. We can therefore conclude that by shifting the time series representing BTC or ETH we still preserve some amount of the valid detrended cross-correlations. The relative dominance of the advanced ($\tau=-1$ min) time series over the lagged ($\tau=1$ min) ones suggest that the remaining part of the market absorbs information that occurred first in the price fluctuations of BTC and ETH with a time needed for this absorption being as long as a minute. An opposite process of information transfer from the less liquid cryptocurrencies to BTC and ETH cannot be detected based on our data set. It must be noted, however, that both the BTC and ETH returns exhibit a detrended autocorrelation with the length of more than 1 min. Such an autocorrelation can artificially produce the delayed detrended cross-correlations which can manifest themselves in a way similar to that observed in Fig.~\ref{fig::BTC.ETH.lagged}. We cannot therefore answer the formulated question decisively.

% start a new page without indent 4.6cm
\clearpage
\end{paracol}
\nointerlineskip
\begin{figure}[H]
\centering
\includegraphics[width=0.9\textwidth]{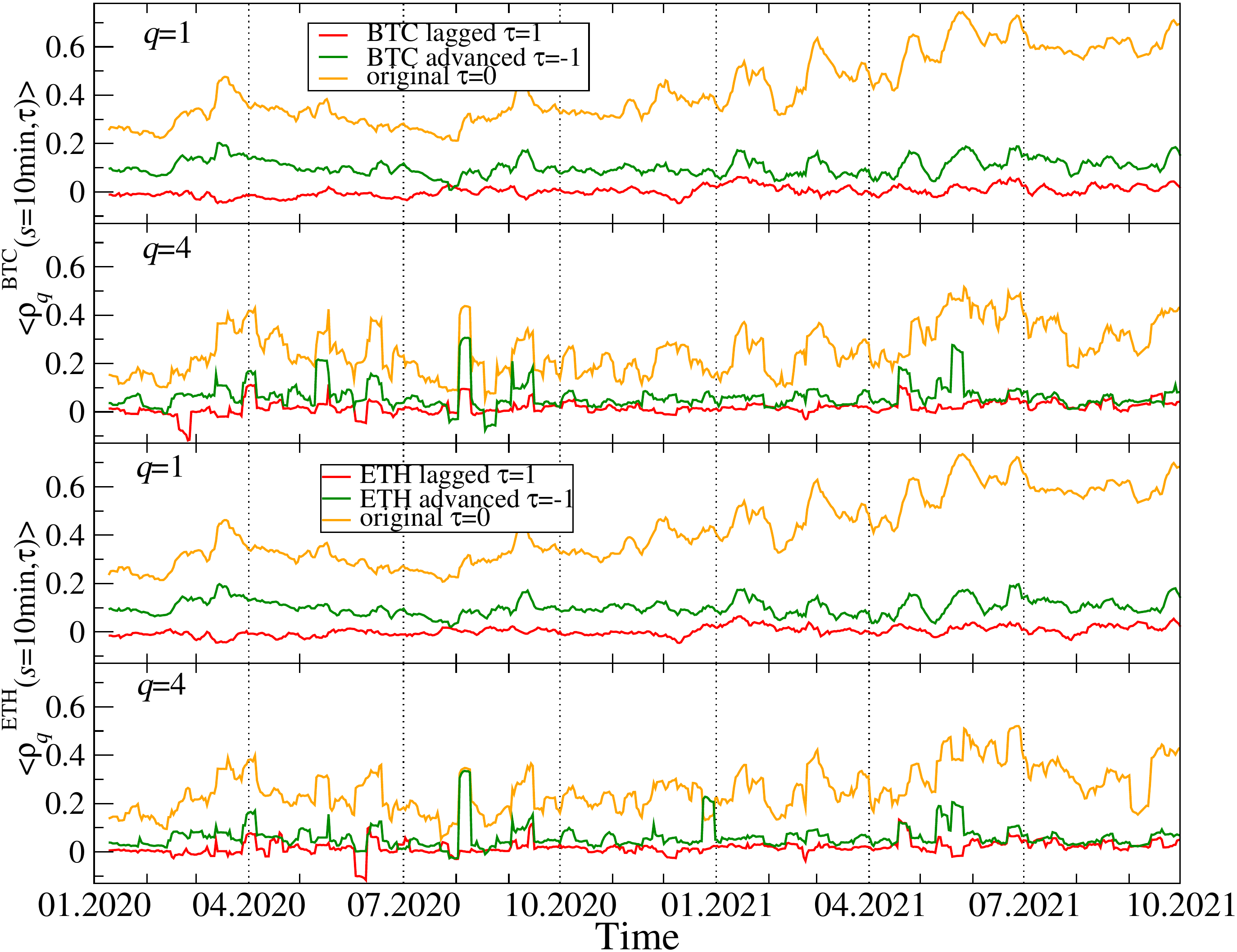}
\caption{Mean lagged $q$-dependent detrended cross-correlation coefficient $\rho_q(s,\tau)$ as a function of time after averaging over all the considered cryptocurrencies other than BTC and ETH. Time series representing BTC and ETH returns have been advanced (green) or delayed (red) by $\tau=1$ min and compared with the original non-shifted time series (orange). Two values of the filtering parameter $q$ are shown: $q=1$ (all fluctuations enter with the same weight, the first and third panels) and $q=4$ (large fluctuations are amplified, the second and fourth panels).}
\label{fig::BTC.ETH.lagged}
\end{figure}
\begin{paracol}{2}
%\linenumbers
\switchcolumn

Our former studies of the cryptocurrency market showed that, recently, it begun to be positively or negatively cross-correlated in some specific periods with the traditional financial markets like the stock market, the currency exchange market, and the commodity markets~\cite{drozdz2020a,watorek2021a}. Among such periods of the statistically significant detrended cross-correlations there was the COVID-19 pandemic in the United States: the very first case on the US territory in the end of January 2020, the first COVID-19 wave outburst in April, and the second wave development in June--July, and the subsequent pandemic slowdown, which brought the across-market rally starting in September 2020. As we have already collected more contemporary data that end in October 2021, we are able to extend our analysis of the detrended cross-correlations between the cryptocurrencies and a few other financial assets. We consider the logarithmic price returns of a few basic cryptocurrencies (BTC, ETH, DASH, EOS, and XMR), the main regular currencies (AUD, CAD, CHF, CNH, CZK, EUR, GBP, JPY, MXN, NOK, NZD, PLN, and ZAR), sample commodities (crude oil, copper, silver, and gold), and the most important stock market indices (S\&P500, NASDAQ100, Russel 2000, DJIA, FTSE, DAX, and NIKKEI). All the assets except the stock market indices are priced in US dollars (data from Dukascopy~\cite{dukascopy}).

% Figure 17

Fig.~\ref{fig::BTC-SP500.high-rho} shows the historical quotes of S\&P500 and BTC together with the distinguished periods of the elevated detrended cross-correlations inside the cryptocurrency markets. One can see that these periods are associated with specific market events that are observed in the historical data: the all-market surge at the COVID-19 pandemic onset in March-April 2020, the second pandemic wave in June--July 2020, a market rally and the following drawdowns in September--October 2020, the cryptocurrency market rally in March--April 2021 and a surge and a subsequent rally in September--October 2021. Looking from a macroscopic perspective, in all these cases the coarse-grained behaviours of S\&P500 and BTC were similar to each other at least for some period of time. 

% start a new page without indent 4.6cm
%\clearpage
\end{paracol}
\nointerlineskip
\begin{figure}[H]
\centering
\includegraphics[width=0.9\textwidth]{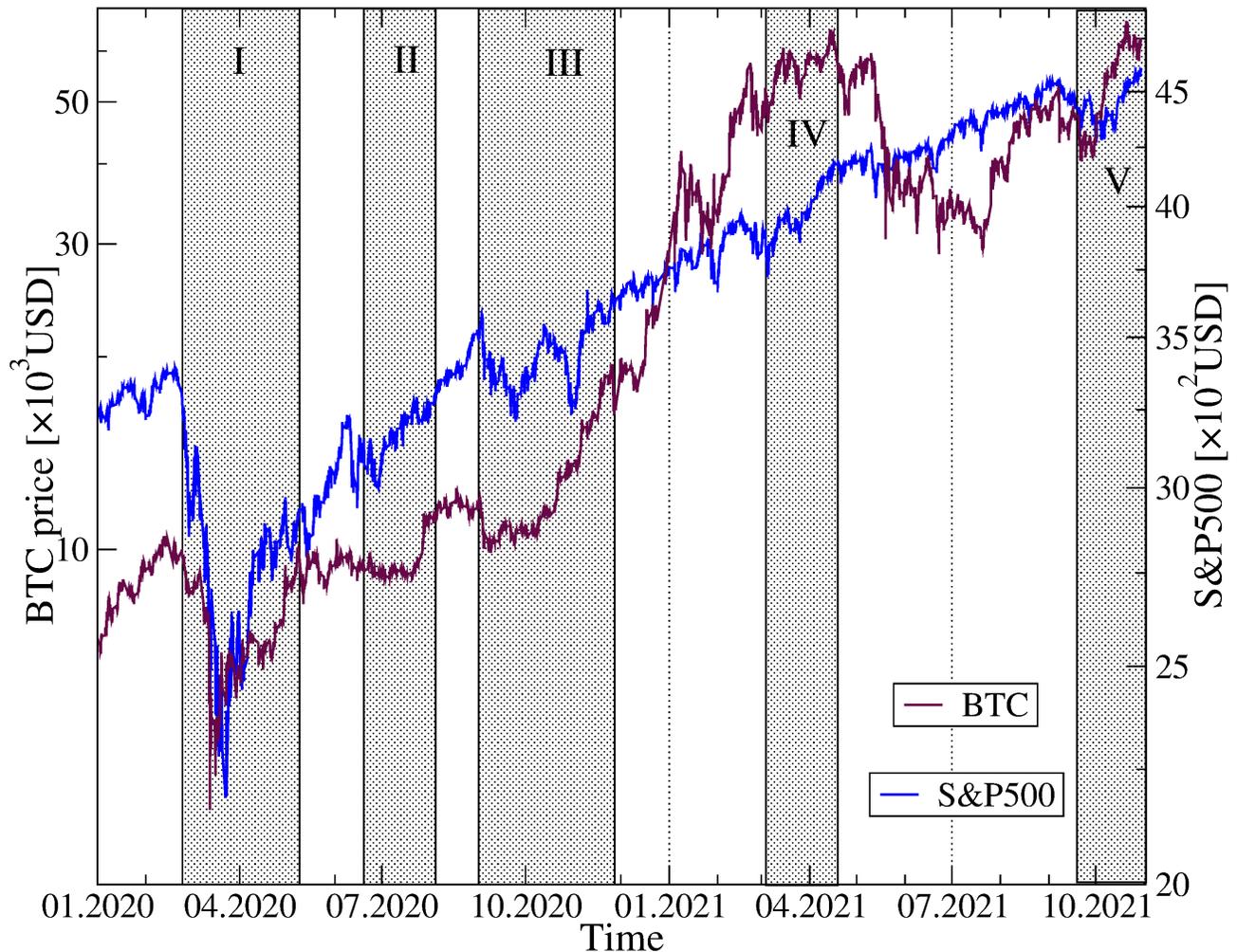}
\caption{Temporal co-evolution of BTC price in USD (maroon) and the S\&P500 index (blue) over the years 2020--2021. Periods, in which $\rho_q(s)$ calculated for these two assets exceed a threshold of 0.25 for $s=360$ min and $q=1$ (see Fig.~\ref{fig::BTC-traditional.rho}), are denoted by grey vertical strips. Specific market events are indicated by Roman numerals: I - the all-market
surge at the COVID-19 pandemic onset in March--April 2020, II - the second pandemic wave in June--July 2020, III - a market rally and the following drawdowns in September--October 2020, IV - the cryptocurrency market rally in March--April 2021, and V - a surge and a subsequent rally in September--October 2021.}
\label{fig::BTC-SP500.high-rho}
\end{figure}
\begin{paracol}{2}
%\linenumbers
\switchcolumn

% Figure 18

To inspect this issue in more detail, we calculated the $q$-dependent detrended cross-correlation coefficients for all the possible pairs of the considered assets. Before we did this, we had to concord all the time series by eliminating the gaps caused by different trading hours. The results for $q=1$ and $q=4$ and for $s=10$ min and $s=360$ min are shown in Fig.~\ref{fig::BTC-traditional.rho}. For both values of the filtering parameter $q$, the cross-correlations are stronger on the long time scale and weaker on the short one. Except for the maximum of $\rho_q(s)$ that occurred for $q=4$ and $s=10$ min in the end of June 2020, which is not present at all for $q=4$ and $s=360$ min and for $q=1$, all the other periods of the amplified cross-correlations can be observed in each case. The maxima of $\rho_q(s)$ calculated for BTC and the traditional assets occur, roughly, over the same periods than the maxima of the inner cross-correlations on the cryptocurrency market.

Different traditional assets reveal different levels of the detrended cross-correlation with BTC: the strongest correlations can be detected for S\&P500 and other stock indices, while the weaker but also significant ones for crude oil, copper, CAD and other regular currencies except for JPY and, to a much smaller extent gold. The Japanese currency is significantly anticorrelated with BTC in the periods, in which the other assets are positively cross-correlated. This means that JPY can be used for the hedging purposes while investing on the cryptocurrency market. After comparing the cross-correlation strength for $q=1$ with that for $q=4$, we may conclude that, during the large fluctuation periods, the traditional assets are less strongly cross-correlated with BTC than during the smaller fluctuation periods. They also need rather long time scales to be fully built up. What can be inferred from these results is that the detrended cross-correlations are weaker in 2021 than they used to be in 2020, but they are still stronger than the corresponding cross-correlations before the COVID-19 pandemic.

% start a new page without indent 4.6cm
%\clearpage
\end{paracol}
\nointerlineskip
\begin{figure}[H]
\centering
\includegraphics[width=0.9\textwidth]{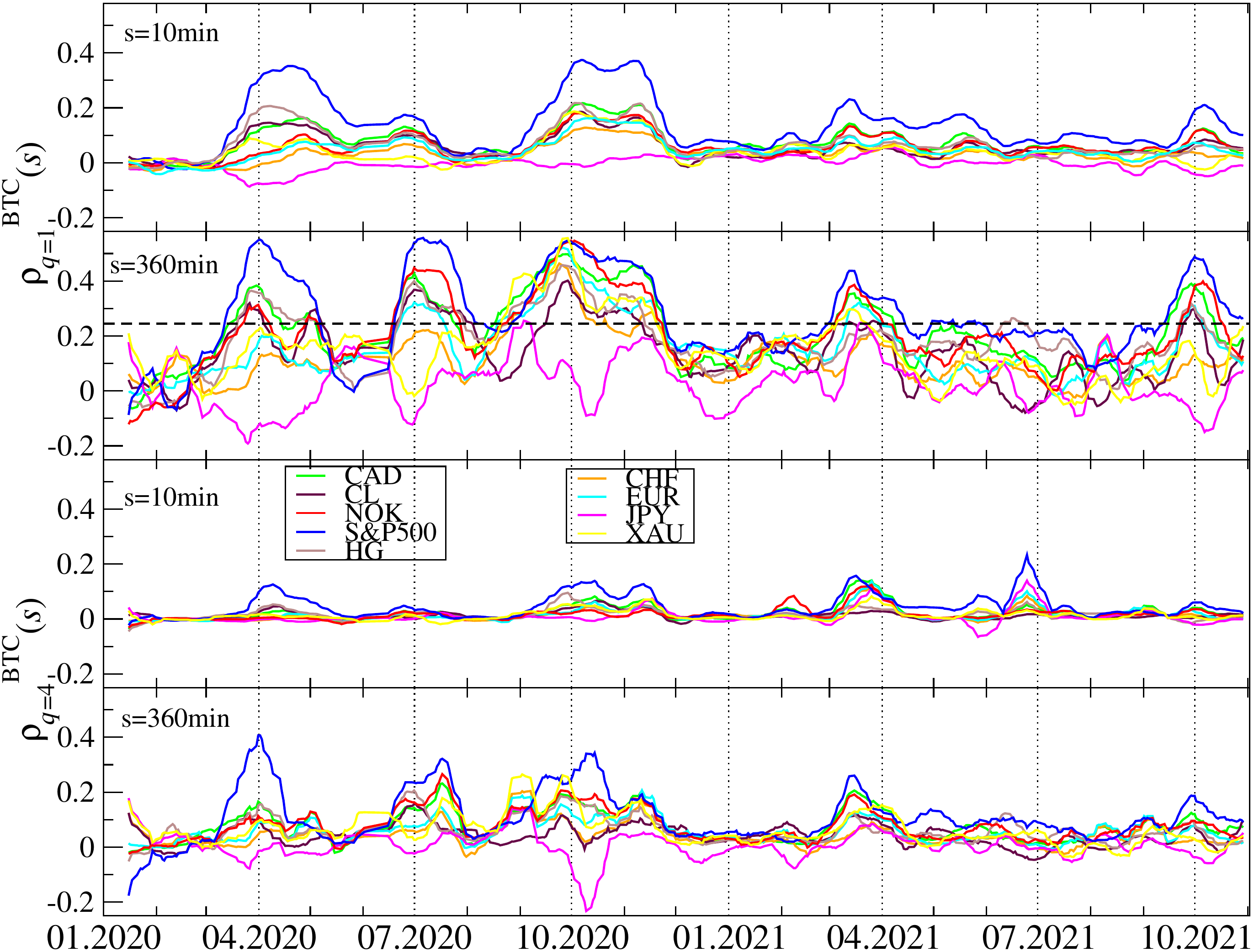}
\caption{The $q$-dependent detrended cross-correlation coefficient $\rho_q(s)$ calculated in 10-day-long moving windows with a 1-day step for BTC and the traditional market assets: the S\&P500 index (blue), crude oil price (CL, black), copper price (HG, brown), gold price (XAU, yellow), and a few regular currencies expressed in the US dollars: euro (EUR, cyan), Swiss franc (CHF, orange), Canadian dollar (CAD, light green), Japanese yen (JPY, magenta), and Norwegian krone (NOK, red). Two temporal scales $s$ ($s=10$ min in the first and third panels, and $s=360$ min in the second and fourth panels) and two filtering parameter $q$ values ($q=1$ in the first and second panels, and $q=4$ in the third and fourth panels) are shown. The horizontal dashed line at $\rho_q(s)=0.25$ in the second panel denotes a discrimination threshold applied to determine the shaded regions in Fig.~\ref{fig::BTC-SP500.high-rho}.}
\label{fig::BTC-traditional.rho}
\end{figure}
\begin{paracol}{2}
%\linenumbers
\switchcolumn

% Figure 19

Based on the coefficients $\rho_q^{(i,j)}(s)$, where $i$ and $j$ labels the cryptocurrencies and traditional assets, we created the related minimal spanning trees. A few sample trees for specific moving window positions are presented in Fig.~\ref{fig::MSTtrad}. It is easy to notice that the detrended cross-correlation strength between BTC and the traditional markets, the closest ones being the stock markets and not the currency markets is much smaller than the analogous strength among the traditional assets representing the same market type and even different market types. Topology of the MSTs is heterogeneous with both the significant hubs (S\&P500, AUD, EUR, and some cryptocurrency) and the long branches.

% start a new page without indent 4.6cm
\clearpage
\end{paracol}
\nointerlineskip
\begin{figure}[H]
\widefigure
\includegraphics[width=0.38\textwidth]{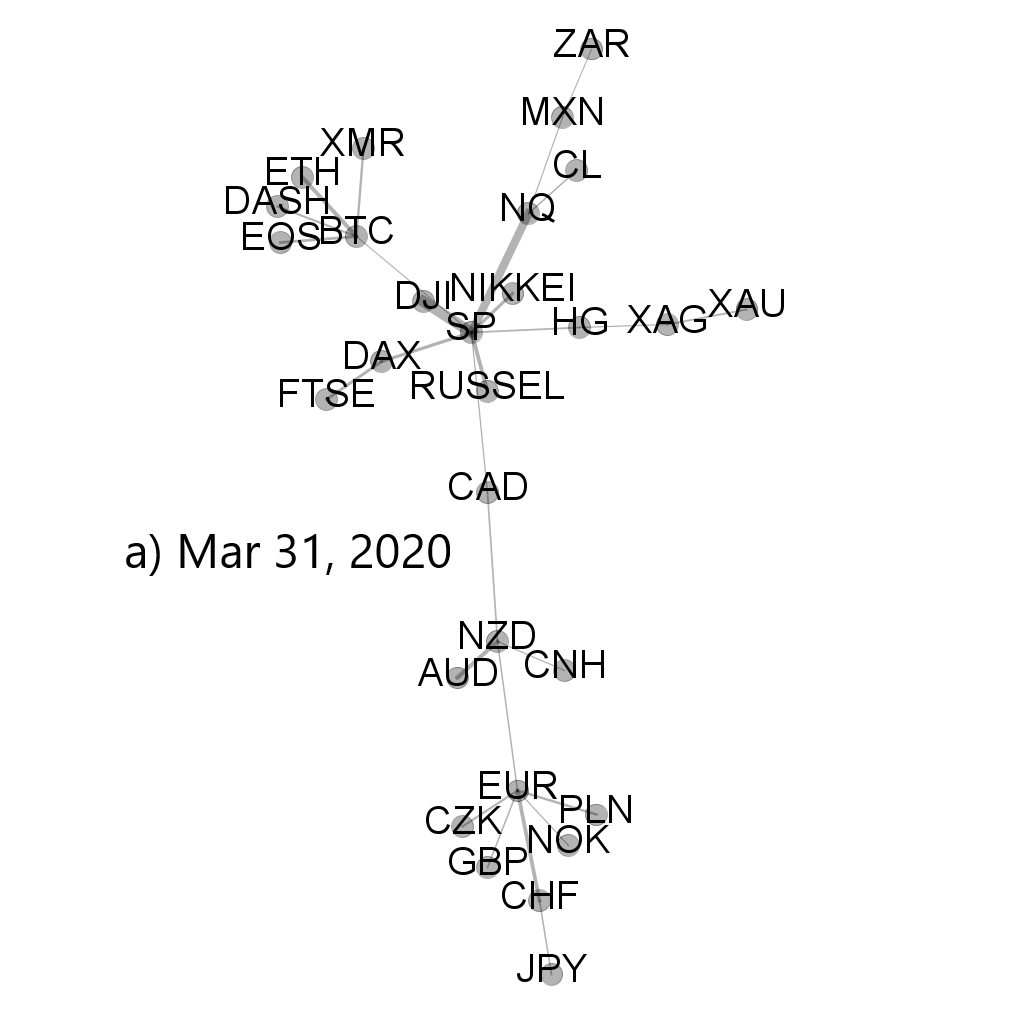}
\includegraphics[width=0.38\textwidth]{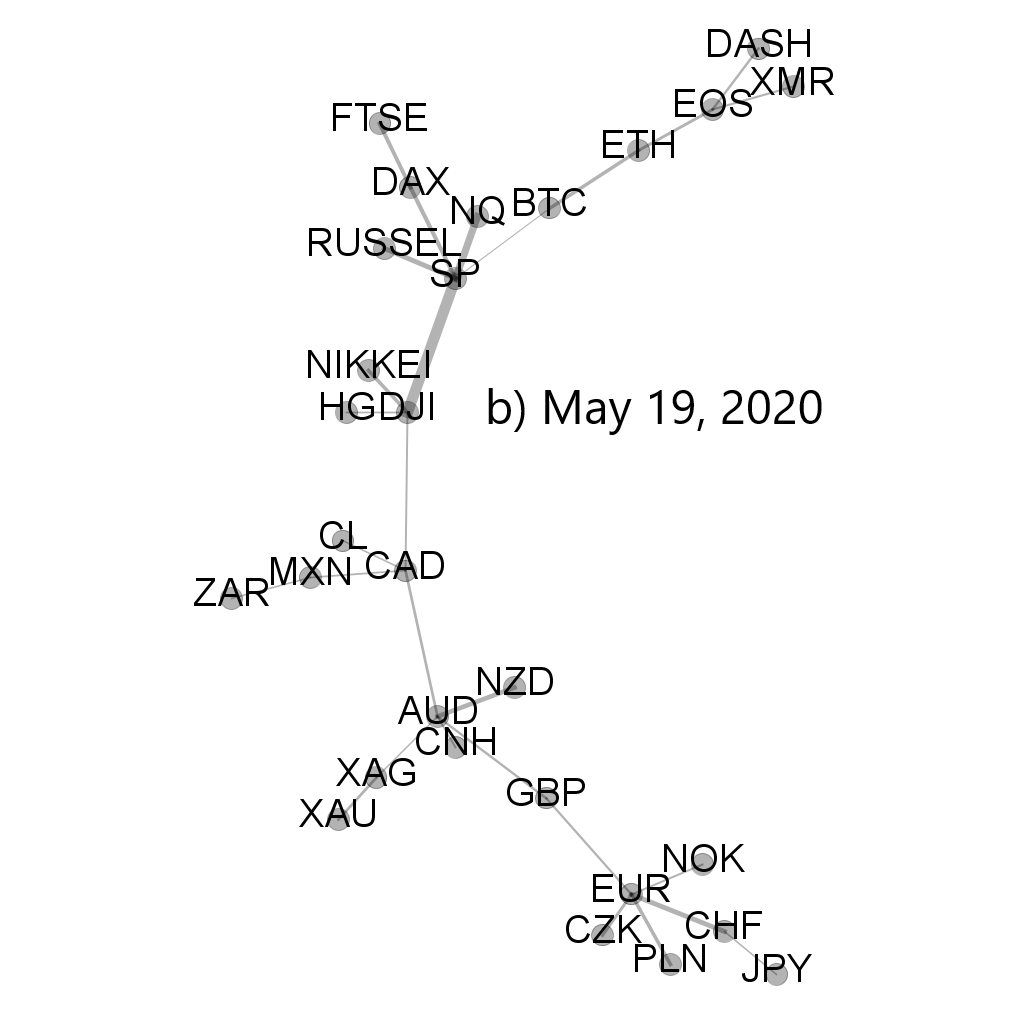}

\includegraphics[width=0.38\textwidth]{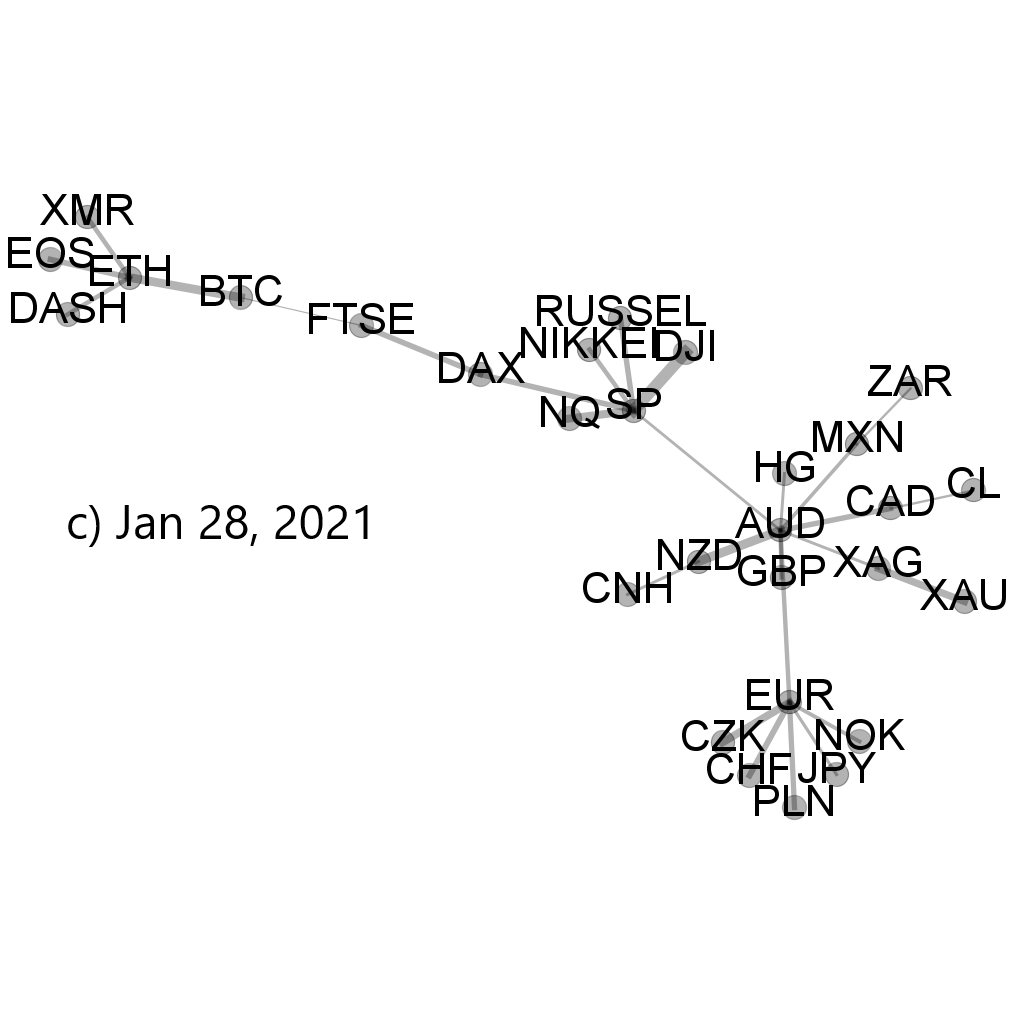}
\includegraphics[width=0.38\textwidth]{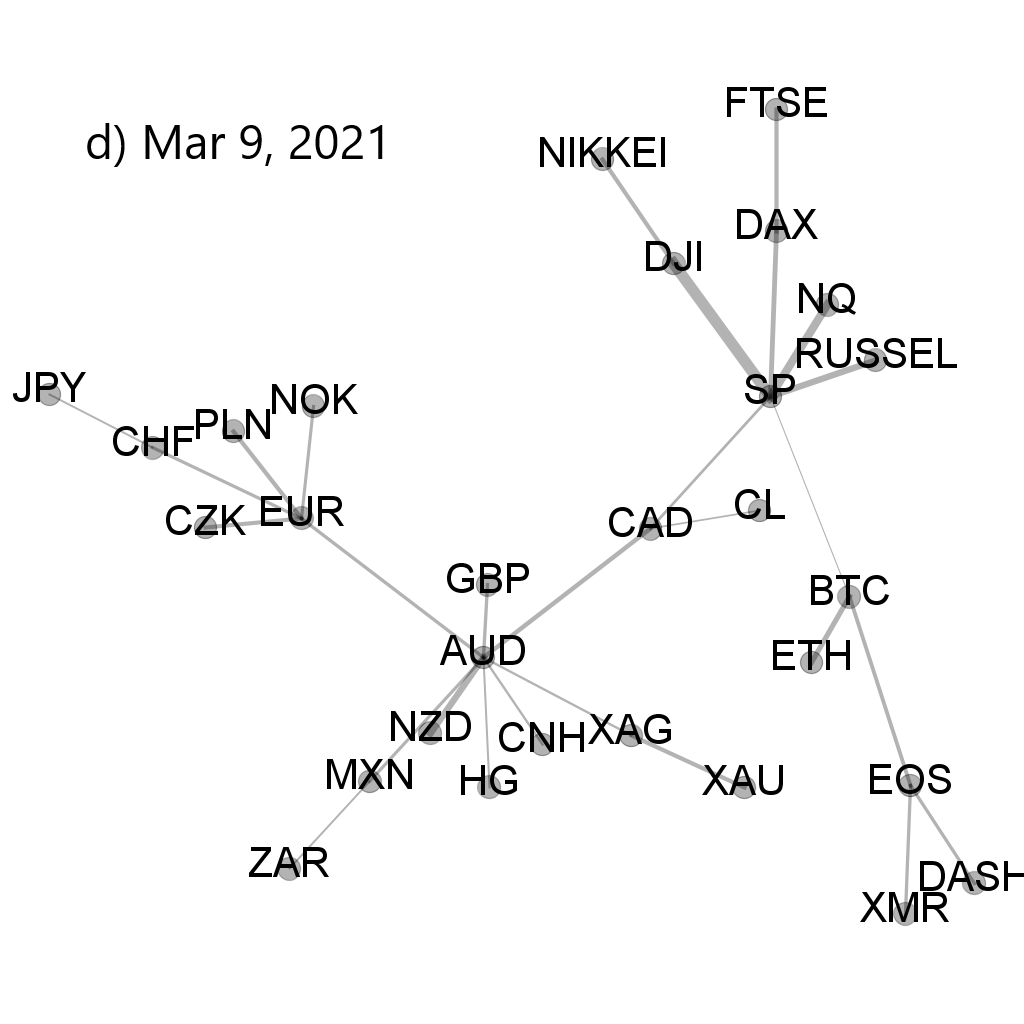}

\includegraphics[width=0.38\textwidth]{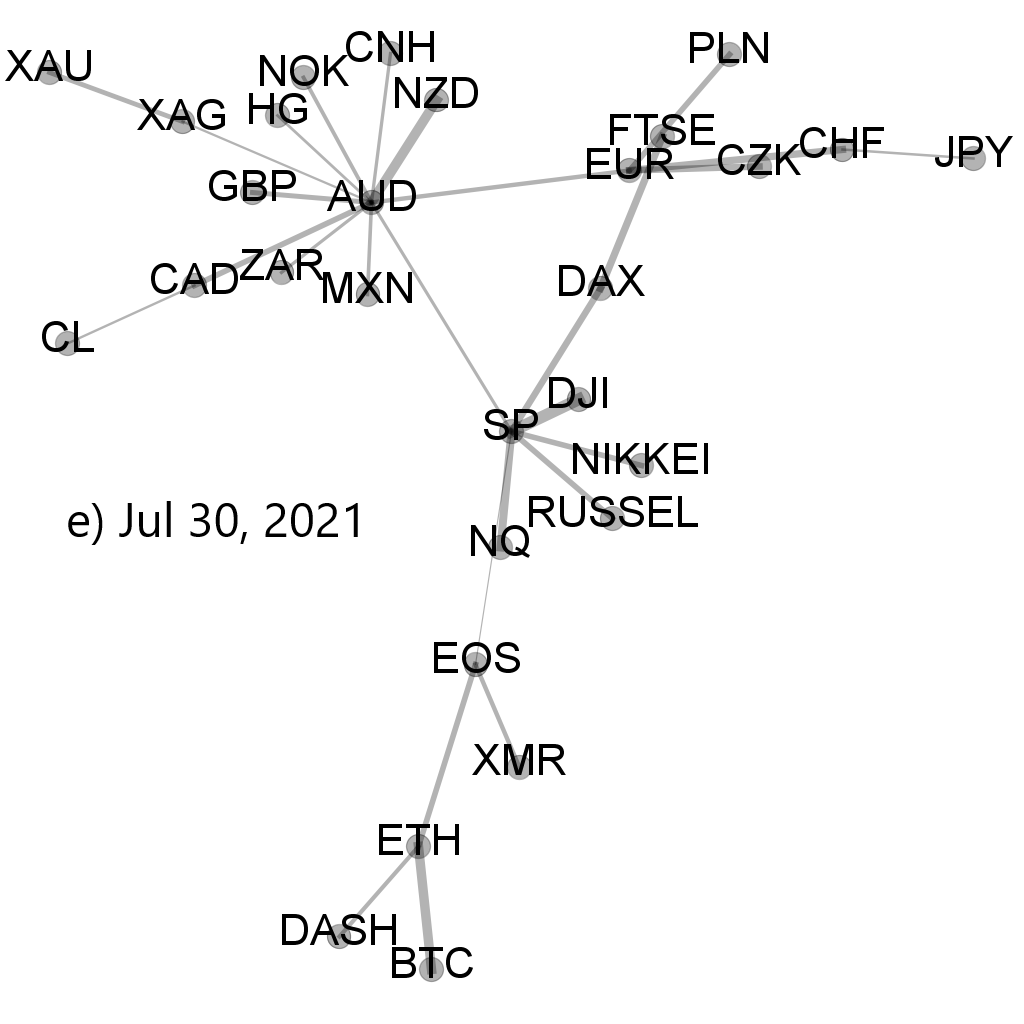}
\includegraphics[width=0.38\textwidth]{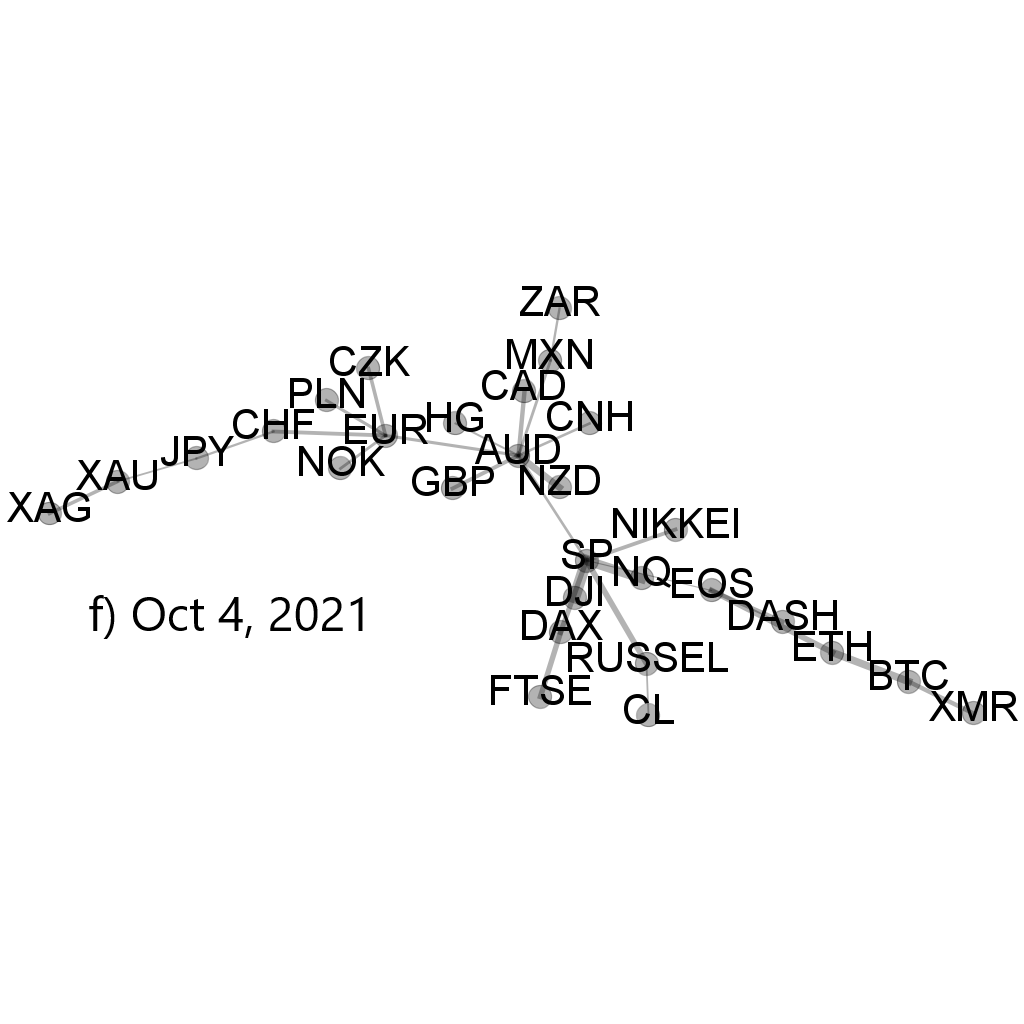}
\caption{Minimal spanning trees calculated from a distance matrix ${\bf D}_q(s)$ based on $\rho_q(s)$ for $q=1$ and $s=10$ min. The data used to create MSTs consists of cryptocurrencies (BTC, ETH, DASH, EOS, and XMR), regular currencies (AUD, GBP, NZD, MXN, ZAR, CNH, EUR, CHF, JPY, CZK, NOK, CAD, and PLN), commodities (gold - XAU, silver - XAG, copper - HG, and crude oil - CL), as well as stock market indices (S\&P500 - SP, NASDAQ100 - NQ, Russel 2000, FTSE, DAX, NIKKEI, and DJIA) in 10-day-long moving windows ended at specific dates: (a) 31 March 2020 (highly correlated markets during the pandemic onset in the United States), (b) 19 May 2020 (maximum cross-market correlations), (c) 28 January 2021 (the GameStop short squeeze related market turbulence accompanied by the cryptocurrency market decoupling), (d) 9 March 2021 (the elevated market cross-correlations), (e) 30 July 2021 (the cryptocurrencies starting a rally phase with minimum cross-market correlations), and (f) 4 October 2021 (the latest phase of the cross-market correlations).}
\label{fig::MSTtrad}
\end{figure}
\begin{paracol}{2}
%\linenumbers
\switchcolumn

\section{Conclusions}

In this paper, we studied the high-frequency time series of price returns representing 80 cryptocurrencies that were the most actively traded on the Binance platform. We focused on the detrended cross-correlation structure of the cryptocurrency market at different time intervals and calculated the $q$-dependent detrended cross-correlation coefficient $\rho_q(s)$ for all the cryptocurrency pairs and in different moving window positions. Based on these coefficients, we analyzed the spectral properties of the detrended correlation matrix and topology of the minimal spanning trees calculated from this matrix.

The main issue that has  been pointed out is that our analysis comprises only a small fraction of all traded cryptocurrencies, whose number exceeds 7500~\cite{statista}. However, the less well-known and less capitalized a cryptocurrency is, the less liquid and less reliable are the related data. This is why restricting our analysis to the most capitalized ones was crucial. Another related issue was the MST construction, and it has already been mentioned in Section~3 that the exact connectivity of the MST links is prone to noise effects, which is the most significant source of possible errors. Fortunately,   the more important these errors are, the weaker the correlations, while they are less effective if the correlations are strong (this is an issue that should  be addressed independently in  future work).

Our principal result is the observation that, over the last year, the cryptocurrency market has gradually become more compact from a topological perspective. This was achieved by the increasing market cohesion expressed by the rising average cross-correlation strength among the cryptocurrencies. Spectrally, it was manifested by the elevated magnitude of the largest correlation matrix eigenvalue $\lambda_1$ after mid-2020, as compared with the earlier periods. $\lambda_1$ is associated with an eigenvector that becomes more and more delocalised with time (as detected by the increasing entropy of its components). The largest component of this eigenvector is suppressed by the delocalisation, and its absolute value decreases significantly. These effects are observed if either the large or the small fluctuation intervals are filtered out by tuning the parameter $q$ in $\rho_q(s)$. In addition, the detrended cross-correlations saturated faster than before (small difference between $\lambda_1$ for different time scales). This is a detrended counterpart of the classic Epps effect, which describes a process of the market consolidation due to the cross-correlations among the assets~\cite{epps1979,kwapien2005,drozdz2010}.

The topological properties of the MSTs are in agreement with the outcomes of the spectral analysis and show that the market becomes more centralized with time. On the short scales, the most connected node nowadays develops  more connections to other nodes than it used to have before. The MST topology in this case is centralized and close to a star-like structure. Usually, the role of a stable central hub is played by BTC or ETH on the short time scales, but on the longer scales (e.g., an hour or longer), the hub is unstable and it frequently switches among the most liquid cryptocurrencies. The corresponding MST topology is distributed without any central hub. By increasing the scales, the mean path length also increases and it indicates that the structure for the longer time scales is more distributed and random than for the short scales. In this case, the market consolidates quickly on the short time scales (e.g., 10 min), but then the fine-grained community structure develops itself owing to the increasing cross-correlations and the average cross-correlation level rises across the network. The structure becomes less centralized, but at this point the market is already strongly coupled and compact.

We also calculated the detrended cross-correlation coefficients for BTC and some selected traditional assets like the stock market indices, commodity prices, and the regular currency exchange rates. We found that during the periods associated with the strongly correlated cryptocurrencies, the inter-market cross-correlations are also stronger than usual. Typically, the inter-market couplings rise in the periods of market instability like the COVID-19-pandemic-related events and fall in the more quiet times. However, even in such periods, the cryptocurrency market is more independent from the other markets than those markets are independent among themselves. As the pandemic becomes a normal component of our reality, the cross-correlations between the cryptocurrency market and the other markets tend to decrease, but this process is more prolonged now than the opposite process that occurred suddenly in early 2020. It is an open issue now whether the cryptocurrencies will at some point return to be an entirely independent market or the correlations that can occur from time to time will remain observable.

The main issue that has to be pointed out is that our analysis comprises only a small fraction of all traded cryptocurrencies, whose number exceeds 7500~\cite{statista}. However, the less well-known and less capitalized a cryptocurrency is, the less liquid and less reliable are the related data. This is why restricting our analysis to the most capitalized ones was crucial. Another issue is related to the MST construction, which has already been mentioned in Section~3: the exact connectivity of the MST links is prone to noise effects, which is the most significant source of possible errors. Fortunately, these errors are the more important, the weaker are the correlations, while they are less effective if the correlations are strong (this is an issue that should independently be addressed in future work).

\vspace{6pt} 

\authorcontributions{Conceptualization, S.D. and M.W.; methodology, S.D., J.K., and M.W.; software, M.W.; validation, S.D., J.K., and M.W.; formal analysis, M.W.; investigation, S.D, J.K., and M.W.; resources, M.W.; data curation, M.W.; writing--original draft preparation, J.K.; writing--review and editing, J.K. and M.W.; visualization, M.W.; supervision, S.D. and J.K. All authors have read and agreed to the published version of the manuscript.}

\conflictsofinterest{The authors declare no conflict of interest.} 

\appendixtitles{no}

%\appendixstart
\appendix
\section{}
\label{sect::app}

\begin{table}[H]
\centering
\caption{List of tickers from Dukascopy and Binance.}
\begin{tabular}{ll|llll|}
\hline
\multicolumn{2}{|c|}{\textbf{Dukascopy}} & \multicolumn{4}{c|}{\textbf{Binance}} \\ \hline
\multicolumn{1}{|l|}{\textbf{Ticker}} & \textbf{Name} & \multicolumn{1}{l|}{\textbf{Ticker}} & \multicolumn{1}{l|}{\textbf{Name}} & \multicolumn{1}{l|}{\textbf{Ticker}} & \textbf{Name} \\ \hline
\multicolumn{1}{|l|}{BTC} & bitcoin & \multicolumn{1}{l|}{BTC} & \multicolumn{1}{l|}{bitcoin} & \multicolumn{1}{l|}{LINK} & chainlink \\ \hline
\multicolumn{1}{|l|}{ETH} & ethereum & \multicolumn{1}{l|}{ADA} & \multicolumn{1}{l|}{cardano} & \multicolumn{1}{l|}{LTC} & litecoin \\ \hline
\multicolumn{1}{|l|}{DASH} & dash & \multicolumn{1}{l|}{ALGO} & \multicolumn{1}{l|}{algorand} & \multicolumn{1}{l|}{MATIC} & polygon \\ \hline
\multicolumn{1}{|l|}{EOS} & eos & \multicolumn{1}{l|}{ANKR} & \multicolumn{1}{l|}{ankr} & \multicolumn{1}{l|}{MFT} & hifi finance \\ \hline
\multicolumn{1}{|l|}{XMR} & monero & \multicolumn{1}{l|}{ARPA} & \multicolumn{1}{l|}{arpa chain} & \multicolumn{1}{l|}{MITH} & mithril \\ \hline
\multicolumn{1}{|l|}{AUD} & Australian dollar & \multicolumn{1}{l|}{ATOM} & \multicolumn{1}{l|}{cosmos} & \multicolumn{1}{l|}{MTL} & metal \\ \hline
\multicolumn{1}{|l|}{EUR} & euro & \multicolumn{1}{l|}{BAND} & \multicolumn{1}{l|}{band protocol} & \multicolumn{1}{l|}{NANO} & nano \\ \hline
\multicolumn{1}{|l|}{GBP} & British pound & \multicolumn{1}{l|}{BAT} & \multicolumn{1}{l|}{basic atention token} & \multicolumn{1}{l|}{NEO} & neo \\ \hline
\multicolumn{1}{|l|}{NZD} & New Zealand dollar & \multicolumn{1}{l|}{BCH} & \multicolumn{1}{l|}{bitcoin cash} & \multicolumn{1}{l|}{NKN} & nkn \\ \hline
\multicolumn{1}{|l|}{CAD} & Canadian dollar & \multicolumn{1}{l|}{BEAM} & \multicolumn{1}{l|}{beam} & \multicolumn{1}{l|}{NULS} & nuls \\ \hline
\multicolumn{1}{|l|}{CHF} & Swiss franc & \multicolumn{1}{l|}{BNB} & \multicolumn{1}{l|}{binance coin} & \multicolumn{1}{l|}{OMG} & omg network \\ \hline
\multicolumn{1}{|l|}{CNH} & offshore renminbi & \multicolumn{1}{l|}{BTT} & \multicolumn{1}{l|}{bittorrent} & \multicolumn{1}{l|}{ONE} & harmony \\ \hline
\multicolumn{1}{|l|}{CZK} & Czech krone & \multicolumn{1}{l|}{BUSD} & \multicolumn{1}{l|}{binance USD} & \multicolumn{1}{l|}{ONG} & ontology gas \\ \hline
\multicolumn{1}{|l|}{JPY} & Japanese yen & \multicolumn{1}{l|}{CELR} & \multicolumn{1}{l|}{celer network} & \multicolumn{1}{l|}{ONT} & ontology \\ \hline
\multicolumn{1}{|l|}{MXN} & Mexican peso & \multicolumn{1}{l|}{CHZ} & \multicolumn{1}{l|}{chiliz} & \multicolumn{1}{l|}{PAX} & pax dollar \\ \hline
\multicolumn{1}{|l|}{NOK} & Norwegian krone & \multicolumn{1}{l|}{COS} & \multicolumn{1}{l|}{contentos} & \multicolumn{1}{l|}{PERL} & perl \\ \hline
\multicolumn{1}{|l|}{PLN} & Polish zloty & \multicolumn{1}{l|}{CTXC} & \multicolumn{1}{l|}{cortex} & \multicolumn{1}{l|}{QTUM} & qtum \\ \hline
\multicolumn{1}{|l|}{ZAR} & South African rand & \multicolumn{1}{l|}{CVC} & \multicolumn{1}{l|}{civic} & \multicolumn{1}{l|}{REN} & ren \\ \hline
\multicolumn{1}{|l|}{NIKKEI} & Nikkei 225 & \multicolumn{1}{l|}{DASH} & \multicolumn{1}{l|}{dash} & \multicolumn{1}{l|}{RLC} & iexec \\ \hline
\multicolumn{1}{|l|}{RUSSEL} & Russell 2000 & \multicolumn{1}{l|}{DENT} & \multicolumn{1}{l|}{dent} & \multicolumn{1}{l|}{RVN} & ravencoin \\ \hline
\multicolumn{1}{|l|}{DAX} & DAX 30 & \multicolumn{1}{l|}{DOCK} & \multicolumn{1}{l|}{dock} & \multicolumn{1}{l|}{STX} & stacks \\ \hline
\multicolumn{1}{|l|}{FTSE} & FTSE 100 & \multicolumn{1}{l|}{DOGE} & \multicolumn{1}{l|}{dogecoin} & \multicolumn{1}{l|}{TFUEL} & theta fuel \\ \hline
\multicolumn{1}{|l|}{DJI} & Dow Jones Industrial Average & \multicolumn{1}{l|}{DUSK} & \multicolumn{1}{l|}{dusk network} & \multicolumn{1}{l|}{THETA} & theta \\ \hline
\multicolumn{1}{|l|}{SP} & S\&P 500 & \multicolumn{1}{l|}{ENJ} & \multicolumn{1}{l|}{enj coin} & \multicolumn{1}{l|}{TOMO} & tomochain \\ \hline
\multicolumn{1}{|l|}{NQ} & NASDAQ 100 & \multicolumn{1}{l|}{EOS} & \multicolumn{1}{l|}{eos} & \multicolumn{1}{l|}{TROY} & troy \\ \hline
\multicolumn{1}{|l|}{XAG} & silver & \multicolumn{1}{l|}{ETC} & \multicolumn{1}{l|}{ethereum classic} & \multicolumn{1}{l|}{TRX} & tron \\ \hline
\multicolumn{1}{|l|}{XAU} & gold & \multicolumn{1}{l|}{ETH} & \multicolumn{1}{l|}{ethereum} & \multicolumn{1}{l|}{TUSD} & trueusd \\ \hline
\multicolumn{1}{|l|}{HG} & high-grade copper & \multicolumn{1}{l|}{FET} & \multicolumn{1}{l|}{fetch} & \multicolumn{1}{l|}{USDC} & USD coin \\ \hline
\multicolumn{1}{|l|}{CL} & crude oil & \multicolumn{1}{l|}{FTM} & \multicolumn{1}{l|}{fantom} & \multicolumn{1}{l|}{VET} & vechain \\ \hline
\multicolumn{1}{|l|}{} &  & \multicolumn{1}{l|}{FTT} & \multicolumn{1}{l|}{ftx token} & \multicolumn{1}{l|}{VITE} & vite \\ \hline
\multicolumn{1}{|l|}{} &  & \multicolumn{1}{l|}{FUN} & \multicolumn{1}{l|}{funtoken} & \multicolumn{1}{l|}{WAN} & wanchain \\ \hline
 &  & \multicolumn{1}{l|}{GTO} & \multicolumn{1}{l|}{gifto} & \multicolumn{1}{l|}{WAVES} & waves \\ \cline{3-6} 
 &  & \multicolumn{1}{l|}{HBAR} & \multicolumn{1}{l|}{hedera} & \multicolumn{1}{l|}{WIN} & winklink \\ \cline{3-6} 
 &  & \multicolumn{1}{l|}{HOT} & \multicolumn{1}{l|}{holo} & \multicolumn{1}{l|}{XLM} & stellar \\ \cline{3-6} 
 &  & \multicolumn{1}{l|}{ICX} & \multicolumn{1}{l|}{icon} & \multicolumn{1}{l|}{XMR} & ripple \\ \cline{3-6} 
 &  & \multicolumn{1}{l|}{IOST} & \multicolumn{1}{l|}{iost} & \multicolumn{1}{l|}{XRP} & monero \\ \cline{3-6} 
 &  & \multicolumn{1}{l|}{IOTA} & \multicolumn{1}{l|}{miota} & \multicolumn{1}{l|}{XTZ} & tezos \\ \cline{3-6} 
 &  & \multicolumn{1}{l|}{IOTX} & \multicolumn{1}{l|}{iotex} & \multicolumn{1}{l|}{ZEC} & zcash \\ \cline{3-6} 
 &  & \multicolumn{1}{l|}{KAVA} & \multicolumn{1}{l|}{kava} & \multicolumn{1}{l|}{ZIL} & zilliqa \\ \cline{3-6} 
 &  & \multicolumn{1}{l|}{KEY} & \multicolumn{1}{l|}{key} & \multicolumn{1}{l|}{ZRX} & 0x \\ \cline{3-6} 
\end{tabular}
\label{tab::ticker_list}
\end{table}

\vspace{6pt}

\end{paracol}
\reftitle{References}

\end{document}